\PassOptionsToPackage{hyphens}{url}
\documentclass[letterpaper,twocolumn,10pt]{article}

\usepackage{usenix2020_SOUPS}

\usepackage{graphicx}
\usepackage{tikz,pgfplots,pgfplotstable} 
\pgfplotsset{compat=1.7} 

\usepackage{xcolor}

\usepackage{amsmath}
\usepackage{booktabs}
\usepackage{multirow}
\usepackage{subcaption}
\usepackage{enumitem}

\usepackage{balance}       
\usepackage{graphics}      
\usepackage[T1]{fontenc}   
\usepackage{txfonts}
\usepackage{mathptmx}
\usepackage{cleveref}
\usepackage{color}
\usepackage{booktabs}
\usepackage{textcomp}
\usepackage{xspace}

\usepackage{microtype}        
\usepackage{ccicons}          

\hypersetup{%
  pdflang={en-US},
  pdftitle={Knock, Knock. Who's There? On the Security of LG's Knock Codes},
  pdfauthor={Raina Samuel, Philipp Markert, Adam J. Aviv, Iulian Neamtiu},
  pdfsubject={Knock, Knock. Who's There?},
  pdfkeywords={Authentication, Knock Code, Blocklist, Mobile, Smartphone, Security, Usability},
  pdfdisplaydoctitle=true, 
  pdfstartview={FitH},
  breaklinks
}

\newcommand{\new}[1]{\textcolor{black}{#1}}

\newcommand{\shepherd}[1]{\textcolor{black}{#1}}
\newcommand{\shep}[1]{\textcolor{black}{#1}}

\def\control{control 2x2}
\def\controlshort{con-2x2}
\def\blocklist{blocklist informed 2x2}
\def\blocklistshort{bla-2x2}
\def\big{larger 2x3}
\def\bigshort{big-2x3}

\usepackage{pgffor} 
\usepackage{xstring} 
\newcommand{\knockcodesmall}[1]
{\hspace{-.4em}
    \foreach \knock in {#1} 
    {
        \IfEqCase{\knock}{%
            {0}{\includegraphics[scale=0.03]{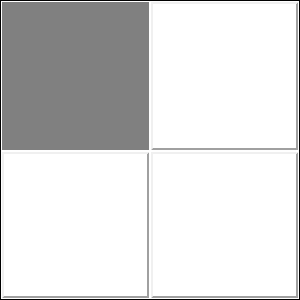}}%
            {1}{\includegraphics[scale=0.03]{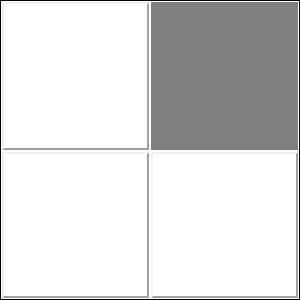}}%
            {2}{\includegraphics[scale=0.03]{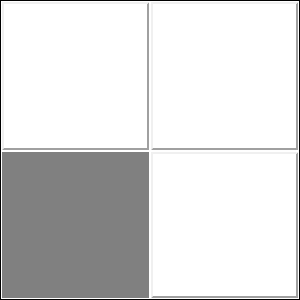}}%
            {3}{\includegraphics[scale=0.03]{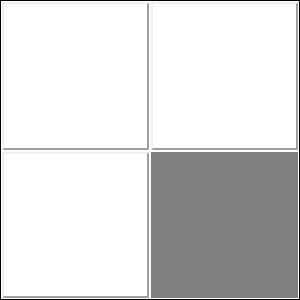}}%
        }
        [\textcolor{red}{\textbf{KNOCKCODE\_SMALL\_ERROR}}]  
    }
\hspace{-1ex}}

\newcommand{\knockcodebig}[1]
{\hspace{-.4em}
    \foreach \knock in {#1} 
    {
        \IfEqCase{\knock}{%
            {0}{\includegraphics[scale=0.02]{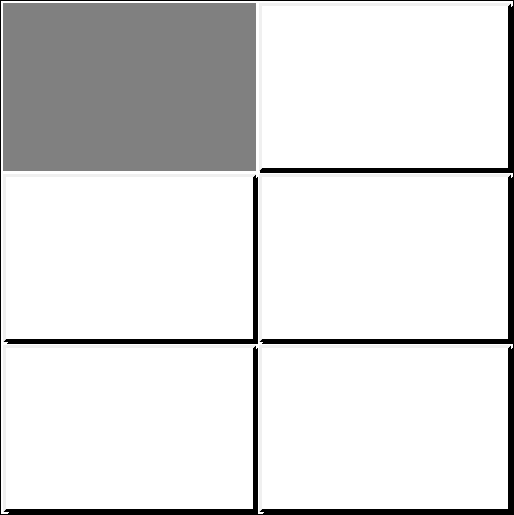}}%
            {1}{\includegraphics[scale=0.02]{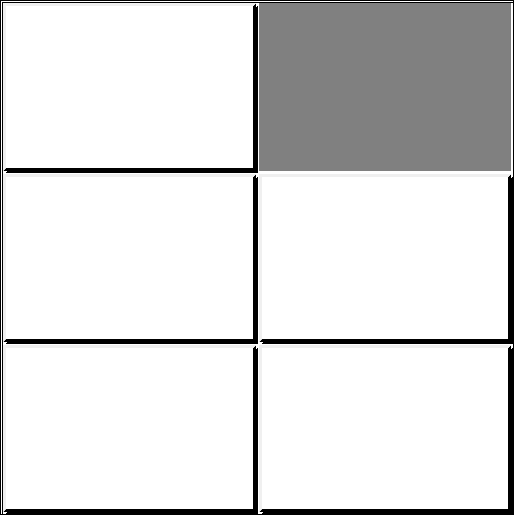}}%
            {2}{\includegraphics[scale=0.02]{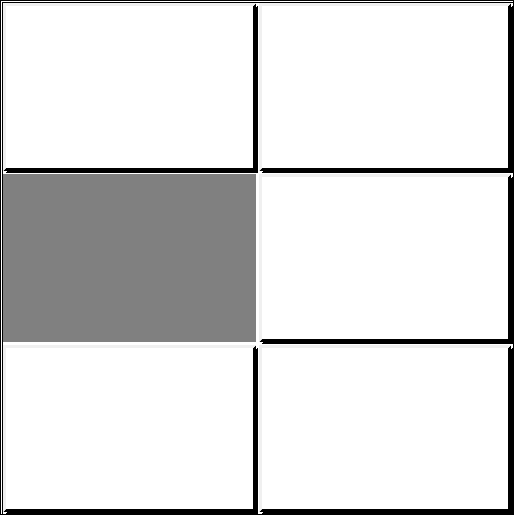}}%
            {3}{\includegraphics[scale=0.02]{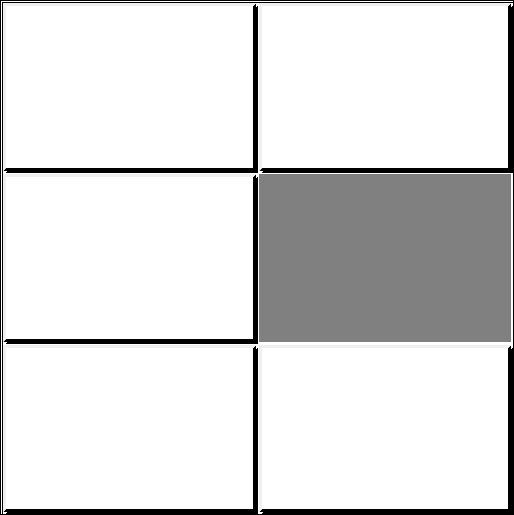}}%
            {4}{\includegraphics[scale=0.02]{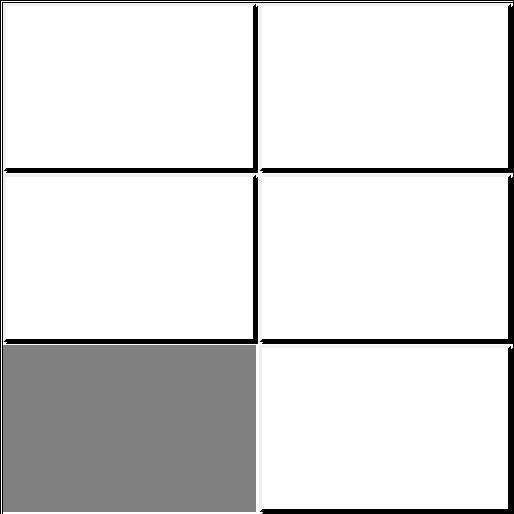}}%
            {5}{\includegraphics[scale=0.02]{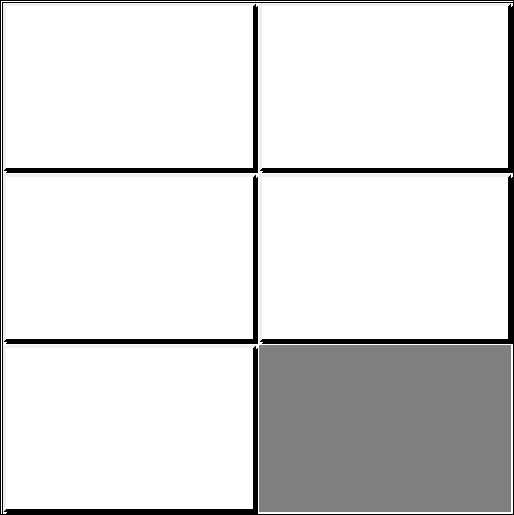}}%
        }
        [\textcolor{red}{\textbf{KNOCKCODE\_LARGE\_ERROR}}]   
    }
\hspace{-1ex}}

\crefformat{footnote}{#2\footnotemark[#1]#3}

\date{}

\title{Knock, Knock. Who's There? On the Security of LG's Knock Codes\thanks{A version of this article appears in the proceedings of the Symposium of Usable Privacy and Security (SOUPS) 2020.}}

\def\plainauthor{Raina Samuel, Philipp Markert, Adam J. Aviv, and Iulian Neamtiu}
\author{
  {\rm Raina Samuel}\\
  New Jersey Institute of Technology\\ \vspace{.5em}
  {\tt \small res9@njit.edu}\\
  {\rm Adam J. Aviv}\\
  The George Washington University  \\
  {\tt \small aaviv@gwu.edu}
  \and
  {\rm Philipp Markert}\\
  Ruhr University Bochum\\ \vspace{.5em}
  {\tt \small philipp.markert@rub.de}\\
  {\rm Iulian Neamtiu}\\
  New Jersey Institute of Technology\\
  {\tt \small ineamtiu@njit.edu}
}

\begin{document}

\maketitle

\begin{abstract}
Knock Codes are a knowledge-based unlock authentication scheme used on LG smartphones where a user enters a code by tapping or ``knocking'' a sequence on a 2x2 grid. \new{While a lesser-used authentication method, as compared to PINs or Android patterns, there is likely a large number of Knock Code users; we estimate, 700,000--2,500,000 in the US alone.} In this paper, we studied Knock Codes security asking participants in an online study to select codes on mobile devices in three settings: a control treatment, a blocklist treatment, and a treatment with a larger, 2x3 grid. We find that Knock Codes are significantly weaker than other deployed authentication, e.g., PINs or Android patterns. In a simulated attacker setting, 2x3 grids offered no additional security.  Blocklisting, on the other hand, was more beneficial, making Knock Codes' security similar to Android patterns. Participants expressed positive perceptions of Knock Codes, yet usability was challenged. SUS values were ``marginal'' or ``ok'' across treatments. Based on these findings, we recommend deploying blocklists for selecting a Knock Code because they improve security but have a limited impact on usability perceptions. 
\end{abstract}





\newcommand{\tabdemo}{

\begin{table}
  \centering
  \caption {Overall demographics of the participants from the main study. Note, zero responses are not shown.}
  \label{tab:demo}
  \resizebox{\linewidth}{!}{
  \begin{tabular}{rrrrrrrrrr}
  \toprule
    && \multicolumn{2}{c}{\textbf{Male}}  & \multicolumn{2}{c}{\textbf{Female}} & \multicolumn{2}{c}{\textbf{Other}} & \multicolumn{2}{c}{\textbf{Total}} \\
   \cmidrule(lr){3-4} \cmidrule(lr){5-6} \cmidrule(lr){7-8} \cmidrule(lr){9-10} && \multicolumn{1}{c}{\textbf{No.}} & \multicolumn{1}{c}{\textbf{\%}} & \multicolumn{1}{c}{\textbf{No.}} & \multicolumn{1}{c}{\textbf{\%}} & \multicolumn{1}{c}{\textbf{No.}} & \multicolumn{1}{c}{\textbf{\%}} & \multicolumn{1}{c}{\textbf{No.}} & \multicolumn{1}{c}{\textbf{\%}} \\
   \midrule
      \parbox[t]{1mm}{\multirow{6}{*}{\rotatebox[origin=c]{90}{\small Age}}} & $18-24$ & 25  & 7\,\%  & 10 & 3\,\%\ & 1 & 0\,\% & 36  & 10\,\% \\
      & $25-34$ & 131 & 37\,\% & 64 & 18\,\% & 2 & 1\,\% & 197 & 56\,\% \\
      & $35-44$ & 46  & 13\,\% & 31 & 9\,\%  & 0 & 0\,\% & 77  & 22\,\% \\
      & $45-54$ & 19  & 6\,\%  & 13 & 3\,\%  & 0 & 0\,\% & 32  & 9\,\% \\
      & $55-64$ & 2   & 1\,\%  & 6  & 2\,\%  & 0 & 0\,\% & 8   & 3\,\% \\
      & Prefer not to say & 0 & 0\,\% & 0 & 0\,\% & 1 & 0\,\% & 1 & 0\,\% \\
    \midrule
	\parbox[t]{1mm}{\multirow{4}{*}{\rotatebox[origin=c]{90}{\small Dexterity}}} & Left-handed  &  31 & 9\,\%  & 14  & 4\,\%  & 0 & 0\,\% &  45 & 13\,\% \\
 	& Right-handed      & 182 & 52\,\% & 103 & 29\,\% & 3 & 1\,\% & 288 & 82\,\%\\
	& Ambidextrous      &   9 & 3\,\%  & 7   & 2\,\%  & 0 & 0\,\% & 16  & 5\,\%\\
	& Prefer not to say &   1 & 0\,\%  & 0   & 0\,\%  & 1 & 0\,\% & 2   & 0\,\% \\
    \midrule
    \parbox[t]{1mm}{\multirow{4}{*}{\rotatebox[origin=c]{90}{\small Location}}} & Urban & 91 & 26\,\% & 44 & 12\,\% & 0 & 0\,\% & 135 & 38\,\% \\
    & Suburban          & 99 & 29\,\% & 57 & 16\,\% & 1 & 0\,\% & 157 & 45\,\% \\
    & Rural             & 33 & 9\,\%  & 23 & 7\,\%  & 2 & 0\,\% & 58  & 17\,\% \\
    & Prefer not to say & 0  & 0\,\%  & 0  & 0\,\%  & 1 & 0\,\% & 1   & 0\,\% \\
    \midrule
     \parbox[t]{1mm}{\multirow{9}{*}{\rotatebox[origin=c]{90}{\small Education}}} & High School & 36 & 10\,\% & 6  & 2\,\%  & 1 & 0\,\% & 43  & 12\,\% \\
     & Some College      & 45 & 13\,\% & 25 & 7\,\%  & 0 & 0\,\%  & 70 & 20\,\%  \\
     & Training          & 8  & 3\,\%  & 9  & 3\,\%  & 0 & 0\,\% & 17  & 6\,\% \\
     & Associates        & 22 & 7\,\%  & 17 & 5\,\%  & 1 & 0\,\% & 40  & 12\,\% \\
     & Bachelor's        & 91 & 26\,\% & 55 & 16\,\% & 1 & 0\,\% & 147 & 42\,\% \\
     & Master's          & 19 & 6\,\%  & 10 & 2\,\%  & 0 & 0\,\% & 29  & 8\,\% \\
     & Professional      & 1  & 0\,\%  & 1  & 0\,\%  & 0 & 0\,\% & 2   & 0\,\% \\
     & Doctorate         & 1  & 0\,\%  & 1  & 0\,\%  & 0 & 0\,\% & 2   & 0\,\%  \\
     & Prefer not to say & 0  & 0\,\%  & 0  & 0\,\%  & 1 & 0\,\% & 1   & 0\,\% \\
    \midrule
     \parbox[t]{1mm}{\multirow{3}{*}{\rotatebox[origin=c]{90}{\small Backgrnd.}}} & Technical & 102 & 30\,\% & 28 & 8\,\% & 2 & 0\,\% & 132 & 38\,\% \\
     & Non Technical           & 110 & 31\,\% & 94 & 27\,\% & 1 & 0\,\% & 205 & 58\,\%\\
     & Prefer not to say & 11  & 3\,\% & 2  & 0\,\% & 1 & 0\,\% & 14 & 4\,\% \\
     \midrule
& {\bf Total} & 223 & 64\,\% & 124 & 35\,\% & 4 & 1\,\% &  351 & 100\,\% \\
    \bottomrule
        \end{tabular}}
  \end{table}  
}
\newcommand{\tabdemotech}{
\begin{table}
  \centering
  \caption{Answers of the participants from the main study regarding their device usage.}
  \label{tab:background}
  \resizebox{\linewidth}{!}{
  \begin{tabular}{rrrrrrrrrr}
  \toprule
    && \multicolumn{2}{c}{\textbf{Male}}  & \multicolumn{2}{c}{\textbf{Female}} & \multicolumn{2}{c}{\textbf{Other}} & \multicolumn{2}{c}{\textbf{Total}} \\
   \cmidrule(lr){3-4} \cmidrule(lr){5-6} \cmidrule(lr){7-8} \cmidrule(lr){9-10} && \multicolumn{1}{c}{\textbf{No.}} & \multicolumn{1}{c}{\textbf{\%}} & \multicolumn{1}{c}{\textbf{No.}} & \multicolumn{1}{c}{\textbf{\%}} & \multicolumn{1}{c}{\textbf{No.}} & \multicolumn{1}{c}{\textbf{\%}} & \multicolumn{1}{c}{\textbf{No.}} & \multicolumn{1}{c}{\textbf{\%}} \\
   \midrule
     \parbox[t]{1mm}{\multirow{4}{*}{\rotatebox[origin=c]{90}{\small No. Devices}}} &One device           & 144 & 41\,\% & 86 & 24\,\% & 2 & 0\,\% & 232 & 66\,\%\\
     &Two devices          & 61  & 18\,\% & 34 & 10\,\% & 1 & 0\,\% & 96  & 28\,\% \\
     &Three devices        & 14  & 4\,\%  &  4 & 1\,\%  & 0 & 0\,\% & 18  & 5\,\% \\
     &Four or more devices & 4   & 1\,\%  &  0 & 0\,\%  & 1 & 0\,\% & 5   & 1\,\% \\
    \midrule
    \parbox[t]{1mm}{\multirow{8}{*}{\rotatebox[origin=c]{90}{\small Device Usage}}} &Apple         & 23  & 5\,\%  & 13 & 3\,\%  & 0 & 0\,\% & 36  & 8\,\% \\
    &Google        & 26  & 6\,\%  & 11 & 2\,\%  & 0 & 0\,\% & 37  & 8\,\% \\
    &Huawei        & 9   & 2\,\%  & 4  & 1\,\%  & 1 & 0\,\% & 13  & 3\,\% \\
    &LG            & 51  & 11\,\% & 26 & 6\,\%  & 0 & 0\,\% & 77  & 22\,\% \\
    &Motorola      & 40  & 9\,\%  & 16 & 4\,\%  & 0 & 0\,\% & 56  & 13\,\% \\
    &Samsung       & 115 & 25\,\% & 77 & 17\,\% & 2 & 0\,\% & 194 & 43\,\% \\
    &ZTE           & 7   & 1\,\%  & 4  & 1\,\%  & 1 & 0\,\% & 12  & 2\,\% \\
    &Miscellaneous & 23  & 5\,\%  & 6  & 1\,\%  & 0 & 0\,\% & 29  & 6\,\% \\
    \midrule
    \parbox[t]{1mm}{\multirow{8}{*}{\rotatebox[origin=c]{90}{\small Authentication Usage}}} &4 digit PIN        & 121 & 21\,\% & 67 & 13\,\% & 2 & 0\,\% & 190 & 34\,\% \\
    &6 digit PIN        & 19  & 3\,\%  & 10 & 2\,\% & 1 & 0\,\% & 30  & 5\,\% \\
    &6+ digit PIN       & 12  & 2\,\%  & 5  & 1\,\% & 0 & 0\,\% & 17  & 3\,\% \\
    &Android pattern    & 69  & 12\,\% & 22 & 4\,\% & 0 & 0\,\% & 91  & 16\,\% \\
    &Knock Code         & 9   & 2\,\%  & 4  & 1\,\% & 0 & 0\,\% & 13  & 3\,\% \\
    &Fingerprint        & 96  & 17\,\% & 41 & 7\,\% & 2 & 0\,\% & 139 & 24\,\% \\
    &Facial Recognition & 33  & 6\,\%  & 14 & 3\,\% & 0 & 0\,\% & 47  & 9\,\% \\
    &Other              & 0   & 0\,\%  & 1  & 0\,\% & 0 & 0\,\% & 1   & 0\,\% \\
    &No Authentication  & 17  & 2\,\%  & 20 & 4\,\% & 1 & 0\,\% & 38  & 6\,\% \\
    \bottomrule
  \end{tabular}}
  \end{table}  
}


\newcommand{\tabSUSaverages}{
  
  }
\newcommand{\tabbeginquad}{
 \begin{table}[t]
	\centering
	\small
	\begin{tabular}{l | c c c c c}
 	Scenario &  0 & 1 & 2 & 3 \\
	\hline
	Device Unlock & 74.4\% & 11.9\% & 8.7\% & 5.0\%\\
	Banking App & 69.3\% & 15.8\%  & 7.0\% & 7.9\%\\
	Shopping Cart & 67.3\% & 14.4\% & 9.6\% & 8.7\%\\
	\end{tabular}
 	\caption{Beginning Quadrant Frequencies} 
	\end{table}
}

\newcommand{\tabendquad}{
 \begin{table}[t]
	\centering
	\small
	\begin{tabular}{l | c c c c c}
 	Scenario &  0 & 1 & 2 & 3 \\
	\hline
	Device Unlock & 11.5 \% & 34.9 \% & 27.1\% & 26.6 \%\\
	Banking App & 13.3\% & 35.4\%  & 27.4\% & 23.9\%\\
	Shopping Cart & 9.6\% & 34.6\% & 26.9\% & 28.8\% \\
	\end{tabular}
 	\caption{Ending Quadrant Frequencies} 
	\end{table}
}

\newcommand{\tabstrength}{
  \begin{table}[t]
    \centering
    \scriptsize
    \resizebox{\linewidth}{!}{%
    \begin{tabular}{r | r c c c c c }
      \toprule
       & \multicolumn{1}{c}{Min} & \multicolumn{1}{c}{1st Quart.} & \multicolumn{1}{c}{Median} & \multicolumn{1}{c}{Mean} & \multicolumn{1}{c}{3rd Quart.} & \multicolumn{1}{c}{Max} \\
      \midrule
      {\bf All Control 2x2} & {\bf 8.37} & {\bf 11.39} & {\bf 14.53} & {\bf 14.20} & {\bf 16.78} & {\bf 20.00} \\
            Device Unlock &8.55 & 11.51 & 14.50 & 14.11 & 16.27 & 19.99 \\
            Banking App & 10.80 & 11.89 & 13.94 & 14.73 & 17.20 & 19.99 \\
            Shopping Cart & 7.57 & 11.39 & 13.43 & 13.29 & 15.84 & 19.37 \\
      \midrule
       {\bf All Blocklist 2x2} & {\bf 10.62} & {\bf 13.81} & {\bf 16.10} & {\bf 15.91} & {\bf 17.82} & {\bf 20.00} \\
            Device Unlock &11.27 & 13.67 & 15.39 & 15.37 & 16.91 & 20.00 \\
            Banking App & 11.84 & 13.61 & 15.45 & 15.69 & 17.61 & 19.96 \\
            Shopping Cart & 11.02 & 14.19 & 15.54 & 15.52 & 16.77 & 19.92 \\
      \midrule      
      {\bf All Large 2x3} & {\bf 9.66} & {\bf 12.01} & {\bf 15.42} & {\bf 15.11} & {\bf 17.98} & {\bf 20.00} \\
      Device Unlock & 10.55 & 12.13 & 15.04 & 14.86 & 17.26 & 20.00 \\
            Banking App & 12.63 & 15.17 & 16.39 & 16.36 & 17.51 & 20.00 \\
            Shopping Cart &10.72 & 13.89 & 15.95 & 15.67 & 17.51 & 19.99 \\
     \bottomrule


    \end{tabular}%
}
    \caption{Security rating using Equation~\ref{eq:sec} based on a Markov model likelihood measure.}
\label{tab:strength}
\end{table}

}

%
%
\newcommand{\tabperfectknowledgetreatments}{
\begin{table}[t]
    \centering 
	\caption{Comparison of the guessing metrics for a perfect-knowledge attacker between the treatments and other schemes. A comparison between the scenarios is shown in Appendix~\ref{sec:figs}.}
	\label{tab:treatments-perfectknowledge} 
    \resizebox{1.0\columnwidth}{!}{\begin{tabular}{l|rrrrr|cccc}
    \toprule
    \multicolumn{1}{c|}{} & \multicolumn{5}{c|}{\textbf{Online Guessing (Success \%)}} & \multicolumn{4}{c}{\textbf{Offline Guessing (bits)}}\\
    \multicolumn{1}{l|}{\textbf{Dataset}} & 
    \multicolumn{1}{c}{$\lambda_{3}$} && 
    \multicolumn{1}{c}{$\lambda_{10}$} && 
    \multicolumn{1}{c|}{$\lambda_{30}$} & 
    \textbf{$H_{\infty}$} & $\widetilde{G}_{0.1}$ & $\widetilde{G}_{0.2}$ & $\widetilde{G}_{0.5}$\\
    \midrule
    All Control 2x2 & 14.2\,\% && 28.0\,\% && 51.3\,\% & 3.86 & 4.20 & 4.79 & 5.69 \\
    All Blocklist 2x2 & 6.9\,\% && 16.0\,\% && 35.4\,\% & 5.27 & 5.79 & 6.03 & 6.72 \\
    All Large 2x3\textsuperscript{$\dagger$} & 12.9\,\% && 31.5\,\% && 53.4\,\% & 4.40 & 4.53 & 4.70 & 5.54 \\
    \midrule
    All First-Entry 2x2\textsuperscript{$\dagger$} & 10.8\,\% && 22.8\,\% && 43.1\,\% & 4.40 & 4.79 & 5.35 & 6.19 \\
    \midrule
    3x3 Pattern~\cite{aviv2015isbigger}\textsuperscript{$\dagger$} & 8.6\,\% && 19.4\,\% && 36.6\,\% & 4.69 & 5.21 & 5.72 & 6.76 \\
    4x4 Pattern~\cite{aviv2015isbigger}\textsuperscript{$\dagger$} & 7.8\,\% && 18.1\,\% && 32.3\,\% & 5.05 & 5.47 & 5.92 & 7.00 \\
    4-digit PINs~\cite{amitay-11-iphone-pins}\textsuperscript{$\dagger$} & 9.5\,\% && 17.2\,\% && 28.0\,\% & 4.40 & 5.14 & 6.05 & 7.21 \\
    6-digit PINs~\cite{wang2017understanding}\textsuperscript{$\dagger$} & 13.4\,\% && 16.8\,\% && 25.4\,\% & 3.10 & 3.10 & 6.38 & 7.32 \\
    \bottomrule
    \end{tabular}}
    \begin{flushleft}
    {\scriptsize{$\dagger$: For a fair comparison we downsampled all marked datasets to the size of Control and Blocklist~(232~Knock Codes).\\}}
    \end{flushleft}
\end{table}
}

\newcommand{\tabperfectknowledgescenarios}{
\begin{table}[t]
    \centering
\caption{Comparison of the guessing metrics for a perfect-knowledge attacker between the scenarios.}
\label{tab:scenarios-perfectknowledge}  
    \resizebox{.55\linewidth}{!}{\begin{tabular}{rl|rrrrr|cccc}
    \toprule
    \multicolumn{2}{c|}{} & \multicolumn{5}{c|}{\textbf{Online Guessing (Success \%)}} & \multicolumn{4}{c}{\textbf{Offline Guessing (bits)}}\\
    \multicolumn{2}{c|}{\textbf{Dataset}} & 
    \multicolumn{1}{c}{$\lambda_{3}$} && 
    \multicolumn{1}{c}{$\lambda_{10}$} && 
    \multicolumn{1}{c|}{$\lambda_{30}$} & 
    \textbf{$H_{\infty}$} & $\widetilde{G}_{0.1}$ & $\widetilde{G}_{0.2}$ & $\widetilde{G}_{0.5}$\\
    \midrule
    \parbox[t]{1mm}{\multirow{3}{*}{\rotatebox[origin=c]{90}{\normalsize Control}}} & Device Unlock\textsuperscript{$\ast$} & 17.9\,\% && 37.5\,\% && 73.2\,\% & 3.81 & 3.81 & 4.10 & 4.93 \\
    & Banking App. & 10.7\,\% && 30.4\,\% && 66.1\,\% & 4.81 & 4.81 & 4.81 & 5.31 \\
    & Shopping Cart\textsuperscript{$\ast$} & 21.4\,\% && 42.9\,\% && 78.6\,\% & 2.81 & 2.81 & 3.75 & 4.63 \\
    \midrule
    \parbox[t]{1mm}{\multirow{3}{*}{\rotatebox[origin=c]{90}{\normalsize Blocklist}}} & Device Unlock\textsuperscript{$\ast$} & 10.7\,\% && 25.0\,\% && 60.7\,\% & 4.81 & 4.81 & 5.19 & 5.53 \\
    & Banking App.\textsuperscript{$\ast$} & 8.9\,\% && 21.4\,\% && 57.1\,\% & 4.22 & 5.20 & 5.52 & 5.67 \\
    & Shopping Cart\textsuperscript{$\ast$} & 12.5\,\% && 26.8\,\% && 62.5\,\% & 4.22 & 4.58 & 4.99 & 5.45 \\
    \midrule
    \parbox[t]{1mm}{\multirow{3}{*}{\rotatebox[origin=c]{90}{\normalsize Large}}} & Device Unlock\textsuperscript{$\ast$} & 17.9\,\% && 41.1\,\% && 76.8\,\% & 3.81 & 3.99 & 4.20 & 4.79 \\
    & Banking App.\textsuperscript{$\ast$} & 12.5\,\% && 32.1\,\% && 67.9\,\% & 4.22 & 4.58 & 4.68 & 5.21 \\
    & Shopping Cart & 16.1\,\% && 38.0\,\% && 73.2\,\% & 3.81 & 3.99 & 4.40 & 4.96 \\
    \midrule
    \parbox[t]{1mm}{\multirow{3}{*}{\rotatebox[origin=c]{90}{\normalsize 1st-Entry}}} & Device Unlock\textsuperscript{$\ast$} & 16.1\,\% && 33.9\,\% && 69.6\,\% & 3.81 & 3.99 & 4.40 & 5.12 \\
    & Banking App.\textsuperscript{$\ast$} & 12.5\,\% && 28.6\,\% && 64.3\,\% & 4.22 & 4.58 & 4.81 & 5.38 \\
    & Shopping Cart.\textsuperscript{$\ast$} & 16.1\,\% && 33.9\,\% && 69.6\,\% & 3.81 & 3.99 & 4.40 & 5.12 \\
    \bottomrule
    \end{tabular}}
    \begin{center}
    {\scriptsize{$\ast$: For a fair comparison we downsampled all marked datasets to the size of the smallest datasets~(56~Knock Codes).\\}}
    \end{center}
\end{table}
}

\newcommand{\tabguessing}{

\begin{table}[t]
\centering
\caption{Guessing performance of a simulated attacker.}
\label{tab:guessing}
\resizebox{\linewidth}{!}{
\begin{tabular}{l|crc|rrrrrr}
    \toprule
    \multicolumn{1}{l|}{} & & \multicolumn{2}{c|}{\textbf{Blocklist Hits}} & \multicolumn{2}{c}{\textbf{3 Guesses}} & \multicolumn{2}{c}{\textbf{10 Guesses}} & \multicolumn{2}{c}{\textbf{30 Guesses}} \\
    \multicolumn{1}{l|}{\textbf{Dataset}} & \textbf{Codes} & \multicolumn{1}{r}{\textbf{No.}} & \multicolumn{1}{c|}{\textbf{\%}}
    & \textbf{No.} & \multicolumn{1}{c}{\textbf{\%}} 
    & \textbf{No.} & \multicolumn{1}{c}{\textbf{\%}}
    & \textbf{No.} & \multicolumn{1}{c}{\textbf{\%}} \\
  \midrule
  {\bf All Control 2x2} & \textbf{232} & \multicolumn{2}{c|}{-} & \textbf{33} & \textbf{14\,\%} & \textbf{44} & \textbf{19\,\%} & \textbf{85} & \textbf{37\,\%} \\
  Device Unlock & 116 & \multicolumn{2}{c|}{-} & 20 & 17\,\% & 28 & 24\,\% & 42 & 36\,\% \\
  Banking App. & ~\,56 & \multicolumn{2}{c|}{-} & 0 & 0\,\% & 4 & 7\,\% & 8 & 14\,\% \\
  Shopping Cart & ~\,60 & \multicolumn{2}{c|}{-} & 9 & 15\,\% & 11 & 18\,\% & 23 & 38\,\% \\
  \midrule
  {\bf All Blocklist 2x2} & \textbf{232} & \textbf{53} & \textbf{23\,\%} & \textbf{9} & \textbf{4\,\%} & \textbf{14} & \textbf{6\,\%} & \textbf{45} & \textbf{19\,\%} \\
  Device Unlock & 116 & 40 & 35\,\% & 1 & 1\,\% & 1 & 1\,\% & 5 & 4\,\% \\
  Banking App. & ~\,57 & ~\,8 & 14\,\% & 3 & 5\,\% & 3 & 5\,\% & 3 & 5\,\% \\
  Shopping Cart & ~\,59 & ~\,5 & ~~9\,\% & 3 & 5\,\% & 3 & 5\,\% & 5 & 9\,\% \\
  \midrule
  {\bf All Large 2x3} & \textbf{238} & \multicolumn{2}{c|}{-} & \textbf{24} & \textbf{10\,\%} & \textbf{62} & \textbf{26\,\%} & \textbf{97} & \textbf{41\,\%} \\
  Device Unlock & 119 & \multicolumn{2}{c|}{-} & 6 & 5\,\% & 37 & 31\,\% & 44 & 37\,\% \\
  Banking App. & ~\,63 & \multicolumn{2}{c|}{-} & 1 & 2\,\% & 6 & 10\,\% & 15 & 23\,\% \\
  Shopping Cart & ~\,56 & \multicolumn{2}{c|}{-} & 5 & 9\,\% & 10 & 18\,\% & 15 & 27\,\% \\
  \midrule
  {\bf All First-Entry 2x2} & \textbf{464} & \multicolumn{2}{c|}{-} & \textbf{42} & \textbf{9\,\%} & \textbf{83} & \textbf{18\,\%} & \textbf{127} & \textbf{27\,\%} \\
  Device Unlock & 232 & \multicolumn{2}{c|}{-} & 31 & 13\,\% & 47 & 20\,\% & 84 & 36\,\% \\
  Banking App. & 113 & \multicolumn{2}{c|}{-} & 5 & 4\,\% & 16 & 14\,\% & 27 & 24\,\% \\
  Shopping Cart & 119 & \multicolumn{2}{c|}{-} & 12 & 10\,\% & 20 & 17\,\% & 29 & 24\,\% \\
\bottomrule
\end{tabular}}
\end{table}
}

\newcommand{\tabmemcode}[0]{
}

\newcommand{\tablefreq}{
\begin{figure*}[p]
\centering
\footnotesize
\begin{tabular}{c c c c c c }
013201&
012301&
031203&
102310&
023023& 
021302 \\ 

\includegraphics[width=0.13\linewidth]{images/top30/013201.pdf}&
\includegraphics[width=0.13\linewidth]{images/top30/012301.pdf}&
\includegraphics[width=0.13\linewidth]{images/top30/031203.pdf}&
\includegraphics[width=0.13\linewidth]{images/top30/102310.pdf}&
\includegraphics[width=0.13\linewidth]{images/top30/023023.pdf}& 
\includegraphics[width=0.13\linewidth]{images/top30/021302.pdf} \\ 

freq=28&
freq=25&
freq=19&
freq=7&
freq=7& 
freq=7 \\ 

\midrule

001133&
001122&
012332&
003311&
032103& 
032102 \\

\includegraphics[width=0.13\linewidth]{images/top30/001133.pdf}&
\includegraphics[width=0.13\linewidth]{images/top30/001122.pdf}&
\includegraphics[width=0.13\linewidth]{images/top30/012332.pdf}&
\includegraphics[width=0.13\linewidth]{images/top30/003311.pdf}&
\includegraphics[width=0.13\linewidth]{images/top30/032103.pdf}& 
\includegraphics[width=0.13\linewidth]{images/top30/032102.pdf} \\ 

freq=7&
freq=7&
freq=6&
freq=6&
freq=5& 
freq=5 \\ 

\midrule

012323&
003312&
002233&
00113322&
001132& 
30122103 \\ 

\includegraphics[width=0.13\linewidth]{images/top30/012323.pdf}&
\includegraphics[width=0.13\linewidth]{images/top30/003312.pdf}&
\includegraphics[width=0.13\linewidth]{images/top30/002233.pdf}&
\includegraphics[width=0.13\linewidth]{images/top30/00113322.pdf}&
\includegraphics[width=0.13\linewidth]{images/top30/001132.pdf}& 
\includegraphics[width=0.13\linewidth]{images/top30/30122103.pdf} \\

freq=5&
freq=5&
freq=5&
freq=5&
freq=5& 
freq=4 \\

\midrule

132013&
012321&
012012&
010123&
221103& 
110033 \\

\includegraphics[width=0.13\linewidth]{images/top30/132013.pdf}&
\includegraphics[width=0.13\linewidth]{images/top30/012321.pdf}&
\includegraphics[width=0.13\linewidth]{images/top30/012012.pdf}&
\includegraphics[width=0.13\linewidth]{images/top30/010123.pdf}&
\includegraphics[width=0.13\linewidth]{images/top30/221103.pdf}& 
\includegraphics[width=0.13\linewidth]{images/top30/110033.pdf} \\ 

freq=4&
freq=4&
freq=4&
freq=4&
freq=3& 
freq=3 \\ 

\midrule

021323&
012303&
01230123&
010120&
003322& 
000132 \\ 

\includegraphics[width=0.13\linewidth]{images/top30/021323.pdf}&
\includegraphics[width=0.13\linewidth]{images/top30/012303.pdf}&
\includegraphics[width=0.13\linewidth]{images/top30/01230123.pdf}&
\includegraphics[width=0.13\linewidth]{images/top30/010120.pdf}&
\includegraphics[width=0.13\linewidth]{images/top30/003322.pdf}& 
\includegraphics[width=0.13\linewidth]{images/top30/000132.pdf} \\

freq=3&
freq=3&
freq=3&
freq=3&
freq=3& 
freq=3 \\

\end{tabular}
\caption{Top 30 most frequent Knock Codes that had at least 3 occurrences.}
\label{tab:topfreq}
\end{figure*}
}
    
\newcommand{\tablegramfreq}{

\begin{figure*}[!t]
  \centering
  \scriptsize
  
  \begin{tabular}{ c c c c c c c}

0123&
0132&
3201&
1320&
0011&
2301&
1230\\

\includegraphics[width=0.1\linewidth]{images/4grams/0123.pdf}&
\includegraphics[width=0.1\linewidth]{images/4grams/0132.pdf}&
\includegraphics[width=0.1\linewidth]{images/4grams/3201.pdf}&
\includegraphics[width=0.1\linewidth]{images/4grams/1320.pdf}&
\includegraphics[width=0.1\linewidth]{images/4grams/0011.pdf}&
\includegraphics[width=0.1\linewidth]{images/4grams/2301.pdf}&
\includegraphics[width=0.1\linewidth]{images/4grams/1230.pdf}\\

freq=68&
freq=62&
freq=47&
freq=47&
freq=41&
freq=37&
freq=37\\

\midrule

0312&
3120&
1203&
0033&
0113&
3210&
0321\\

\includegraphics[width=0.1\linewidth]{images/4grams/0123.pdf}&
\includegraphics[width=0.1\linewidth]{images/4grams/0132.pdf}&
\includegraphics[width=0.1\linewidth]{images/4grams/3201.pdf}&
\includegraphics[width=0.1\linewidth]{images/4grams/1320.pdf}&
\includegraphics[width=0.1\linewidth]{images/4grams/0011.pdf}&
\includegraphics[width=0.1\linewidth]{images/4grams/2301.pdf}&
\includegraphics[width=0.1\linewidth]{images/4grams/1230.pdf}\\

freq=31&
freq=27&
freq=26&
freq=26&
freq=24&
freq=23&
freq=23\\

\midrule

2013&
2130&
1122&
0231&
0213&
1133&
3322\\

\includegraphics[width=0.1\linewidth]{images/4grams/2013.pdf}&
\includegraphics[width=0.1\linewidth]{images/4grams/2130.pdf}&
\includegraphics[width=0.1\linewidth]{images/4grams/1122.pdf}&
\includegraphics[width=0.1\linewidth]{images/4grams/0231.pdf}&
\includegraphics[width=0.1\linewidth]{images/4grams/0213.pdf}&
\includegraphics[width=0.1\linewidth]{images/4grams/1133.pdf}&
\includegraphics[width=0.1\linewidth]{images/4grams/3322.pdf}\\

freq=20&
freq=19&
freq=19&
freq=19&
freq=18&
freq=17&
freq=16\\

\midrule

2310&
1023&
1233&
1302&
0331&
0112&
2103\\

\includegraphics[width=0.1\linewidth]{images/4grams/2310.pdf}&
\includegraphics[width=0.1\linewidth]{images/4grams/1023.pdf}&
\includegraphics[width=0.1\linewidth]{images/4grams/1233.pdf}&
\includegraphics[width=0.1\linewidth]{images/4grams/1302.pdf}&
\includegraphics[width=0.1\linewidth]{images/4grams/0331.pdf}&
\includegraphics[width=0.1\linewidth]{images/4grams/0112.pdf}&
\includegraphics[width=0.1\linewidth]{images/4grams/2103.pdf}\\

freq=16&
freq=16&
freq=15&
freq=14&
freq=14&
freq=14&
freq=13\\

  \end{tabular}
  \caption{Top 28 4-Grams with at least 13 occurrences.}
  \label{fig:grams}
\end{figure*}
}

\newcommand{\tablefreqpilot}{
\begin{table}[t]
\centering
\caption{Top 30 most frequent Knock Codes from the preliminary study, which were used as the blocklist in the bla-2x2 treatment of the main study.}
\label{tab:topfreqpilot}
\label{tab:topfreqpreliminary}
\resizebox{0.7\linewidth}{!}{
\begin{tabular}{ r l c c }
    \toprule
    \textbf{Rank} & \multicolumn{1}{c}{\textbf{Knock Code}} & \textbf{No.} & \textbf{\%} \\
    \midrule
    1 & \knockcodesmall{0,1,3,2,0,1} & 28 & 6.4\,\% \\
    2 & \knockcodesmall{0,1,2,3,0,1} & 25 & 5.7\,\% \\
    3 & \knockcodesmall{0,3,1,2,0,3} & 19 & 4.4\,\% \\
    4 & \knockcodesmall{1,0,2,3,1,0} & 7 & 1.6\,\% \\
     & \knockcodesmall{0,2,3,0,2,3} & 7 & 1.6\,\% \\
     & \knockcodesmall{0,2,1,3,0,2} & 7 & 1.6\,\% \\
     & \knockcodesmall{0,0,1,1,3,3} & 7 & 1.6\,\% \\
     & \knockcodesmall{0,0,1,1,2,2} & 7 & 1.6\,\% \\
    9 & \knockcodesmall{0,1,2,3,3,2} & 6 & 1.4\,\% \\
     & \knockcodesmall{0,0,3,3,1,1} & 6 & 1.4\,\% \\
    11 & \knockcodesmall{0,3,2,1,0,3} & 5 & 1.1\,\% \\
     & \knockcodesmall{0,3,2,1,0,2} & 5 & 1.1\,\% \\
     & \knockcodesmall{0,1,2,3,2,3} & 5 & 1.1\,\% \\
     & \knockcodesmall{0,0,3,3,1,2} & 5 & 1.1\,\% \\
     & \knockcodesmall{0,0,2,2,3,3} & 5 & 1.1\,\% \\
     & \knockcodesmall{0,0,1,1,3,3,2,2} & 5 & 1.1\,\% \\
     & \knockcodesmall{0,0,1,1,3,2} & 5 & 1.1\,\% \\
    18 & \knockcodesmall{3,0,1,2,2,1,0,3} & 4 & 0.9\,\% \\
     & \knockcodesmall{1,3,2,0,1,3} & 4 & 0.9\,\% \\
     & \knockcodesmall{0,1,2,3,2,1} & 4 & 0.9\,\% \\
     & \knockcodesmall{0,1,2,0,1,2} & 4 & 0.9\,\% \\
     & \knockcodesmall{0,1,0,1,2,3} & 4 & 0.9\,\% \\
    23 & \knockcodesmall{2,2,1,1,0,3} & 3 & 0.7\,\% \\
     & \knockcodesmall{1,1,0,0,3,3} & 3 & 0.7\,\% \\
     & \knockcodesmall{0,2,1,3,2,3} & 3 & 0.7\,\% \\
     & \knockcodesmall{0,1,2,3,0,3} & 3 & 0.7\,\% \\
     & \knockcodesmall{0,1,2,3,0,1,2,3} & 3 & 0.7\,\% \\
     & \knockcodesmall{0,1,0,1,2,0} & 3 & 0.7\,\% \\
     & \knockcodesmall{0,0,3,3,2,2} & 3 & 0.7\,\% \\
     & \knockcodesmall{0,0,0,1,3,2} & 3 & 0.7\,\% \\
    \bottomrule
\end{tabular}}
\end{table}
}

\newcommand{\tablefreqnew}{
\begin{table*}[ht]
\caption{Top 30 most frequent Knock Codes in all three treatments.}
\label{tab:topfreqnew}
\resizebox{\textwidth}{!}{
\centering
\begin{tabular}{ l l c c | l l c c | l l c c }
    \toprule
    \multicolumn{4}{c|}{\textbf{All Control 2x2}} & 
    \multicolumn{4}{c|}{\textbf{All Blocklist 2x2}} & 
    \multicolumn{4}{c}{\textbf{All Large 2x3}} \\
    \textbf{Rank} & \multicolumn{1}{c}{\textbf{Knock Code}} & \textbf{No.} & \textbf{\%} & 
    \textbf{Rank} & \multicolumn{1}{c}{\textbf{Knock Code}} & \textbf{No.} & \textbf{\%} &
    \textbf{Rank} & \multicolumn{1}{c}{\textbf{Knock Code}} & \textbf{No.} & \textbf{\%} \\
    \midrule
    1 & \knockcodesmall{0,1,2,3,0,1} & 16 & 6.9\,\% & 
    1 & \knockcodesmall{0,0,2,2,1,1} & 6 & 2.6\,\% &
    1 & \knockcodebig{0,1,2,3,4,5} & 11 & 4.6\,\% \\
    2 &\knockcodesmall{0,1,3,2,0,1} & 9 & 3.9\,\% & 
    2 & \knockcodesmall{1,1,2,2,0,3} & 5 & 2.2\,\% &
    2 & \knockcodebig{0,3,4,5,2,1} & 10 & 4.2\,\% \\
    3 & \knockcodesmall{0,0,1,1,2,2} & 8 & 3.5\,\% & 
      & \knockcodesmall{0,0,3,3,1,1,2,2} & 5 & 2.2\,\% &
    3 & \knockcodebig{0,3,4,1,2,5} & 9 & 3.8\,\% \\
    4 & \knockcodesmall{0,2,1,3,0,2} & 6 & 2.6\,\% & 
    4 & \knockcodesmall{3,2,0,1,3,2} & 3 & 1.3\,\% &
      & \knockcodebig{0,2,4,5,3,1} & 9 & 3.8\,\% \\
    5 & \knockcodesmall{0,3,1,2,0,3} & 5 & 2.2\,\% & 
      & \knockcodesmall{3,1,0,2,3,1} & 3 & 1.3\,\% &
    5 & \knockcodebig{0,1,4,5,2,3} & 8 & 3.4\,\% \\
      & \knockcodesmall{0,0,3,3,1,2} & 5 & 2.2\,\% & 
      & \knockcodesmall{2,0,1,2,0,1} & 3 & 1.3\,\% &
    6 & \knockcodebig{0,0,3,3,4,4} & 7 & 2.9\,\% \\
    6 & \knockcodesmall{3,0,2,1,3,0} & 4 & 1.7\,\% & 
      & \knockcodesmall{1,2,0,3,1,2} & 3 & 1.3\,\% &
      & \knockcodebig{0,2,4,1,3,5} & 6 & 2.5\,\% \\
      & \knockcodesmall{2,0,1,1,0,2} & 4 & 1.7\,\% & 
      & \knockcodesmall{0,2,3,3,2,0} & 3 & 1.3\,\% &
    8 & \knockcodebig{1,3,5,4,2,0} & 5 & 2.1\,\% \\
      & \knockcodesmall{0,2,3,1,0,2} & 4 & 1.7\,\% & 
      & \knockcodesmall{0,2,3,1,0,2} & 3 & 1.3\,\% &
      & \knockcodebig{0,3,4,0,3,4} & 5 & 2.1\,\% \\
      & \knockcodesmall{0,1,3,0,1,3} & 4 & 1.7\,\% & 
      & \knockcodesmall{0,2,1,3,0,2,1,3} & 3 & 1.3\,\% &
      & \knockcodebig{0,1,5,4,2,3} & 5 & 2.1\,\% \\
      & \knockcodesmall{0,1,2,3,0,1,2,3} & 4 & 1.7\,\% & 
      & \knockcodesmall{0,1,3,2,3,1} & 3 & 1.3\,\% &
   11 & \knockcodebig{1,2,5,4,3,0} & 4 & 1.7\,\% \\
      & \knockcodesmall{0,0,2,2,3,3} & 4 & 1.7\,\% & 
      & \knockcodesmall{0,1,3,2,0,1,3} & 3 & 1.3\,\% &
      & \knockcodebig{0,5,4,1,2,3} & 4 & 1.7\,\% \\
   13 & \knockcodesmall{1,3,2,0,1,3,2,0} & 3 & 1.3\,\% & 
      & \knockcodesmall{0,1,2,3,0,1,2,3,0,1} & 3 & 1.3\,\% &
      & \knockcodebig{0,0,2,2,4,4} & 4 & 1.7\,\% \\
      & \knockcodesmall{1,3,0,2,1,3} & 3 & 1.3\,\% & 
      & \knockcodesmall{0,0,3,3,2,1} & 3 & 1.3\,\% &
   14 & \knockcodebig{5,2,1,0,3,4} & 3 & 1.3\,\% \\
      & \knockcodesmall{1,1,3,3,2,2} & 3 & 1.3\,\% & 
      & \knockcodesmall{0,0,1,1,2,3} & 3 & 1.3\,\% &
      & \knockcodebig{4,1,5,0,2,3} & 3 & 1.3\,\% \\
      & \knockcodesmall{0,3,2,1,0,3,2,1} & 3 & 1.3\,\% & 
   16 & \knockcodesmall{3,3,2,2,1,0} & 2 & 0.9\,\% &
      & \knockcodebig{1,2,5,0,3,4} & 3 & 1.3\,\% \\
      & \knockcodesmall{0,3,2,1,0,3} & 3 & 1.3\,\% & 
      & \knockcodesmall{3,3,2,2,0,0,1,1} & 2 & 0.9\,\% &
      & \knockcodebig{0,2,4,0,2,4} & 3 & 1.3\,\% \\
      & \knockcodesmall{0,3,1,2,0,3,1,2} & 3 & 1.3\,\% & 
      & \knockcodesmall{3,0,2,1,3,0} & 2 & 0.9\,\% &
      & \knockcodebig{0,2,1,3,4,5} & 3 & 1.3\,\% \\
      & \knockcodesmall{0,1,2,3,0,3} & 3 & 1.3\,\% & 
      & \knockcodesmall{2,0,1,3,2,0} & 2 & 0.9\,\% &
      & \knockcodebig{0,0,1,1,3,3} & 3 & 1.3\,\% \\
      & \knockcodesmall{0,1,2,0,1,2} & 3 & 1.3\,\% & 
      & \knockcodesmall{1,2,3,0,1,2} & 2 & 0.9\,\% &
      & \knockcodebig{5,4,0,1,5,4} & 3 & 1.3\,\% \\
      & \knockcodesmall{0,0,3,3,1,1} & 3 & 1.3\,\% & 
      & \knockcodesmall{1,1,2,2,3,3} & 2 & 0.9\,\% &
   21 & \knockcodebig{5,3,1,4,2,0} & 2 & 0.8\,\% \\
      & \knockcodesmall{0,0,2,2,1,1} & 3 & 1.3\,\% & 
      & \knockcodesmall{1,1,0,0,3,3,2,2} & 2 & 0.9\,\% &
      & \knockcodebig{4,5,3,2,1,0} & 2 & 0.8\,\% \\
  23  & \knockcodesmall{3,2,3,2,0,1} & 2 & 0.9\,\% &
      & \knockcodesmall{1,1,0,0,2,2} & 2 & 0.9\,\% &
      & \knockcodebig{4,3,0,1,2,5} & 2 & 0.8\,\% \\
      & \knockcodesmall{2,0,2,0,2,0,1} & 2 & 0.9\,\% &
      & \knockcodesmall{0,3,1,2,0,3,1,2} & 2 & 0.9\,\% &
      & \knockcodebig{4,2,0,1,3,5} & 2 & 0.8\,\% \\
      & \knockcodesmall{1,2,3,0,1,2} & 2 & 0.9\,\% &
      & \knockcodesmall{0,2,1,3,3,0} & 2 & 0.9\,\% &
      & \knockcodebig{1,2,5,1,2,5} & 2 & 0.8\,\% \\
      & \knockcodesmall{1,2,0,3,1,2} & 2 & 0.9\,\% &
      & \knockcodesmall{0,2,1,0,2,1} & 2 & 0.9\,\% &
      & \knockcodebig{1,1,3,3,5,5} & 2 & 0.8\,\% \\
      & \knockcodesmall{1,0,2,3,1,0} & 2 & 0.9\,\% &
      & \knockcodesmall{0,1,3,2,0,3} & 2 & 0.9\,\% &
      & \knockcodebig{0,4,1,5,2,3} & 2 & 0.8\,\% \\
      & \knockcodesmall{0,3,2,1,3,2} & 2 & 0.9\,\% &
      & \knockcodesmall{0,1,2,3,1,0} & 2 & 0.9\,\% &
      & \knockcodebig{0,3,2,4,5,3} & 2 & 0.8\,\% \\
      & \knockcodesmall{0,3,2,1,0,0} & 2 & 0.9\,\% &
      & \knockcodesmall{0,1,2,2,3,0} & 2 & 0.9\,\% &
      & \knockcodebig{0,2,4,5,5,5} & 2 & 0.8\,\% \\
      & \knockcodesmall{0,3,1,2,2,3} & 2 & 0.9\,\% &
      & \knockcodesmall{0,0,3,2,1,0} & 2 & 0.9\,\% &
      & \knockcodebig{0,2,1,0,2,1} & 2 & 0.8\,\% \\
    \bottomrule
\end{tabular}}
\end{table*}
}

\newcommand{\tablegramfreqpilot}{
\begin{table}[t]
\centering
\begin{tabular}{ r l c c }
    \toprule
    \textbf{Rank} & \multicolumn{1}{c}{\textbf{4-Gram}} & \textbf{No.} & \textbf{\%} \\
    \midrule
    1 & \knockcodesmall{0,1,2,3} & 68 & 9.1\,\% \\
    2 & \knockcodesmall{0,1,3,2} & 62 & 8.3\,\% \\
    3 & \knockcodesmall{3,2,0,1} & 47 & 6.3\,\% \\
      & \knockcodesmall{1,3,2,0} & 47 & 6.3\,\% \\
    5 & \knockcodesmall{0,0,1,1} & 41 & 5.5\,\% \\
    6 & \knockcodesmall{2,3,0,1} & 37 & 4.9\,\% \\
      & \knockcodesmall{1,2,3,0} & 37 & 4.9\,\% \\
    8 & \knockcodesmall{0,3,1,2} & 31 & 4.1\,\% \\
    9 & \knockcodesmall{3,1,2,0} & 27 & 3.6\,\% \\
   10 & \knockcodesmall{1,2,0,3} & 26 & 3.5\,\% \\
      & \knockcodesmall{0,0,3,3} & 26 & 3.5\,\% \\
   12 & \knockcodesmall{0,1,1,3} & 24 & 3.2\,\% \\
   13 & \knockcodesmall{3,2,1,0} & 23 & 3.1\,\% \\
      & \knockcodesmall{0,3,2,1} & 23 & 3.1\,\% \\
   15 & \knockcodesmall{2,0,1,3} & 20 & 2.7\,\% \\
   16 & \knockcodesmall{2,1,3,0} & 19 & 2.5\,\% \\
      & \knockcodesmall{1,1,2,2} & 19 & 2.5\,\% \\
      & \knockcodesmall{0,2,3,1} & 19 & 2.5\,\% \\
   19 & \knockcodesmall{0,2,1,3} & 18 & 2.4\,\% \\
   20 & \knockcodesmall{1,1,3,3} & 17 & 2.3\,\% \\
   21 & \knockcodesmall{3,3,2,2} & 16 & 2.1\,\% \\
      & \knockcodesmall{2,3,1,0} & 16 & 2.1\,\% \\
      & \knockcodesmall{1,0,2,3} & 16 & 2.1\,\% \\
   24 & \knockcodesmall{1,2,3,3} & 15 & 2.0\,\% \\
   25 & \knockcodesmall{1,3,0,2} & 14 & 1.9\,\% \\
      & \knockcodesmall{0,3,3,1} & 14 & 1.9\,\% \\
      & \knockcodesmall{0,1,1,2} & 14 & 1.9\,\% \\
   28 & \knockcodesmall{2,1,0,3} & 13 & 1.8\,\% \\
    \bottomrule
\end{tabular}
\caption{Top 28 4-Grams with at least 13 occurrences.}
\label{tab:topfreqpilot}
\label{tab:topfreqpeliminary}
\end{table}
}

\newcommand{\tablerecall}{
\begin{table}[t]
\centering
\begin{tabular}{  c l c c | }
\toprule
treatment & recalled & failed \\
\midrule
con &206  (202.83)  [0.05]  & 26  (29.17)  [0.34]\\
bla & 187  (202.83)  [1.24]	  & 45  (29.17)  [8.60]\\
big & 219  (206.33)  [0.78]	  & 17  (29.67)  [5.41]\\
\hline
con& 206  (196.50)  [0.46]&	26  (35.50)  [2.54]\\				
bla&187  (196.50)  [0.46]&45  (35.50)  [2.54]\\
\hline
con	& 206  (210.68)  [0.10]	&26  (21.32)  [1.03]\\				
big	& 219  (214.32)  [0.10]	&17  (21.68)  [1.01]\\
\hline
bla	& 187  (201.26)  [1.01] &	45  (30.74)  [6.62]\\				
big &	219  (204.74)  [0.99]	& 17  (31.26)  [6.51]\\
\bottomrule
\end{tabular}
\caption{Chi Square Test on Recall Rates}
\label{tab: recall rates}
\end{table}

}


\newcommand{\tabtimes}{
\begin{table}
  \centering
  \caption{The average (median) time and number of attempts required for setup and recall.}
  \label{tab:timing}
  \resizebox{\linewidth}{!}{
  \begin{tabular}{r|ccc|rcc}
  \toprule
  & \multicolumn{3}{c|}{\textbf{Setup}} & \multicolumn{3}{c}{\textbf{Recall}} \\
  \textbf{Treatment} & \textbf{Time} & \textbf{Attempts} & \textbf{Time/Attempt} & \multicolumn{1}{c}{\textbf{Time}} & \textbf{Attempts} & \textbf{Time/Attempt} \\
  \midrule
  Control 2x2 &		$16.2\,s \ (14.2\,s)$ & $1.1$ & $14.7\,s \ (12.9\,s)$ & $8.8\,s \ (7.6\,s)$ &	$1.1$  & $8.0\,s \ (6.9\,s)$ \\
  Blocklist 2x2 &	$22.5\,s \ (18.4\,s)$ & $1.5$ & $15.0\,s \ (12.3\,s)$ & $11.3\,s \ (9.3\,s)$ &	$1.2$ & $9.4\,s \ (7.8\,s)$ \\
  Large 2x3 &		$18.4\,s \ (14.7\,s)$ & $1.1$ & $16.7\,s \ (13.4\,s)$ & $8.4\,s \ (7.0\,s)$ &	$1.1$ & $7.6\,s \ (6.4\,s)$ \\
  \bottomrule
  \end{tabular}}
  \end{table}  
}

\newcommand{\tabtimeslarge}{
\begin{table*}
  \centering
  \caption{The average time and number of attempts required for setup and recall. The standard deviation is shown in brackets.}
  \label{tab:timing}
  \resizebox{0.65\linewidth}{!}{
  \begin{tabular}{r|ccc|rcc}
  \toprule
  & \multicolumn{3}{c|}{\textbf{Setup}} & \multicolumn{3}{c}{\textbf{Recall}} \\
  \textbf{Treatment} & \textbf{Time} & \textbf{Attempts} & \textbf{Time/Attempt} & \multicolumn{1}{c}{\textbf{Time}} & \textbf{Attempts} & \textbf{Time/Attempt} \\
  \midrule
  Control 2x2 &		$16.2\,s \ \ \, \, (\textit{7.7\,s})$ & $1.1 \ \ (\textit{0.4})$ & $15.6\,s \ \ (\textit{7.6\,s})$ & $8.8\,s \ \ (\textit{4.5\,s})$ &	$1.1 \ \ (\textit{0.7})$  & $7.2\,s \ \ (\textit{2.7\,s})$ \\
  Blocklist 2x2 &	$22.5\,s \ (\textit{13.7\,s})$ & $1.5 \ \ (\textit{1.1})$ & $18.3\,s \ \ (\textit{8.6\,s})$ & $11.3\,s \ \ (\textit{6.7\,s})$ &	$1.2 \ \ (\textit{0.8})$ & $7.4\,s \ \ (\textit{2.6\,s})$ \\
  Large 2x3 &		$18.4\,s \ (\textit{11.0\,s})$ & $1.1 \ \ (\textit{0.5})$ & $17.4\,s \ \  (\textit{8.4\,s})$ & $8.4\,s \ \ (\textit{4.1\,s})$ &	$1.1 \ \  (\textit{0.6})$ & $7.1\,s \ \ (\textit{2.6\,s})$ \\
  \bottomrule
  \end{tabular}}
  \end{table*}  
}


\newcommand{\figquantsec}{
\begin{figure}[t]
\centering
\includegraphics[width=\linewidth]{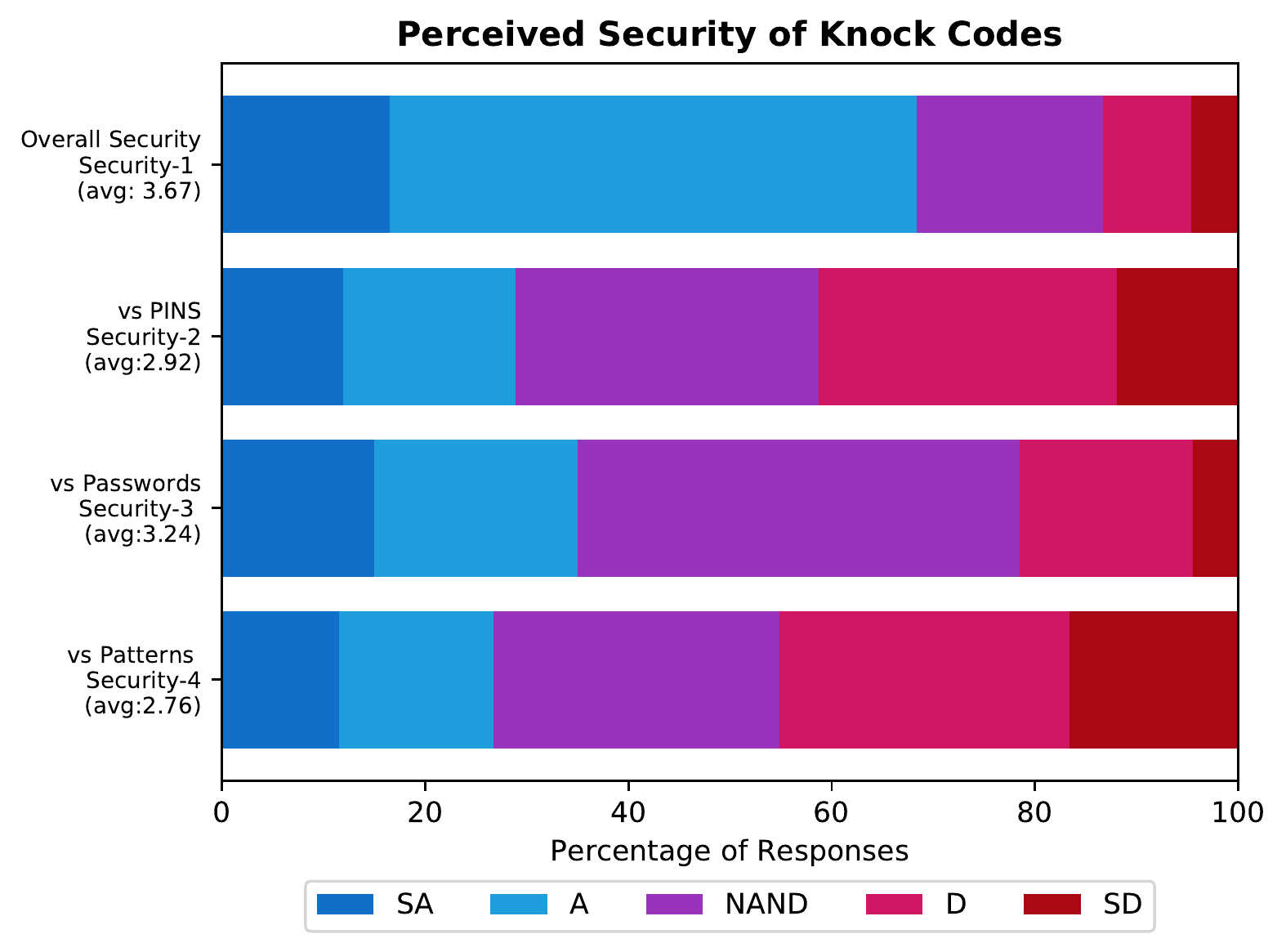}
\caption{Quantitative responses between ``Strongly Agree'' (SA) and ``Strongly
  Disagree'' (SD) for the perceived security of Knock Codes (``Overall
  Security'') and compared to (i.e., Knock codes are more secure than \ldots)
  other mobile authentication techniques, PINs, patterns, and alpha-numeric
  passwords. Averages are reported inset in labels.}
\label{fig:quantsec}
\end{figure}
 }

\newcommand{\figsus}{
\begin{figure}[t]
\centering
\includegraphics[width=\linewidth]{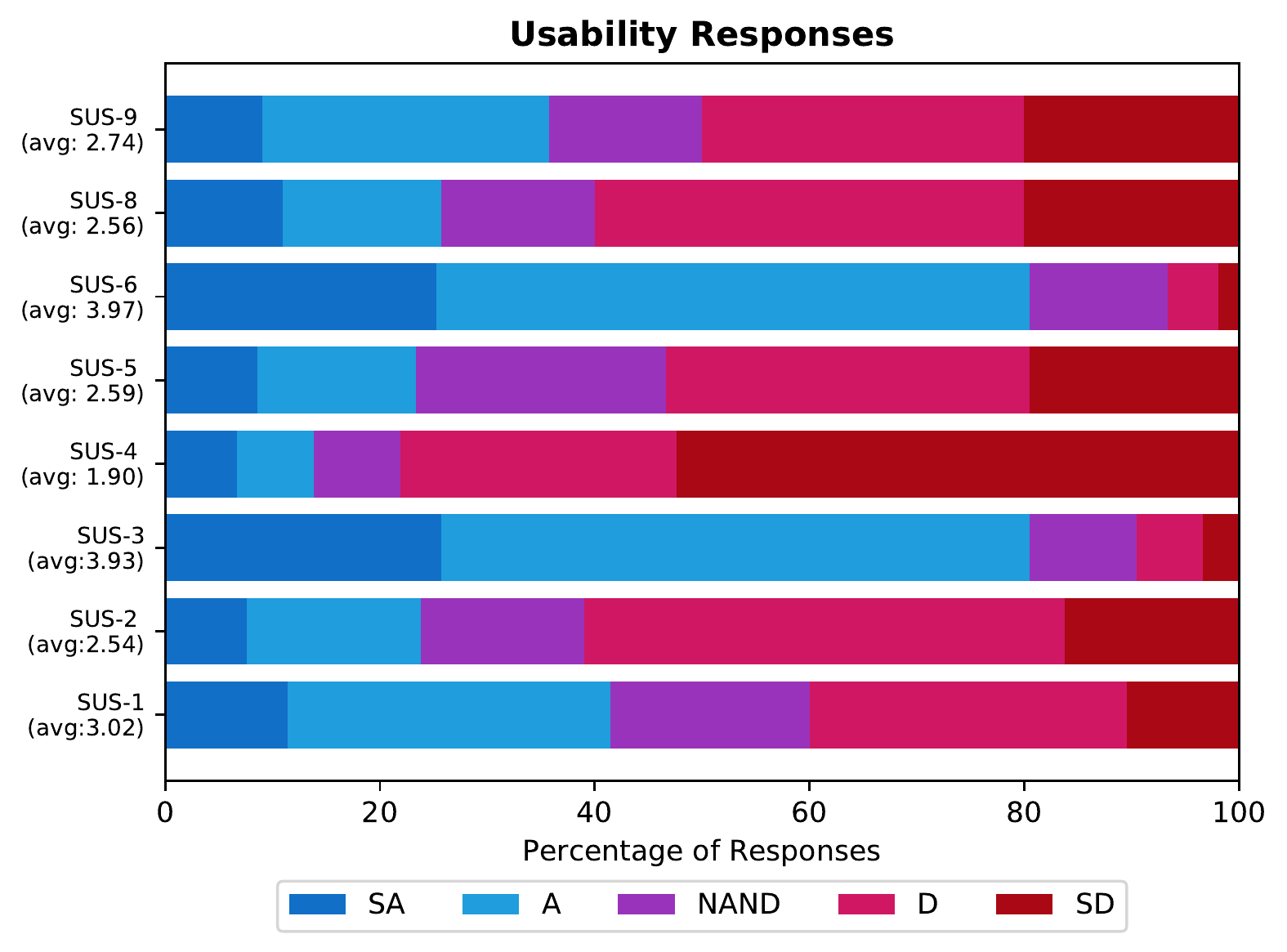}
\caption{Quantitative responses between ``Strongly Agree'' (SA) and ``Strongly
  Disagree'' (SD) for the perceived usability of Knock Codes. See
  Appendix~\ref{sec:survey} for specific phrasing of questions under the
  $\mathsf{SUS}$ header.}
\label{fig:sus}
\end{figure}
 }

\newcommand{\figstartendpilot}{

\begin{figure*}[t]
\centering
\begin{minipage}{0.49\linewidth}
\centering
\begin{subfigure}[t]{0.24\textwidth}
\centering
\includegraphics[width=0.8\linewidth]{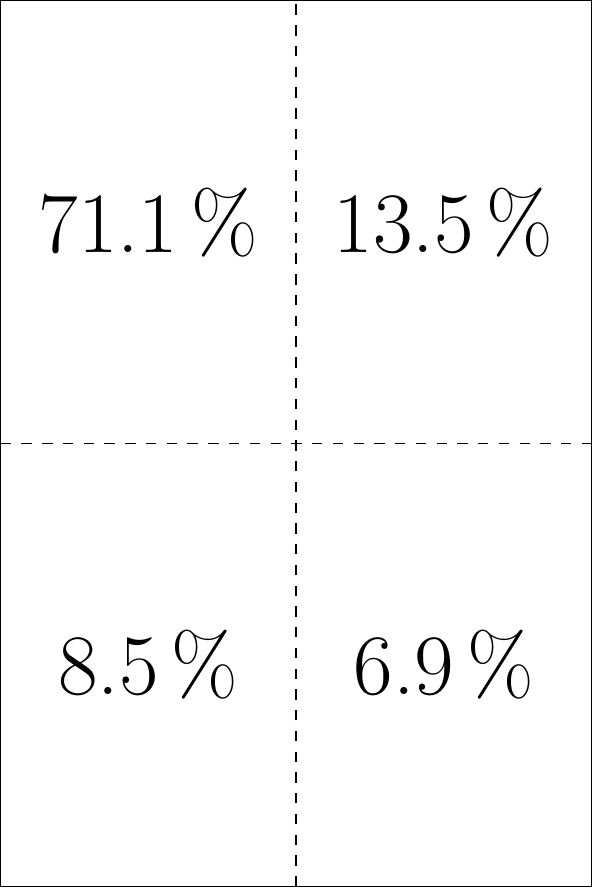}
\caption{\scriptsize Preliminary Study}
\end{subfigure}
\begin{subfigure}[t]{0.24\textwidth}
\centering
\includegraphics[width=0.8\linewidth]{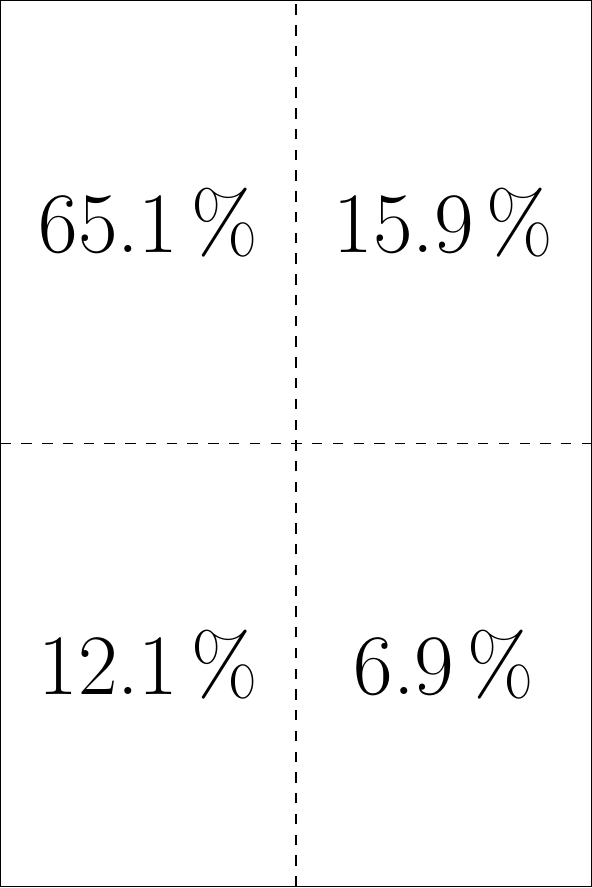}
\caption{\scriptsize Control 2x2}
\end{subfigure}
\begin{subfigure}[t]{0.24\textwidth}
\centering
\includegraphics[width=0.8\linewidth]{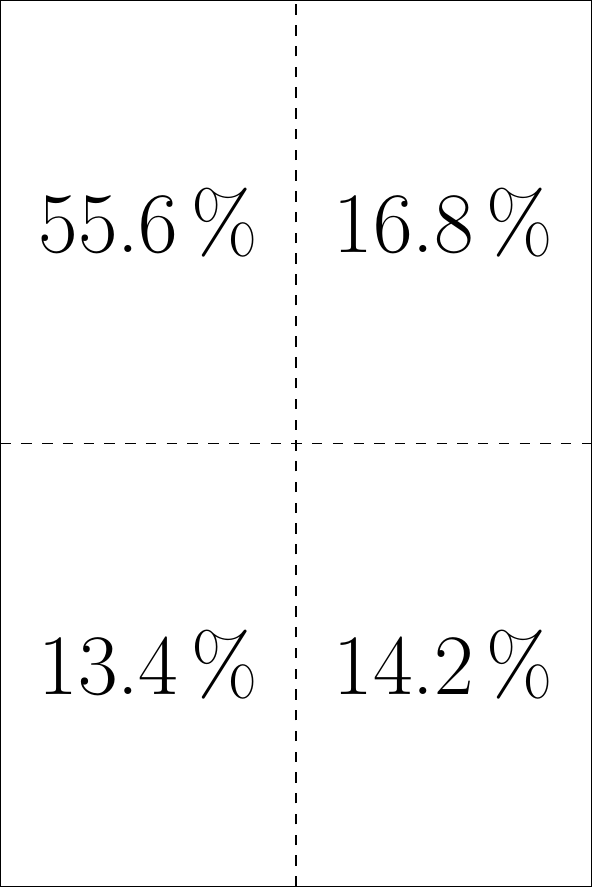}
\caption{\scriptsize Blocklist 2x2}
\end{subfigure}
\begin{subfigure}[t]{0.24\textwidth}
\centering
\includegraphics[width=0.8\linewidth]{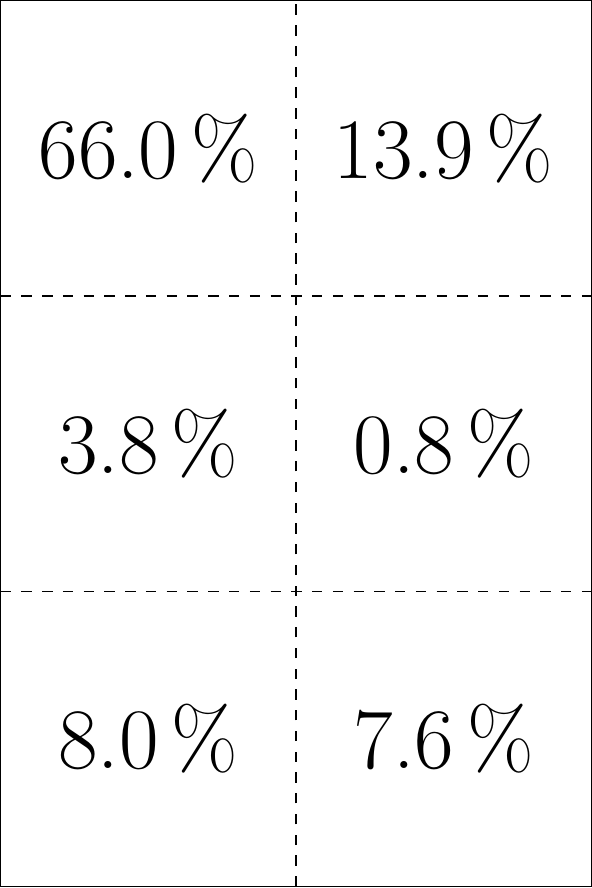}
\caption{\scriptsize Large 2x3}
\end{subfigure}
\caption{Frequency of start quadrants per treatment combined, across all scenarios. More detailed figures with frequencies for every single scenario can be found in the Appendices.}
\label{fig:startfreq}
\end{minipage}
\hfill
\begin{minipage}{0.49\linewidth}
\centering
\begin{subfigure}[t]{0.24\textwidth}
\centering
\includegraphics[width=0.8\linewidth]{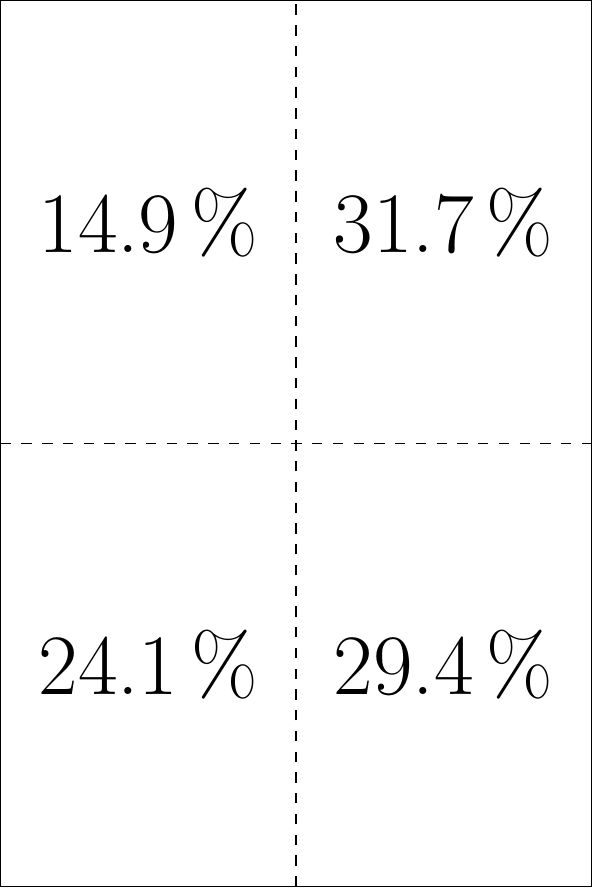}
\caption{\scriptsize Preliminary Study}
\end{subfigure}
\begin{subfigure}[t]{0.24\textwidth}
\centering
\includegraphics[width=0.8\linewidth]{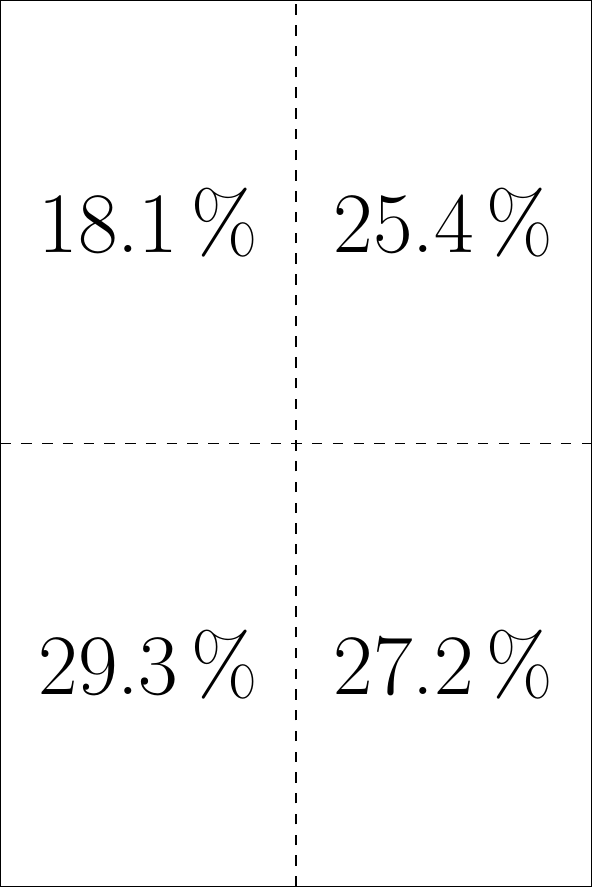}
\caption{\scriptsize Control 2x2}
\end{subfigure}
\begin{subfigure}[t]{0.24\textwidth}
\centering
\includegraphics[width=0.8\linewidth]{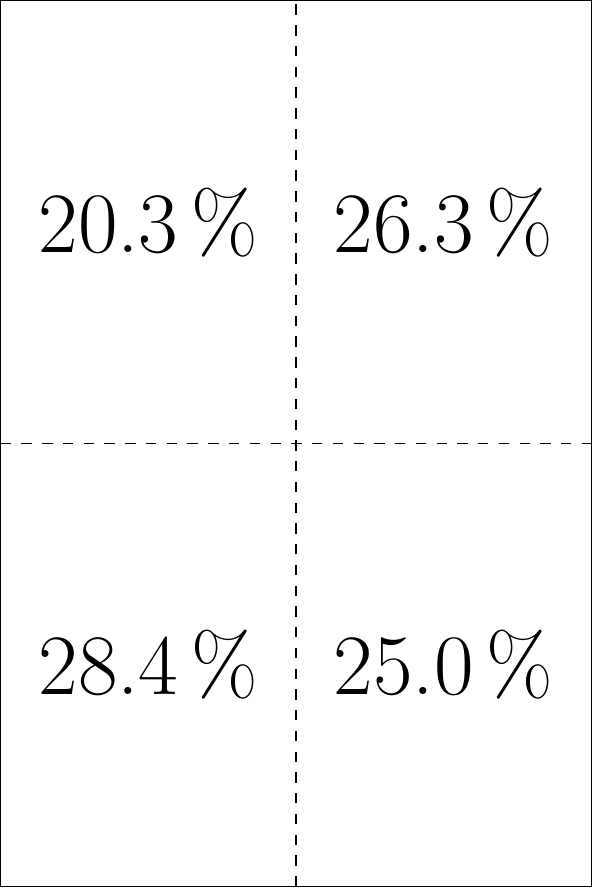}
\caption{\scriptsize Blocklist 2x2}
\end{subfigure}
\begin{subfigure}[t]{0.24\textwidth}
\centering
\includegraphics[width=0.8\linewidth]{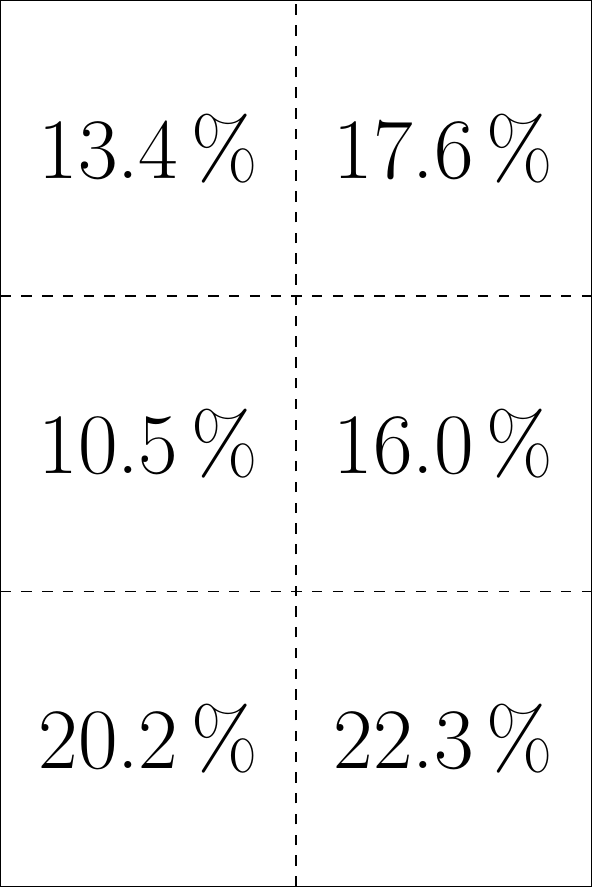}
\caption{\scriptsize Large 2x3}
\end{subfigure}
\caption{Frequency of end quadrants per treatment across all scenarios. More detailed figures with frequencies for every single scenario can be found in the Appendices.}
\label{fig:endfreq}
\end{minipage}

\end{figure*}

}

\newcommand{\figguessing}{
  \begin{figure}[t]
  \begin{center}
      \begin{subfigure}[c]{1\linewidth}
          \includegraphics[width=.95\linewidth]{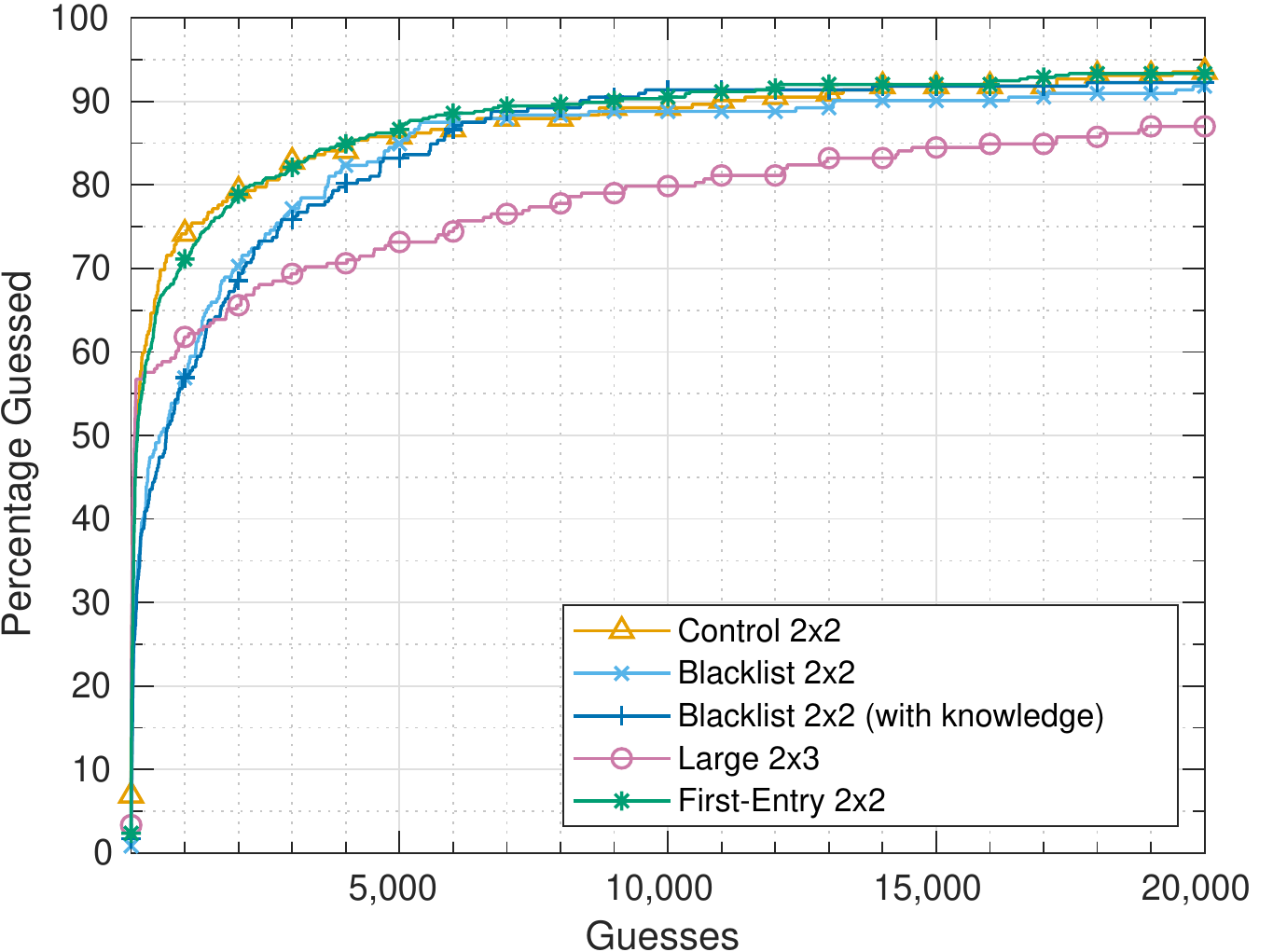}
          \caption{Guessing performance of an \textit{unthrottled} attacker.}
          \label{fig:guessing2k}
      \end{subfigure}
      \begin{subfigure}[c]{1\linewidth}
          \hspace{.2em} 
          \includegraphics[width=.9\linewidth]{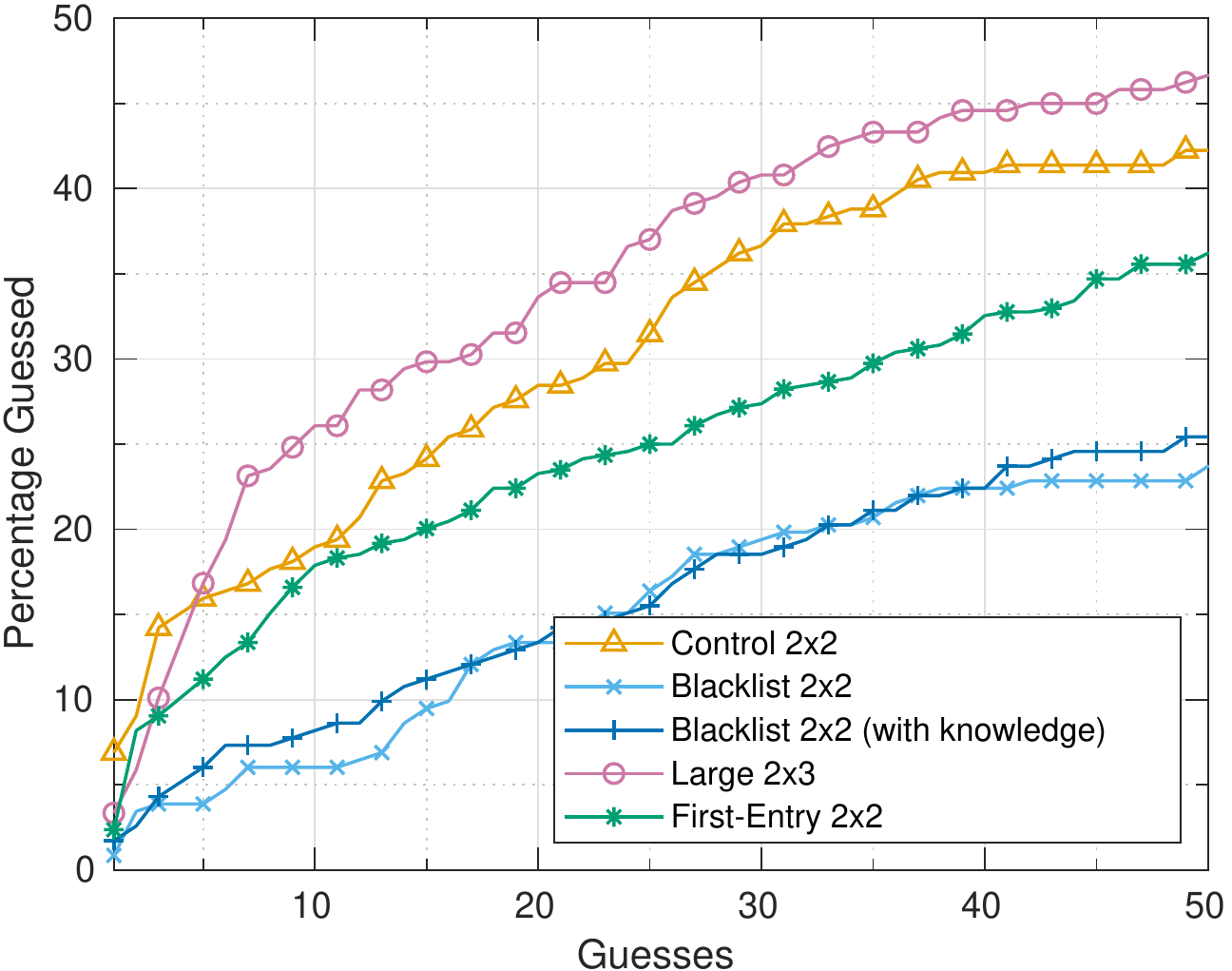}
          \caption{Guessing performance of a \textit{throttled} attacker.}
          \label{fig:guessing30}
      \end{subfigure}
  \caption{Guessing performance of a simulated attacker against the different treatments based on the numbers of guesses.}
  \label{fig:guessing}
\end{center}
\end{figure}

}
  

\section{Introduction}




Mobile device unlock authentication has many variations and there have been
extensive user-based studies on the security of knowledge-based mobile
authentication, including Android graphical unlock patterns~\cite{uellenbeck, aviv2015isbigger}, PINs~\cite{bonneau2012birthday,
  wang2017understanding, markert-20-pin-blocklist}, as well as using passwords on mobile
devices~\cite{melicher-16-mobile-passwords}. The conclusion of most of this work
is that mobile device users, much like with traditional password
selection~\cite{kelley-12-again, mazurek-13-guessability-uni,
  cubrilovic-09-rockyou}, opt for predictable and easily guessed
authenticators. Additionally several physical attacks have been proposed on
knowledge-based mobile authentication, such as smudge
attacks~\cite{aviv2010smudge}, sensor
attacks~\cite{aviv2012accel,cai2011touchlogger}, vision
attacks~\cite{ye-17-pattern-five-attempts}, acoustic
signals~\cite{zhou2018patternlistener}, and shoulder
surfing~\cite{deluca2014nowyouseeme,eiband2017understanding,aviv-17-shoulder-surfing-baseline}.

\shepherd{Into} this space, LG developed {\em Knock Codes} as a new \shepherd{mobile} authentication system that is
designed to combat some of these
attacks\footnote{\label{first}\url{https://youtu.be/0Imk5JILUc0} (as accessed on \today)} and provide, per LG's advertising,\footnote{\url{https://youtu.be/NRInfu-Lhnc} (as accessed on \today)}
``perfect security.'' Knock Codes require a user to recall a pre-selected series of \shep{at least 6 and at most 10} knocks\footnote{\shep{In earlier models, like the 2014 LG G2~\cite{tofel-2014-LG-G2-knock-code}, where this method first appeared, codes required at least 3 and at most 8. Newer models require 6 to 10 knocks occurring in at least 3 quadrants}.}  (or taps)
on a $2 \times 2$ quadrant which is displayed upon setup and can be entered with the phone screen on or off. \shep{Knock Codes are used less frequently than PINs or Android patterns, but we estimate that there is a large number of Knock Code users, 700,000--2,500,000 in the US alone.}







\shep{To evaluate the security and usability of Knock Codes,} we conducted two \shep{online} user studies on \shep{Amazon Mechanical Turk}: a preliminary study ($n=218$) and a main study ($n=351$)\shep{, analyzing} a total of 1,138 Knock Codes (436 in the preliminary study and 702 in the main study). 
In the main study, we evaluated three between-group treatments: a control treatment, where participants used the current 2x2 Knock Code interface; a blocklist treatment, where participants selected 2x2 Knock Codes with some popular codes, as measured in the preliminary study, being disallowed; and finally, a big grid treatment, where participants selected Knock Codes on a larger, 2x3 grid.

%
%
%


We analyzed the selected Knock Codes across treatments and scenarios for security using \new{standard guessing metrics}, considering both an offline attacker with unlimited guesses and an online attacker with a limited number of guesses. \shep{We find that Knock Codes, as currently deployed, offer \textit{worse} security (51.3\,\% guessed after 30 attempts) as compared to other widely available unlock authentication schemes, e.g., 4-digit PINs (28.0\,\%), 6-digit PINs (25.4\,\%) and Android unlock patterns (36.6\,\%).}

%


\shep{While it seems like a straightforward attempt to increase security, an expanded Knock Code grid to 2x3 does not increase, and sometimes worsens, security as compared to 2x2 Knock Codes.} After 30 attempts, a simulated attacker correctly guesses {\em more} 2x3 Knock Codes compared to 2x2 (41\,\% vs. 37\,\%). However, blocklisting common Knock Codes \shep{(as collected in the preliminary study)} is more effective at improving guessing security
: only 19\,\% of these codes were guessed within 30 attempts in simulation.


%
Overall, participants \shep{perceived} Knock Codes (across treatments) as secure;
%
however, among all treatments, participants were more hesitant to rate Knock Codes as {\em more secure} than PINs, Android Unlock Patterns, or alphanumeric passwords.
Despite the fact that participants reported Knock Codes as ``simple'' and ``memorable'',
responses to the SUS~\cite{brooke1996sus} questions
averaged to ``marginal'' or ``ok'' usability (69.8, 68.1, and 64.3, for the control 2x2 treatment, the larger 2x3 treatment, and the blocklist informed 2x2 treatment, respectively). \shep{Entry and recall times for Knock Codes were also much slower than what was reported for PINs and Android patterns~\cite{harbach-14-hard-lock-life, markert-20-pin-blocklist}, suggesting lower usability.}

%



Based on the survey and analysis, we make the following contributions and
findings:
\begin{itemize}[itemsep=0.5pt]
\item We conducted a user study of Knock Codes that considers usability and security analysis.

\item We find that Knock Codes, as currently deployed, offer worse security compared to other available methods, both in terms of an online and offline guessing analysis. 
  
\item We evaluated different designs for Knock Codes, finding that larger grid sizes offer no benefits (and might actually be less secure) while blocklisting offers promise for improving security.

\item We analyzed both qualitative and quantitative feedback of the perceptions of security and usability of Knock Codes, finding that while there are some features of Knock Codes that users like the overall usability was ``ok'' or ``marginal'' and the security perceptions were weak compared to other available schemes.
  
\end{itemize}

These results indicate that users are interested in new forms of mobile authentication, in particular ones that have options for unlocking with the display off. However, given the usability and security challenges of Knock Codes, {\em we would not recommend further deployment as currently configured.} For users and developers who wish to continue to use Knock Codes, we would recommend using a blocklist to inform selection as it provides increased security with small effects on usability.


\section{Related Work and Background}
\new{While Knock Codes have not been broadly studied in the community,} other
 mobile authentication methods have been investigated widely, namely
PINs~\cite{clarke2005authentication, de2007evaluation},
patterns~\cite{uellenbeck,aviv2015isbigger,song-15-pattern-psm}, passwords~\cite{klein1990foiling,
  lindsey1981survey}, and biometrics \cite{prabhakar2003biometric}\shep{, as well as adoption rates~\cite{harbach-14-hard-lock-life} and authentication times~\cite{harbach-16-the-anatomy}.}

Research on user-chosen authentication has shown that users tend towards predictable and popular choices, regardless
of the authentication method. For instance, Bonneau et al.~\cite{bonneau2012birthday} studied 4-digit PINs
and concluded that
while 4-digit PINs fare better in user management and choices, guessing the birthday is an effective strategy to access a user's
account. Wang et al. showed that 6-digit PINs have
marginally better security than 4-digit PINs, yet both English and Chinese users fall into certain patterns when choosing PINs~\cite{wang2017understanding}.

\shep{Markert et al. collected PINs specifically primed for mobile authentication and demonstrated that 6-digit PINs offer little (and perhaps worse) benefit than 4-digit PINs against a throttled attacker. Moreover, non-enforcing blocklists (as deployed by iOS) do not increase security~\cite{markert-20-pin-blocklist}. We use an enforcing blocklist in our data collection, as recommend by Markert et al., and compare Knock Codes to the same RockYou~\cite{cubrilovic-09-rockyou} and Amitay~\cite{amitay-11-iphone-pins} datasets used by Wang et al. and Markert et al.}

Patterns, or graphical passwords, have been studied in multiple contexts, including smudge attacks~\cite{aviv2010smudge}, shoulder-surfing~\cite{forget2010eyegaze, deluca2014nowyouseeme, man2003shoulder, aviv-17-shoulder-surfing-baseline}, and
user strength perceptions~\cite{andriotis2013pilot, andriotis2014complexity}. The selection process has also been studied~\cite{uellenbeck, aviv2015isbigger, song-15-pattern-psm}, and in all cases, users choices are predictable. We compare our results to those from  Uellenbeck et al.~\cite{uellenbeck} and Aviv et al.~\cite{aviv2015isbigger}.

There have also been proposals for incorporating more tactile interaction into mobile authentication. For example, Deyle and Roth suggested using ``tactile pins''~\cite{deyle2006accessible}.
Kuber et al.~\cite{kuber2010toward,kuber2010feasibility,kuber2010tactile} studied tactile stimuli: a special mouse with a 4x4 matrix of PINs
for selecting a ``tactile password.''
Krombholz et al. considered extra touch interactions through pressure-sensitive touches on iPhones to enhance PINs~\cite{krombholz2016use}. However, these user interaction modalities are very different from Knock Codes. \shep{Similar to Knock Codes, "personal identifiable chords" (PIC) for smartwatches (a multi-touch PIN entered on a 2x2 grid) have been proposed~\cite{oakley2018personal}; these differ in setting (smartwatches) and input type (multi-touch), but the approach could be used to improve Knock Codes by adding multi-touch.}



Along with security, usability is an important facet regarding the
adoption of authentication methods, thus, quantifying user feedback of
such methods is pertinent~\cite{schaub2012password}. Regarding
biometric adoption and perceptions, users considered biometrics
to be more secure than PINs according to Bhagavatula et al.~\cite{bhagavatula2015biometric}. In addition,
usability factors (such as poor lighting for facial recognition) contributed to
users' negative feedback and reluctance to adopt this method versus a more convenient method such as fingerprint
recognition. Even with biometrics, this can lead
to users choosing weaker forms of knowledge-based
authenticators~\cite{cherapau2015impact}.




\section{Methodology}

We collected data via Amazon Mechanical Turk \shep{(MTurk)}
using an online survey whereby
participants were directed to use their mobile devices (checked via the user-agent) to select {\em two} Knock Codes as well as answer
general questions about Knock Codes and their demographics. The two Knock Codes
were primed based on different security scenarios, as informed by prior work of
Loge et al.~\cite{loge-16-pattern-user-choice}. We found some, but minor,  differences between Knock Codes in each scenario, similar to Loge et al.'s findings for Android patterns.

We conducted two studies: a preliminary study and a main study which is based on the preliminary study and presented here. The main difference between the two studies is that the main study was focused on participants using mobile devices while the preliminary allowed participants to use traditional computers. From the preliminary study, we were able to refine the main study as well as develop a blocklist of the 30 most common Knock Codes selected in the preliminary study (see Table~\ref{tab:topfreqpreliminary}). We provide all study material in the Appendices. Both studies were approved by our institutional review board (IRB).

We found that usage and awareness of Knock Codes are relatively uncommon. Only 3\% of our participants in the main study responded that they use Knock Codes, see Table~\ref{tab:background} and only 1\% reported so in our preliminary study. \shep{Despite the low percentages, this suggests that 700K-2.5M users may deploy Knock Codes in the US alone, and we would ideally focus our study just on these users. This is unfortunately not feasible due to the low concentration on MTurk, and as such, we consider a broader set of study participants who may (or may not) be aware of Knock Codes. For those unaware of Knock Codes, our survey would simulate their first experience, as would be the case if they were selecting Knock Codes for the first time on a new device.}



\begin{figure}[t]
  \centering
  \includegraphics[width=0.7\linewidth]{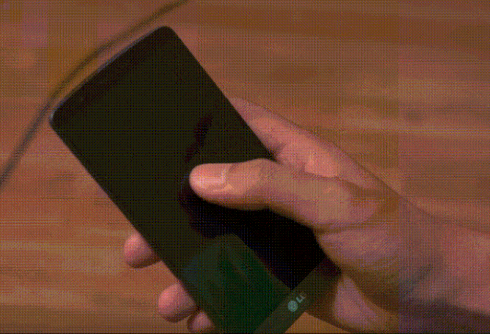}
  \caption{Screenshot of a video exploring Knock
    Codes (\url{https://youtu.be/tPYypLe8LEU}) where a
    user enters a Knock Code with the screen off to unlock the phone. This was
    used to provide instructions and background information to users on Knock Codes.}
  \label{fig:knock-code-gif}
\end{figure}

\paragraph{Detailed description of the survey.}
The survey consisted of 12 parts as described below. 
Please see Appendix~\ref{sec:survey} for the exact questions and wording on the pages.
We refer to specific questions within a survey part using the page name and question number.

\begin{enumerate}[itemsep=0.5pt]
\item {\em Overview and Informed Consent:} Upon starting the survey,
  participants were informed about the nature of the research (per the requirements
  of our IRB), and provided general instructions for proceedings.
  
\item {\em Device Usage Questions:}  
  Participants reported on the number of mobile devices (as defined by a smartphone but excluding tablet computers and laptops) they own, the brands they use, and which types of mobile authentication they use on those devices. \new{We use this data, normalized to US census data, to estimate Knock Code usage.} 
  
\item {\em Instructions:} As we could not expect participants to be familiar with  Knock Codes,
  we provided detailed instructions of Knock Codes. This
  included a GIF animation of a user entering a Knock Code (see
  Figure~\ref{fig:knock-code-gif}), a display of the entry screen used later in
  the survey (see Figure~\ref{fig:knock-code-entry}), and requirements of Knock
  Codes (use at least 3 different regions and at least 6 total knocks). 
  We also introduced the size of the grid, 2x2 for participants who were assigned to the control or blocklist treatment, and 2x3 for the group that tested a larger grid. Those in the blocklist treatment were {\em not} informed of the existence of the blocklist. A detailed description of the treatments is given later in this section. 
  
\item {\em Practice:} After the instructions, participants could practice selecting a sample Knock Code and familiarize themselves with the interface, before proceeding to the actual Knock Code selection. 
It was clearly stated that this stage was for practice purposes only. 
Participants practiced on the appropriate grid size for their treatment and for those in the blocklist treatment, there was no blocklist in place yet, i.e., no indication that a code would or would not be allowed.

\item {\em Scenario Overview:} In addition to a treatment, each participant was assigned to two scenarios under which they would select  Knock Codes for protection. 
The first of the scenarios was always {\em Device Unlock}; the other was either {\em Banking App} or
  {\em Shopping Cart}. 
  These scenarios were adapted from prior work of Loge
  et al.~\cite{loge-16-pattern-user-choice} for collecting Android patterns. Participants were made aware of {\em both} scenarios before proceeding and the order in which they would be asked to select Knock Codes.
\shepherd{On this page, we also highlighted that the selected Knock Code will have to be recalled later, hence, participants were asked to ``choose something that is secure and memorable.''}

  \begin{figure}[t]
      \begin{center}
      \begin{subfigure}[b]{0.38\linewidth}
      \includegraphics[width=\textwidth]{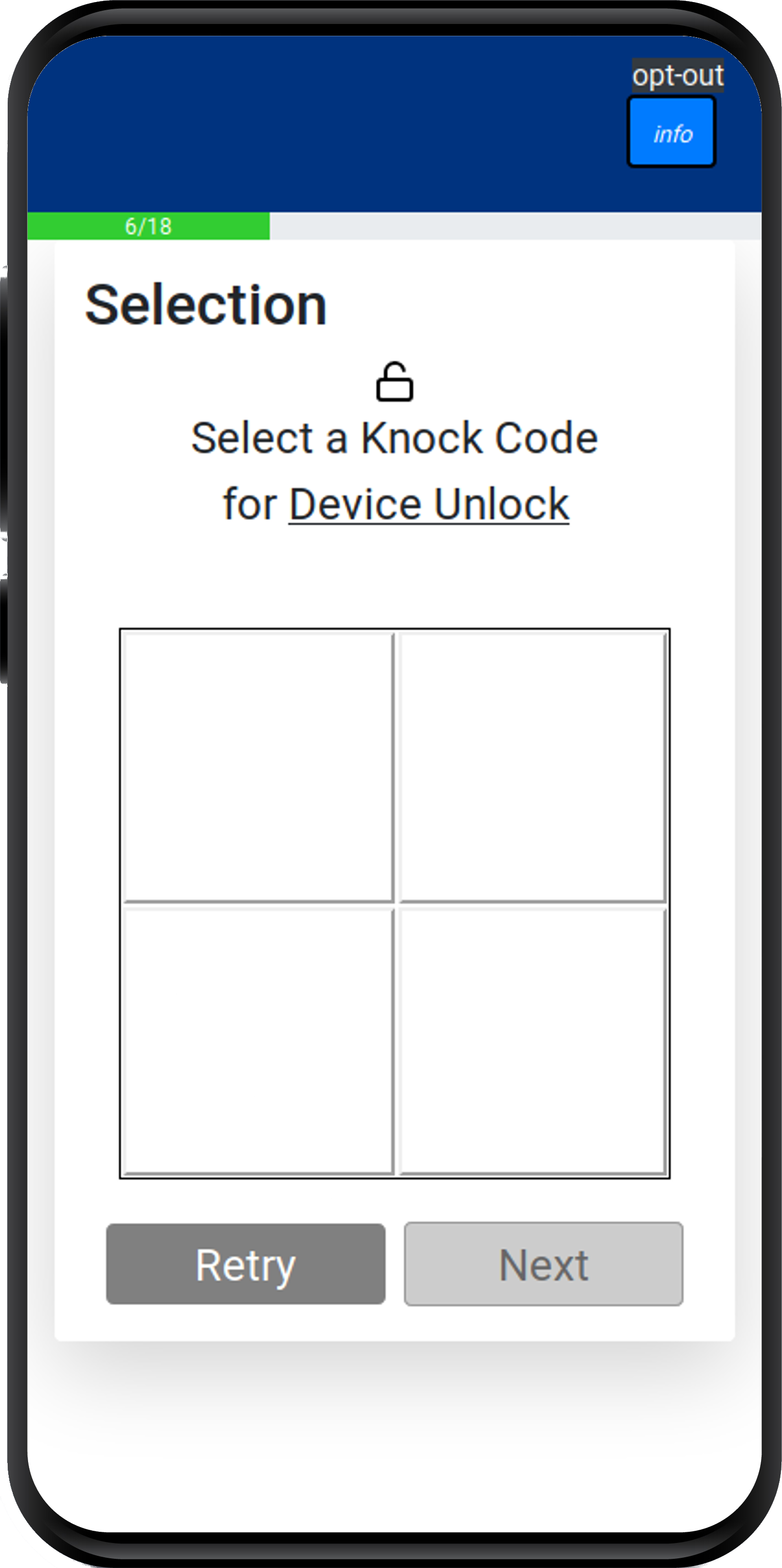}
        \caption{\controlshort{} \& \blocklistshort}
    \label{fig:knock-code-entry-2x2}
    \end{subfigure}
    \quad
      \begin{subfigure}[b]{0.38\linewidth}
     \ \includegraphics[width=\textwidth]{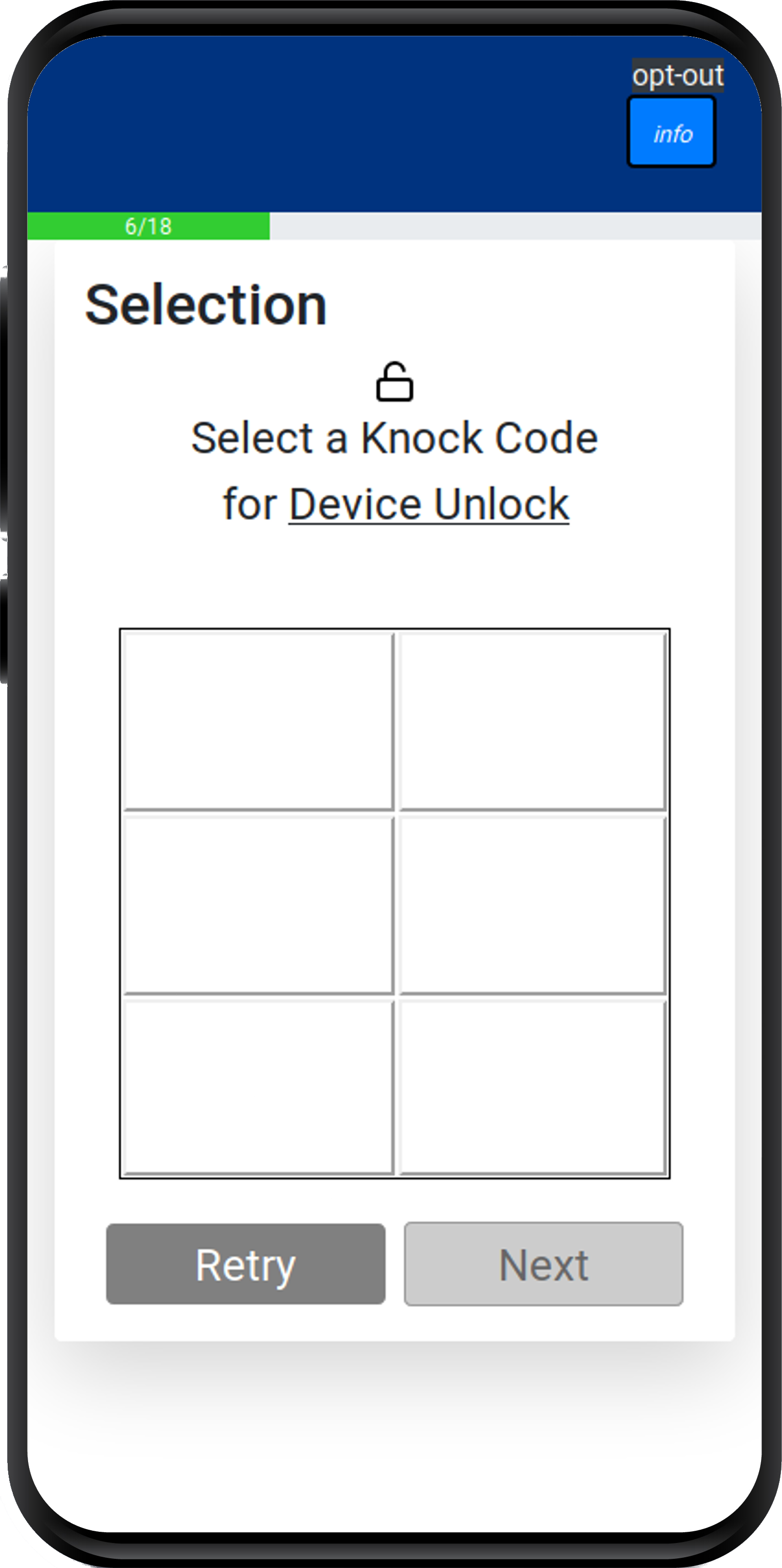}
        \caption{\bigshort}
    \label{fig:knock-code-entry-2x3}
    \end{subfigure}
  \end{center}
  \caption{(a) Interface for selecting 2x2 Knock Codes  and (b) interface for selecting 2x3 Knock Codes. Both designs mimic the look and feel of LG's Knock Code implementation.}
  \label{fig:knock-code-entry}
\end{figure}

\begin{figure}[t]
  \begin{center}
    \includegraphics[width=0.43\linewidth]{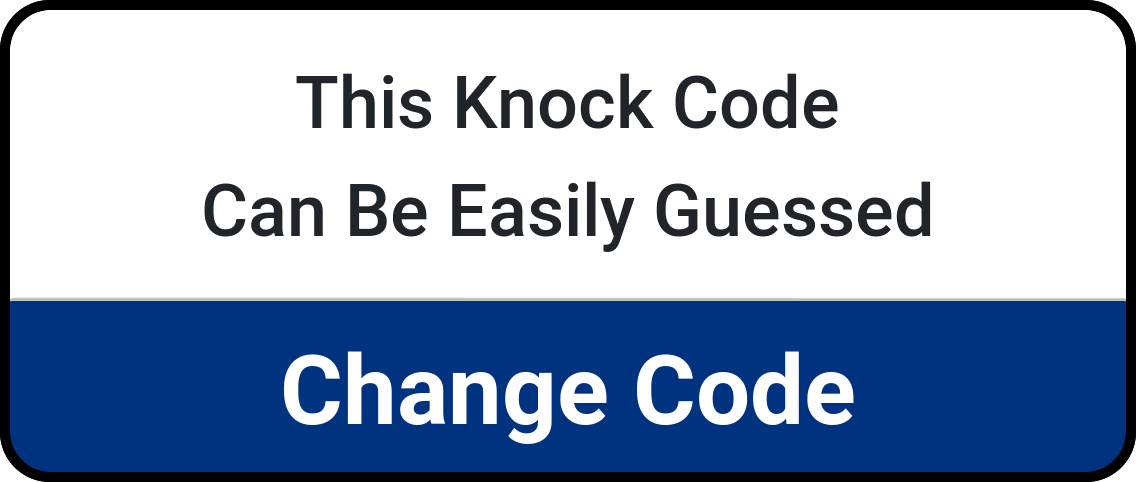}
    \caption{Blocklist warning display, which mimics blocklist warnings as used by iOS for PINs.}
    \label{fig:blwarn}
    \end{center}
  \end{figure}

\item {\em Select and Confirm (2x):} Participants were prompted to select a Knock Code for the scenario, and confirm it before proceeding. The respective pages are shown in Figure~\ref{fig:knock-code-entry}. 
  Participants of the blocklist treatment saw the warning message shown in Figure~\ref{fig:blwarn} if any selection was disallowed. Table~\ref{tab:topfreqpreliminary} contains the list of blocklisted codes as collected in the preliminary study. 
  
\item {\em Selection Feedback (2x):}  
  After selecting and confirming a Knock Code, participants were asked for feedback about their views on the security of their code and any difficulties in selecting a secure and usable code. Data was collected in both Likert agreement and through open answer forms.
  
\item {\em Security Prompts:} 
  Now with more familiarity with Knock Codes, participants answered
  questions about the perceived security of Knock Codes, and also compared it
  to PINs and Android Unlock Patterns. Participants also provided qualitative
  feedback on their security likes and dislikes related to Knock Codes in general.

\item {\em Usability Prompts:}  
  We asked the 10 System Usability Scale questions~\cite{brooke1996sus} related to Knock Codes (plus an attention test).

\item {\em Recall (2x):} Participants were asked to recall their selected Knock Codes. We allowed up to three guesses for each of the scenarios and forwarded participants if they were not able to recall their Knock Code within this limit.
  
\item {\em Demographic Questions:} 
Participants answered basic demographic questions about their age, gender, dominant hand, educational background, and technology background.
We also included another attention check question on this page. 

\item {\em Submission:} The survey ended with participants answering an honesty question (i.e., indicated yes/no to ``I honestly participated in this survey and followed instructions completely.''). Negative responses were removed from the results, however, all participants were compensated for their work. 
\end{enumerate}

\paragraph{Treatments.}
As part of the study, we assigned participants to one of three treatments. 
In addition to the standard implementation of LG's Knock Code, which we refer to as \textbf{\control} or \textbf{\controlshort} throughout this paper, we tested two additional ones. 


We first include a blocklist treatment (\textbf{\blocklist} or \textbf{\blocklistshort}) which differs from the \control{} treatment by the fact that we blocklisted 30 Knock Codes. 
These codes were the most frequently used as measured in the preliminary study (see Table~\ref{tab:topfreqpreliminary}). \
The blocklist warning, shown in cases of a blocklist hit, is depicted in Figure~\ref{fig:blwarn} and is a copy of a warning used by Apple on iOS devices to warn users about an insecure PIN choice.  

We conjecture that by disallowing participants from selecting these common codes, the Knock Codes they eventually select would be stronger (harder to guess). There is a risk with blocklists as they may increase frustration during the selection process by having to perform selection multiple times. But as setting up an authentication method is a one-time event, we wished to understand if blocklists can improve the security of Knock Codes.

As another method for increasing security, we considered a modification to the Knock Code interface. The \textbf{\big{}} treatment (\textbf{\bigshort}) uses a 2x3 instead of 2x2 grid and provides participants with more options for creating a Knock Code.
Theoretically, this increase makes a substantial difference with 72,520,440 possible 2x3 Knock Codes of length 6-to-10, as compared to 1,384,872 2x2 Knock Codes of similar length. The layout is shown in Figure~\ref{fig:knock-code-entry-2x3}.

We decided to use a 2x3 grid rather than a horizontal extension (3x2) or making a square (3x3) because of the form factor of the phone's screen, which is taller than it is wide. The 2x3 grid offers a natural extension that fits within the form factor of the screen and mirrors the same interface.

\tabdemo{}

\paragraph{Recruitment.}
The survey was distributed as an Amazon Mechanical Turk task, paying \$1.25. On
average, it took our participants 8.5 minutes to complete the survey. We ran the survey over the course of two days in June 2019. We recruited 351
participants, each creating two Knock Codes, for a total of 702 selected and confirmed Knock Codes, \new{but also additional Knock Codes that were not confirmed, either due to memorability or the blocklists.}
We do not consider the practice Knock Codes in our analysis.

The demographics and backgrounds of the participants are listed in
Table~\ref{tab:demo} and~\ref{tab:background}. As usual for Amazon Mechanical
Turk, the participants tended to be younger and predominantly male, but there was
diversity in other categories. A number of our participants reported using Knock
Codes on their devices as part of their authentication choice. As Knock Codes were a new interface to many participants, our design models the scenario where a user acquires and first uses an LG phone to perform the initial Knock Code set-up.

\tabdemotech{}

\paragraph{Estimating US Knock Code Usage.}
\new{We generalized our participants' device usage and authentication methods based on age and 
normalized it to the US population using census data~\cite{agebreakdown2018, censusdata2018}. We saw that LG's market share in the US had a range between 8\% to 12\% among the estimated 285,300,000 smartphone users~\cite{lgshares2019, smartphonemarketshare19}. Using that, as well as a 95\% confidence interval, as our lower and upper bounds, we conclude that there are potentially many Knock Code users: 728,693 to 2,567,207 in the US alone. We believe, though, that the actual adoption rate is most likely on the lower end. While this may be an optimistic estimate, it still suggests that there is a substantial number of Knock Code users in the general public, particularly worldwide.}

\new{Even though Knock Codes are not as widely adopted as other traditional methods of mobile authentication, it is still important to study user behavior with real-world, deployed authentication systems. In addition, on Google Play many Knock Code apps can be installed on any Android device, thus not limiting Knock Codes to solely LG devices. For instance,  the most highly rated Knock Code app on Android,  ``Knock Lock,'' boasts more than 1 million installations and claims that it is an innovative lock screen that ``will leave intruders baffled''~\cite{knocklockGplay}. This app is just one among the plethora of Knock Code knock-off apps that can be found on Google Play, indicating that this authentication method may have a higher adoption rate and influence on mobile authentication systems than appears initially. 
}




\section{Limitations}
\label{sec-limits}

There are a number of limitations associated with our methodology and survey
design. One such limitation is that the survey's recall component
occurred within a short time frame with minimal distraction tasks. While we can
report on short-term memorability of Knock Codes, we cannot report on the
memorability over extended time periods, e.g., days.



  However, as a mobile unlock
authentication method, users must recall their codes frequently, hence
short-term recall is still relevant. The increased use of biometrics, which
reduces the number of knowledge-based recalls, confounds the issue though, and
more research would be needed to better understand long-term memorability of
Knock Codes.

There are also some limitations on how likely the selected Knock Codes would be
real Knock Codes of real users. We believe that the simple
interface and the nature of the initial device setup suggest that these Knock Codes
would be akin to those used on real devices. Most of our participants were
unfamiliar with Knock Codes when taking the survey and so would be new users of
LG devices setting up their Knock Code for the first time.  It should also be
noted that a few participants who do use Knock Codes (both in the preliminary study and main study) reported that they reused
their Knock Code in the survey.

Nevertheless, we attempted to address this limitation and thus decided to provide different security
scenarios for which participants should create Knock Codes. This technique was
used by Loge et al.~\cite{loge-16-pattern-user-choice} when collecting Android
Unlock Patterns. The motivation is that different scenarios, one always being
device unlock, will help users to be more careful about their choices,
similar to how they may be during device setup. In analyzing the data (\Cref{sec:results}), we did not find significant differences between the Knock
Codes selected under each scenario for the \blocklistshort{} treatment but did see some differences for the \controlshort{} and \big{} treatment. 


\tablefreqpilot{}

\tablefreqnew{}

\section{Statistics of Knock Codes}
The first step in analyzing Knock Codes is to determine the frequency
statistics.
Table~\ref{tab:topfreqnew} displays the 30 most frequent patterns,
combined, across the scenarios for three treatments of the main study.  
The frequencies which we observed in the preliminary study are shown in Table~\ref{tab:topfreqpilot}. The preliminary study codes and the con-2x2 codes have a lot of overlap, with 42.0\% of the Knock Codes from the preliminary study appearing in the top-30 most frequent codes in the Control 2x2 treatment. This helps justify using the most frequent preliminary study codes as the basis of the blocklist for the bla-2x2 treatment.

\figstartendpilot{}

\paragraph{Code frequency.}
The most common Knock Code in our control dataset is \knockcodesmall{0,1,2,3,0,1} ($freq=6.9\,\%$). 
It starts in the upper left corner, follows a left-to-right sequence, and is repeated until the minimum length of 6 is reached.
We observe a similar strategy for the code \knockcodebig{0,1,2,3,4,5} ($freq=4.6\,\%$) which is the most frequent one in the \big{} treatment.
However, participants were able to reach the minimum length without repeating the pattern because of the larger grid.

The second most common Knock Code \knockcodesmall{0,1,3,2,0,1} ($freq=3.9\,\%$) in the \control{} treatment starts in the upper left quadrant, moving clockwise.
In contrast to this, \knockcodebig{0,3,4,5,2,1} ($freq=4.2\,\%$), the second most used code in the \big{} treatment, has different attributes: 
participants proceed diagonally over the grid, going down in a right-left movement for the first diagonal and up in a left-right movement for the second one.
The first half of the third most used Knock Code \knockcodebig{0,3,4,1,2,5} ($freq=3.8\,\%$) is identical, yet, it differs at the second diagonal which follows a top-down movement instead of bottom-up.

The third most used Knock Code in the \control{} treatment (\knockcodesmall{0,0,1,1,2,2}, $freq=3.5\,\%$) pursues a left-to-right sequence again, however, participants used double taps to comply with the required minimum length of 6 knocks. 

Participants of the \blocklist{} treatment used this strategy to an even greater extent: 
the three most used Knock Codes all contain multiple double taps and 51.0\,\% of all codes created for this treatment include one or more repeated taps.
In contrast to this, only 41.0\,\% of the codes in the \control{} treatment and 29.0\,\% of the codes in the \big{} treatment contain at least one repeated tap.
Moreover, the distribution of Knock Codes in the \blocklist{} treatment is more equal compared to the other two. 
The most used Knock Code, \knockcodesmall{0,0,2,2,1,1}, occurs in only $2.6\,\%$ of the cases and  as can be seen in Table~\ref{tab:topfreqnew} the distribution flattens the fastest. 

To summarize, the frequencies of the Knock Codes show different characteristics depending on the assigned treatment, suggesting natural, human tendencies in the selection that can be leveraged in predicting and guessing Knock Codes. We take advantage of this observation when guessing codes.
Participants in the \blocklist{} group use more repeated taps whereas codes created for the 2x3 treatment make use of the larger grid and follow directional patterns.
Knock Codes created for the \control{} depict a mix and follow both strategies equally.

\paragraph{Start/end quadrant frequency.}
Figure~\ref{fig:startfreq} and \ref{fig:endfreq} present the frequency of start and end taps in the Knock Codes.
Clearly, there is a strong tendency to begin codes in the
upper-left. Similar observations were made for Android Graphical
Patterns~\cite{uellenbeck} and is likely due to the left-to-right nature of the English language which is dominant among our participants. 
The least common starting points in the preliminary study as well as the control and blocklist treatment were in the lower row. 
In the \big{} treatment, on the other hand, the middle row is used the least often.

To understand the left/right and up/downshifting of the Knock Codes' start locations we mapped the Cartesian coordinate to each quadrant in the grid, where (-1,1) is the upper left quadrant
\knockcodesmall{0}, (1,1) is the upper right quadrant 
\knockcodesmall{1}, (-1,-1) is the lower left quadrant
\knockcodesmall{2}, and (1,-1) is the lower right quadrant
\knockcodesmall{3}.
Similarly, in the \big{} treatment, we mapped the coordinates (-1,1), (1,1), (-1,0), (1,0), (-1,-1), and (-1,1) to the grid spaces, scanning left to right, top to bottom. We then computed the average $x$ and $y$ coordinate for the start and end taps, across treatments.

A Shapiro Wilk's test ($p < 0.001$) indicated that the generated frequencies are not normally distributed, so a Mann-Whitney \textit{U} test was used to identify any initial significance, followed by a posthoc test with Bonferroni correction. 
We found significant differences between both the \control{} and \big{} treatment ($p < 0.001$) as well as \blocklist{} and \big{} ($p < 0.001$), suggesting that the larger grid size affected how participants chose to start and end their codes.

\paragraph{Code length.}
We also analyzed the Knock Codes with respect to length. The average code length was 
 6.4, 6.5, and 6.2 in each treatment, con-2x2, bla-2x2, and big-2x3, respectively. We observed  statistical
differences using ANOVA ($f= 11.57,\mbox{p < 0.001}$) between the treatments. 
In post hoc analysis, using pairwise $t$-test comparison, the difference lies primarily in the longer big-2x3 Knock Codes, which was statistically different from both bla-2x2 ($p<0.001$) and the con-2x2 ($p<0.001$). Surprisingly, the larger grid size encouraged slightly shorter Knock Codes.
Regardless, the vast majority of Knock Codes were of length
6, which was the median value, or 8, with a few codes of length 10.





\section{Security Analysis} 
\label{sec:results}
We now analyze the security of Knock Codes. 
We start by \shep{describing the threat model which we are considering for the attack}. Afterwards, we analyze the security of Knock Codes by using a perfect knowledge metric in Section~\ref{sec:perfectstrength} to define an upper bound on generic attack performance.
In Section~\ref{sec:simulatedstrength}, we assess the success rate of a simulated attacker to provide a more realistic security estimation.

\paragraph{\shep{Threat Model.}}
\shep{We consider a generic, non-targeted attacker that attempts to access an arbitrary victim's device by guessing the Knock Code without additional knowledge or previous observations of the victim. A targeted attacker who may know the victim's tendencies or previously observed an entry (e.g., via a shoulder surfing attack) would likely perform better than the generic attacker. A generic attacker, though, provides a lower bound on the scope of attacker performance, and it also provides a clear comparison point to other reported results~\cite{markert-20-pin-blocklist, wang2017understanding, bonneau2012birthday, aviv2015isbigger, uellenbeck} which use the same threat model.}

\shep{For the security analysis, we employ two different attacker variations. First is a {\em perfect knowledge attacker}, which assumes that the attacker has complete knowledge of the frequency order Knock Codes, from most to least frequent. This attack is still generic as the same strategy is assumed for every victim, and it allows one to estimate the security of the Knock Codes as selected by users. See Section~\ref{sec:perfectstrength} for more details.}

\shep{Second, a {\em simulated attacker} who knows a subset of the Knock Codes and constructs a model based on that observed distribution. The attacker then attempts to guess a set of arbitrary victims' (unknown) Knock Codes. We use a cross-fold validation to mimic the attacker, whereby the attacker trains on a subset of the data and guesses on an unknown test set.}

\paragraph{First-Entry 2x2 Codes.}
\new{Throughout this section, we refer to a  \textit{First-Entry 2x2} dataset which contains participants' first entered codes in the control and blocklist treatment. These codes may or may not have been confirmed (i.e., on the confirm entry screen) either due to lack of recall or because of the blocklist. 
We include this dataset, as it offers the perspective of an ideal user choice for how the authenticator may have been selected in the absence of external influences. 
As we expected, this dataset is slighter more secure than that of the confirmed control 2x2 codes and offers insights into how users compromise on security to gain more memorable codes.}

\subsection{Perfect Knowledge Strength Estimations} 
\label{sec:perfectstrength}
We consider the guessing strength of Knock Codes against a perfect knowledge attacker as described by Bonneau et al.~\cite{bonneau-12-entropy}. 
A perfect knowledge attack depicts the upper bound for an attack as it assumes that the attacker knows the attacked dataset and always guesses in the ideal order, that is, the Knock Code with the next highest frequency. 
This approach has been regularly applied to analyzing mobile authentication, such as Android Patterns~\cite{uellenbeck,aviv2015isbigger,song-15-pattern-psm} or PINs~\cite{wang2017understanding,markert-20-pin-blocklist}.

We use two different perfect-knowledge guessability metrics to evaluate Knock Codes, one based on an offline attack model and one based on an online (or throttled) attack model. 
An offline attack model assumes that the attacker can guess as many times as possible, while an online attack model assumes an attacker with a limited number of attempts. 
The online attack model better matches the realities of mobile authentication, where users typically have a maximal number of attempts before the device is locked out. 
The offline attack model, on the other hand, provides a more holistic approach to measuring the security of a set of user-chosen passwords.

\tabperfectknowledgetreatments{}

For an offline attack metric, we use {\em partial guessing entropy} or {\em $\alpha$-guesswork} ($\widetilde{G}_{\alpha}$). 
Partial guessing entropy estimates the amount of guesswork that is needed to guess a fraction~$\alpha$ of all codes. 
The Min-entropy $H_{\infty}$ depicts a special case as it is only based on the most frequent Knock Code.
As an online (or throttled) attack metric, we use {\em $\beta$-success rate}. 
It essentially measures what fraction of codes would be guessed if the attacker only had $\beta$ guesses, e.g., $\lambda_{3}$ considers an attack which is limited to  3~guesses.


Table~\ref{tab:treatments-perfectknowledge} shows the guessing results for our three treatments as well as the combined dataset First-Entry 2x2.
As an additional comparison we included datasets from previous studies for Android patterns~\cite{aviv2015isbigger} as well as 4- and 6- digit PINs~\cite{amitay-11-iphone-pins, wang2017understanding}. 
Because the datasets all differ in size which would influence the results, we downsampled all marked datasets to the size of \control{} and blocklist 2x2 (232 entries) and calculated the statistics for the samples. 
To rule out any sampling bias, we repeated this process 500 times, removed outliers using Tukey fences with $k=1.5$, and report the median value of the remaining set in Table~\ref{tab:treatments-perfectknowledge}.
\shepherd{With a 95\,\% confidence level the margin of error is lower than 0.3\,\% for the online guessing and lower than 0.1 bits for the offline case.}

  

Across all comparisons, we find that Knock Codes in the control 2x2 are significantly weaker in terms of their guessability.
This means, Knock Codes as they are currently deployed are more guessable than both 4- and 6-digit PINs as well as Android Patterns. 
When considering the First-Entry dataset, the differences are less distinct, but even in this ideal case, the inferiority of Knock Codes remains. 


Surprisingly, increasing the size of the keyspace by enlarging the grid size to 2x3 offers only little security gain.
Moreover, in some cases increasing the grid size may even {\em decrease} security.
This is most apparent when considering a throttled attacker.
After 10 guesses, 31.5\% of the \big{} codes are guessed compared to 28.0\,\% for the \control{} codes. 
A similar observation can be made after 30 guesses, 53.4\,\% of \big{} codes are guessed compared to  51.3\,\% of \control{} codes.

Future works needs to examine why larger Knock Codes performed so poorly, but a similar phenomenon was observed by Aviv et al. with increasing Android patterns from 3x3 to 4x4 grid sizes~\cite{aviv2015isbigger}. 
Aviv et al. conjectured, and we do so here as well, that there may be a false sense of security that the larger set of choices offers, whereby users believe their individual choice matters less in the face of the increased number of possibilities. 
Analyzing other grid sizes, such as 3x2 or 3x3, would offer additional insight; nevertheless, it is interesting to see that providing more complexity in how to select Knock Codes does not increase the security.

Finally, we observed strong security improvements with the introduction of a blocklist. 
As compared to the \controlshort{}, the blocklist cuts the success rate of an attacker within the first 30 attempts by 30\,\% to 50\,\% and increases the guesswork by \textasciitilde 1.5 bits when considering an offline attacker.
While the blocklist clearly encouraged more diverse choices, it also had the side effect of increasing user frustration and usability, as we describe later in Section~\ref{sec:usability}.

\subsection{Simulated Attacker Strength}
\label{sec:simulatedstrength}
We are also interested in modeling a more realistic, limited-knowledge attacker that has access to a subset of training data and attempts to guess some test set of unknown data: a {\em simulated attacker}.

A simulated attacker must model Knock Codes from a training set to predict a test set.
We used a three-gram Markov model probability estimator for the likelihood of a given Knock Code, based on the empirical observations in the test set. 
This is a standard approach when analyzing user chose secrets, e.g., passwords~\cite{castelluccia-12-adaptive, golla-18-psm}, PINs~\cite{wang2017understanding}, or Android Patterns~\cite{uellenbeck, aviv2015isbigger}.  
In order to encode the start and end transitions, we defined special symbols for transitions to ending/starting nodes. This can be defined more formally:
\[
x = \{x_{-(g-1)},\ldots,x_{-1},x_{0}, x_1, \ldots, x_n, \ldots, x_{n+g-1}\}
\]
where $x$ is the Knock Code of length $n$ with first knock $x_1$, and $g$ is the gram size.
If $i \le 0$ or $i > n$, then this indicates that $x_i$ is a start or end
transition state. These extra states are used to capture the early and late transitions taken by a user, for example, for the following Knock Code
\knockcodesmall{0,1,2,3}
, we would produce the following set of tri-grams, where $\bot$ is a start state and $\top$ is an end state: ($\bot$ \ $\bot$
\knockcodesmall{0}
),
($\bot$%
\knockcodesmall{0,1}
), 
(
\knockcodesmall{0,1,2}
),
(
\knockcodesmall{1,2,3}
),
(
\knockcodesmall{2,3}
$\top$)
(
\knockcodesmall{3}
$\top$ \ $\top$).

Using the transition probabilities, as measured in the training data, the attacker can calculate a likelihood measure of a Knock Code by considering the following
Markov model formulation, 
\begin{equation}\label{eq:markov}
\begin{split}
  P(x) &= P(\mathsf{len}(x)) \cdot P(\mathsf{start}(x)) \cdot P(\mathsf{end}(x)) \cdot \\
       & \prod^{n+(g-1)}_{i=-(g-1)} P(x_{i} \ldots x_{i+g} \ | \ x_{i-1}\ldots x_{i-1+g})
\end{split}
\end{equation}
where $P(\cdot)$ is the probability function, $\mathsf{len}(x)$ is the length function, $\mathsf{start}(x)$ is the start function, and $\mathsf{end}(x)$ is the end function. These are our prior probabilities that capture the likelihood of a given length, start quadrant, and end quadrant. The transition probabilities are captured using the conditional probabilities of each transition between each sub-sequence of length $g$, given the prior state. As not all transitions are represented in our dataset, we used constant smoothing to avoid zero probabilities.

The simulated attackers guessing routine, given a training set, is to (1) create a Markov model of the training data; (2) guess patterns in frequency order of the training set, with ties broken by the likelihood estimation; and (3) guess from a set of additional Knock Codes (not in the training set) sorted based on the likelihood estimation. For (3), we generated a list of all length 6-to-8 Knock Codes for the 2x2 and 2x3 grid sizes, excluding those in our training set that were previously guessed.
This accounted for 1,384,872 and 72,520,440 additional 2x2 and 2x3 Knock Codes that could be guessed, respectively. In our blocklist treatment, we assumed the attacker had knowledge of the blocklist.

\tabguessing{}

\figguessing{}

The results of our simulated attacker are presented in Table~\ref{tab:guessing}, and a graphical representation is provided in Figure~\ref{fig:guessing}. We report on the average of five randomized cross-fold validations. \new{As expected, the simulated attacker performs worse than the perfect-knowledge attacker, but we find similar results comparing across treatments. Notably, the 2x3 Knock Codes offer little, or worse, security while there is marked improvement for the blocklist informed 2x2 Knock Codes.}

\tabmemcode{}


\vspace{-.1in}
\section{Usability of Knock Codes}
\label{sec:usability}
In this section, we focus on the usability metrics of Knock Codes. We first report results on the setup and recall times.
Afterwards, we will focus on memorability and recall rates within our study, followed by the qualitative and quantitative responses to security and usability prompts. 

\tabtimeslarge{}

\subsection{Setup and Recall Times}
\shepherd{Table~\ref{tab:timing} presents the average selection and recall times, as well as the number of attempts, needed to select a Knock Code. 
Outliers were removed using Tukey fences with $k=1.5$.}

\shep{Participants needed on average  16.2\,s and 18.4\,s to select and confirm a 2x2 and 2x3 Knock Code, respectively. This is faster than the blocklist treatment (22.5\,s), where participants also had to make more attempts due to blocklisting (1.5 vs. 1.1 attempts).
In comparison, setting up a 4- or 6-digit PIN takes on average only 7.9 and 11.5 seconds respectively~\cite{markert-20-pin-blocklist} which is significantly faster than Knock Codes. 
While the described discrepancy between Knock Codes and PINs is distinct, the numbers for PINs may be lower since users are presumably more familiar with PINs as compared to Knock Codes. The differences may decrease with increased familiarity with Knock Codes.}

\shepherd{In terms of the recall, which can be compared to unlocking a smartphone, the 2x2 (7.2\,s per attempt) and 2x3 Knock Codes (7.1\,s per attempt) are more efficient than Knock Codes selected with a blocklist (7.4\,s per attempt). With 1.2 attempts per entry, it took participants 11.3 seconds to enter their Knock Codes for the blocklist treatment. 
Compared to entering an Android pattern (3.0\,s) or a PIN (4.7\,s)~\cite{harbach-14-hard-lock-life}, clear usability issues with Knock Codes emerge as entering them is twice as slow. With greater use of Knock Codes, these differences may decrease, but it is unlikely that Knock Codes will be as efficient to enter as patterns or PINs.}

\subsection{Memorability}
\label{sec:memorability}
We will now go into more details on the memorability as it depicts an important benchmark for any authentication method.
We analyzed the memorability of Knock Codes by looking at the recall rates at the end of the survey. While this is an imperfect measure for the memorability, as the survey took most participants less than 10 minutes to complete, it does speak to potential underlying usability issues, particularly if codes were not properly recalled in this short window.


We separated the recall rates based on each treatment.
The con-2x2 treatment participants successfully recalled their codes 88.8\,\% of the time.
The participants with the larger 2x3 grid had higher recall rates of  92.9\,\%, which may suggest an interesting usability vs. security trade-off as this group chose shorter and also some of the weakest Knock Codes. 
However, we did not find significant differences between the con-2x2 and big-2x3 recall rates using a $\chi^2$ test. We would expect long term memorability rates to be equally high, but further study would be needed to confirm that conjecture.

The worst recall rate came from participants in the bla-2x2 treatment: 80.6\,\% successfully recalled their Knock Code, and the result was significantly different from the other two recall rates ($p<0.0001$ for both comparison tests). This could be attributed to the impact of the blocklist, where participants who hit the blocklist had lower recall rates (66.0\,\%) than those that did not (84.9\,\%). Most likely, the blocklist affected users in two ways. First, participants who chose blocklisted codes were forced to select multiple codes until landing on one that was not blocklisted. The average number of blocklisting events per user who hit the blocklist was 1.4. Second, that final Knock Code chosen ended up being more complex (as evident in the prior section), and thus harder to recall. Again, this suggests a clear trade-off between usability and security.
  

We also analyzed the number of attempts to successfully recall a Knock Code. We found no statistical difference across all treatments between the attempts made in recalling the first or second scenario Knock Code correctly.  In the big-2x3 treatment and the con-2x2 treatment, participants took on average 
1.1 attempts when recalling a Knock Code correctly, with 3 attempts as the maximum. For the  bla-2x2 group, users took on average 
1.2 attempts to correctly recall a Knock Code, again having a maximum of 3 attempts.  Again, we find bla-2x2`s result to be significantly different in terms of the number of attempts made in the other treatments ($p<0.001$ vs.  big-2x3  and  $p<0.001$ vs. con-2x2), thus showing that the blocklist has an impact on recalling Knock Codes, even for those participants that eventually correctly do so. 
It is important to note though, that users had a maximum limit of 3 attempts to recall their code before we considered it ``cannot be recalled'' for the purpose of expediting the survey.



We also analyzed how participants failed to recall their Knock Codes by calculating the average edit distances between the submitted code and the true code for both recalls attempts, one for each scenario.  We determined that there was no statistical difference between the average edit distances among treatments. The average edit distance between correct and incorrect recalls was 3.6, suggesting that when users get a code wrong, they get it wrong by a large margin, as the median length Knock Code is 6.

\subsection{User Responses}

Users provided their opinions and insights regarding Knock Codes'
usability and security. We coded these free responses using two
independent reviewers where disputes in coding were resolved until consensus was reached. The specific codes and their frequencies are presented in the Appendices.

Overall Knock Codes were perceived positively by users,
citing that they were ``Easy,'' ``Quick,'' and ``Hard to Guess.''
The uniqueness of Knock Codes also appealed to users who indicated they
especially liked the fact that it is a ``Discreet'' and ``Secure''
authentication method which can be inconspicuous and hidden from others.

For many of the participants, this was a new method of authentication, and they
employed various tactics when choosing their Knock Codes.  We observed such
strategies in determining memorable yet secure codes. To make the Knock Code
more memorable, the majority of users opted to use some sort of ``Pattern'' or ``Variation'' 
that would be ``Simple.'' Other techniques users employed include 
``Directional,'' ``Shape,'' ``Game,''  and ``Repeated.''  Often,
users would create codes based on something ``Personal'' to them, such as
the letter of a word that had meaning to the user.

While many users did not have a specific strategy for security and still focused on making their code  ``Easy to Remember'' as the main
priority, others determined that using ``All Quadrants'' or `` Multiple Regions,''  as well as
making the code  ``Long'' or ``Random'' or being ``Unexpected'' and ``Different'' would secure
their codes. Making their codes ``Hard to Guess'' often included attempts to
obfuscate the number of clicks and the regions, using speed and potentially unpredictable tactics.  Users continued to use similar tactics
for memorability to double as security in their Knock Codes, for instance having
``Repeated'' regions.

\definecolor{strong_agree}{HTML}{549C77} 
\definecolor{agree}{HTML}{ABD4B4} 
\definecolor{neutral}{HTML}{A0A0A0} 
\definecolor{disagree}{HTML}{EF959A} 
\definecolor{strong_disagree}{HTML}{CE2029} 

\begin{figure}[t]
\centering
\renewcommand*{\arraystretch}{0.7}
    \centering
\resizebox{0.8\linewidth}{!}{
    \setlength{\tabcolsep}{0pt}

    \begin{tabular}{r c}
        \toprule
        \textbf{Group} & \textbf{Question}\\
        \midrule

\multicolumn{2}{c}{Knock Codes are a secure authenticator.} \\
\midrule

\raisebox{0.3em}{con-2x2} & \scalebox{0.7}{
\begin{tikzpicture}
\begin{axis}[xbar stacked,bar width=1cm,ytick=\empty,y post scale=0.07,xmin=0,xmax=100,legend style={at={(0.15,-2)},
legend style={cells={align=left, anchor=center, fill}, nodes={inner sep=0.4ex,below=-2ex}},
column sep=0cm, draw=none, anchor=west, legend columns=5},xticklabel style={opacity=0,yshift=12pt}]
\addplot[fill=strong_agree, xbar legend, mark=none] coordinates {(15.52,1)};
\addplot[fill=agree, xbar legend, mark=none] coordinates {(54.31,1)};
\addplot[fill=neutral, xbar legend, mark=none] coordinates {(16.38,1)};
\addplot[fill=disagree, xbar legend, mark=none] coordinates {(10.34,1)};
\addplot[fill=strong_disagree, xbar legend, mark=none] coordinates {(3.45,1)};
\end{axis}
\end{tikzpicture}}\\

\raisebox{0.3em}{bla-2x2} & \scalebox{0.7}{
\begin{tikzpicture}
\begin{axis}[xbar stacked,bar width=1cm,ytick=\empty,y post scale=0.07,xmin=0,xmax=100,legend style={at={(0,-3)},
legend style={cells={align=left, anchor=center, fill}, nodes={inner sep=0.4ex,below=-2ex}},
column sep=0cm, draw=none, anchor=west, legend columns=5},xticklabel style={opacity=0,yshift=12pt}]
\addplot[fill=strong_agree, xbar legend, mark=none] coordinates {(22.41,1)};
\addplot[fill=agree, xbar legend, mark=none] coordinates {(41.38,1)};
\addplot[fill=neutral, xbar legend, mark=none] coordinates {(19.83,1)};
\addplot[fill=disagree, xbar legend, mark=none] coordinates {(15.52,1)};s
\addplot[fill=strong_disagree, xbar legend, mark=none] coordinates {(0.86,1)};
\end{axis}
\end{tikzpicture}}\\

\raisebox{0.3em}{big-2x3} & \scalebox{0.7}{
\begin{tikzpicture}
\begin{axis}[xbar stacked,bar width=1cm,ytick=\empty,y post scale=0.07,xmin=0,xmax=100,legend style={at={(0.15,-2)},
legend style={cells={align=left, anchor=center, fill}, nodes={inner sep=0.4ex,below=-2ex}},
column sep=0cm, draw=none, anchor=west, legend columns=5},xticklabel style={opacity=0,yshift=12pt}]
\addplot[fill=strong_agree, xbar legend, mark=none] coordinates {(16.81,1)};
\addplot[fill=agree, xbar legend, mark=none] coordinates {(46.22,1)};
\addplot[fill=neutral, xbar legend, mark=none] coordinates {(24.37,1)};
\addplot[fill=disagree, xbar legend, mark=none] coordinates {(10.08,1)};
\addplot[fill=strong_disagree, xbar legend, mark=none] coordinates {(2.52,1)};
\end{axis}
\end{tikzpicture}}\\

\midrule
\multicolumn{2}{c}{Knock Codes are more secure than PIN codes.} \\
\midrule

\raisebox{0.3em}{con-2x2} & \scalebox{0.7}{
\begin{tikzpicture}
\begin{axis}[xbar stacked,bar width=1cm,ytick=\empty,y post scale=0.07,xmin=0,xmax=100,legend style={at={(0.15,-2)},
legend style={cells={align=left, anchor=center, fill}, nodes={inner sep=0.4ex,below=-2ex}},
column sep=0cm, draw=none, anchor=west, legend columns=5},xticklabel style={opacity=0,yshift=12pt}]
\addplot[fill=strong_agree, xbar legend, mark=none] coordinates {(11.21,1)};
\addplot[fill=agree, xbar legend, mark=none] coordinates {(22.41,1)};
\addplot[fill=neutral, xbar legend, mark=none] coordinates {(31.03,1)};
\addplot[fill=disagree, xbar legend, mark=none] coordinates {(22.41,1)};
\addplot[fill=strong_disagree, xbar legend, mark=none] coordinates {(12.93,1)};
\end{axis}
\end{tikzpicture}}\\

\raisebox{0.3em}{bla-2x2} & \scalebox{0.7}{
\begin{tikzpicture}
\begin{axis}[xbar stacked,bar width=1cm,ytick=\empty,y post scale=0.07,xmin=0,xmax=100,legend style={at={(0.15,-2)},
legend style={cells={align=left, anchor=center, fill}, nodes={inner sep=0.4ex,below=-2ex}},
column sep=0cm, draw=none, anchor=west, legend columns=5},xticklabel style={opacity=0,yshift=12pt}]
\addplot[fill=strong_agree, xbar legend, mark=none] coordinates {(13.79,1)};
\addplot[fill=agree, xbar legend, mark=none] coordinates {(22.41,1)};
\addplot[fill=neutral, xbar legend, mark=none] coordinates {(27.59,1)};
\addplot[fill=disagree, xbar legend, mark=none] coordinates {(26.72,1)};
\addplot[fill=strong_disagree, xbar legend, mark=none] coordinates {(9.48,1)};
\end{axis}
\end{tikzpicture}}\\

\raisebox{0.3em}{big-2x3} & \scalebox{0.7}{
\begin{tikzpicture}
\begin{axis}[xbar stacked,bar width=1cm,ytick=\empty,y post scale=0.07,xmin=0,xmax=100,legend style={at={(0.15,-2)},
legend style={cells={align=left, anchor=center, fill}, nodes={inner sep=0.4ex,below=-2ex}},
column sep=0cm, draw=none, anchor=west, legend columns=5},xticklabel style={opacity=0,yshift=12pt}]
\addplot[fill=strong_agree, xbar legend, mark=none] coordinates {(10.17,1)};
\addplot[fill=agree, xbar legend, mark=none] coordinates {(16.95,1)};
\addplot[fill=neutral, xbar legend, mark=none] coordinates {(36.44,1)};
\addplot[fill=disagree, xbar legend, mark=none] coordinates {(26.27,1)};
\addplot[fill=strong_disagree, xbar legend, mark=none] coordinates {(10.17,1)};
\end{axis}
\end{tikzpicture}}\\

\midrule
\multicolumn{2}{c}{Knock Codes are more secure than~alphanumeric~passwords.} \\
\midrule

\raisebox{0.3em}{con-2x2} & \scalebox{0.7}{
\begin{tikzpicture}
\begin{axis}[xbar stacked,bar width=1cm,ytick=\empty,y post scale=0.07,xmin=0,xmax=100,legend style={at={(0.15,-2)},
legend style={cells={align=left, anchor=center, fill}, nodes={inner sep=0.4ex,below=-2ex}},
column sep=0cm, draw=none, anchor=west, legend columns=5},xticklabel style={opacity=0,yshift=12pt}]
\addplot[fill=strong_agree, xbar legend, mark=none] coordinates {(6.9,1)};
\addplot[fill=agree, xbar legend, mark=none] coordinates {(13.79,1)};
\addplot[fill=neutral, xbar legend, mark=none] coordinates {(30.17,1)};
\addplot[fill=disagree, xbar legend, mark=none] coordinates {(29.31,1)};
\addplot[fill=strong_disagree, xbar legend, mark=none] coordinates {(19.83,1)};
\end{axis}
\end{tikzpicture}}\\

\raisebox{0.3em}{bla-2x2} & \scalebox{0.7}{
\begin{tikzpicture}
\begin{axis}[xbar stacked,bar width=1cm,ytick=\empty,y post scale=0.07,xmin=0,xmax=100,legend style={at={(0.15,-2)},
legend style={cells={align=left, anchor=center, fill}, nodes={inner sep=0.4ex,below=-2ex}},
column sep=0cm, draw=none, anchor=west, legend columns=5},xticklabel style={opacity=0,yshift=12pt}]
\addplot[fill=strong_agree, xbar legend, mark=none] coordinates {(7.76,1)};
\addplot[fill=agree, xbar legend, mark=none] coordinates {(24.14,1)};
\addplot[fill=neutral, xbar legend, mark=none] coordinates {(21.55,1)};
\addplot[fill=disagree, xbar legend, mark=none] coordinates {(25,1)};
\addplot[fill=strong_disagree, xbar legend, mark=none] coordinates {(21.55,1)};
\end{axis}
\end{tikzpicture}}\\

\raisebox{0.3em}{big-2x3} & \scalebox{0.7}{
\begin{tikzpicture}
\begin{axis}[xbar stacked,bar width=1cm,ytick=\empty,y post scale=0.07,xmin=0,xmax=100,legend style={at={(0.15,-2)},
legend style={cells={align=left, anchor=center, fill}, nodes={inner sep=0.4ex,below=-2ex}},
column sep=0cm, draw=none, anchor=west, legend columns=5},xticklabel style={opacity=0,yshift=12pt}]
\addplot[fill=strong_agree, xbar legend, mark=none] coordinates {(8.85,1)};
\addplot[fill=agree, xbar legend, mark=none] coordinates {(16.81,1)};
\addplot[fill=neutral, xbar legend, mark=none] coordinates {(50.44,1)};
\addplot[fill=disagree, xbar legend, mark=none] coordinates {(18.58,1)};
\addplot[fill=strong_disagree, xbar legend, mark=none] coordinates {(5.31,1)};
\end{axis}
\end{tikzpicture}}\\

\midrule
\multicolumn{2}{c}{\centering Knock Codes are more secure than~Android~patterns.} \\
\midrule

\raisebox{0.3em}{con-2x2} & \scalebox{0.7}{
\begin{tikzpicture}
\begin{axis}[xbar stacked,bar width=1cm,ytick=\empty,y post scale=0.07,xmin=0,xmax=100,legend style={at={(0.15,-2)},
legend style={cells={align=left, anchor=center, fill}, nodes={inner sep=0.4ex,below=-2ex}},
column sep=0cm, draw=none, anchor=west, legend columns=5},xticklabel style={opacity=0,yshift=12pt}]
\addplot[fill=strong_agree, xbar legend, mark=none] coordinates {(8.77,1)};
\addplot[fill=agree, xbar legend, mark=none] coordinates {(21.05,1)};
\addplot[fill=neutral, xbar legend, mark=none] coordinates {(42.98,1)};
\addplot[fill=disagree, xbar legend, mark=none] coordinates {(21.05,1)};
\addplot[fill=strong_disagree, xbar legend, mark=none] coordinates {(6.14,1)};
\end{axis}
\end{tikzpicture}}\\

\raisebox{0.3em}{bla-2x2} & \scalebox{0.7}{
\begin{tikzpicture}
\begin{axis}[xbar stacked,bar width=1cm,ytick=\empty,y post scale=0.07,xmin=0,xmax=100,legend style={at={(0.15,-2)},
legend style={cells={align=left, anchor=center, fill}, nodes={inner sep=0.4ex,below=-2ex}},
column sep=0cm, draw=none, anchor=west, legend columns=5},xticklabel style={opacity=0,yshift=12pt}]
\addplot[fill=strong_agree, xbar legend, mark=none] coordinates {(8.11,1)};
\addplot[fill=agree, xbar legend, mark=none] coordinates {(27.03,1)};
\addplot[fill=neutral, xbar legend, mark=none] coordinates {(44.14,1)};
\addplot[fill=disagree, xbar legend, mark=none] coordinates {(13.51,1)};
\addplot[fill=strong_disagree, xbar legend, mark=none] coordinates {(7.21,1)};
\end{axis}
\end{tikzpicture}}\\

\raisebox{2.3em}{big-2x3} & \scalebox{0.7}{
\begin{tikzpicture}
\begin{axis}[xbar stacked,bar width=1cm,ytick=\empty,y post scale=0.07,xmin=0,xmax=100,legend style={at={(-0.06,-2)},
legend style={cells={align=left, anchor=center, fill}, nodes={inner sep=0.4ex,below=-2ex}},
column sep=0cm, draw=none, anchor=west, legend columns=5}]
\addplot[fill=strong_agree, xbar legend, mark=none] coordinates {(8.85,1)};
\addplot[fill=agree, xbar legend, mark=none] coordinates {(16.81,1)};
\addplot[fill=neutral, xbar legend, mark=none] coordinates {(50.44,1)};
\addplot[fill=disagree, xbar legend, mark=none] coordinates {(18.58,1)};
\addplot[fill=strong_disagree, xbar legend, mark=none] coordinates {(5.31,1)};
\addlegendentry{Strg. Agr.};
\addlegendentry{Agr.};
\addlegendentry{Neut.};
\addlegendentry{Disagr.};
\addlegendentry{Strg. Disagr.};
\end{axis}
\end{tikzpicture}}\\

    \bottomrule
    \end{tabular}}
    
  \caption{Likert response to comparisons to other mobile authentication methods.}
  \label{fig:likert2}
  \vspace{-.2in}
\end{figure}
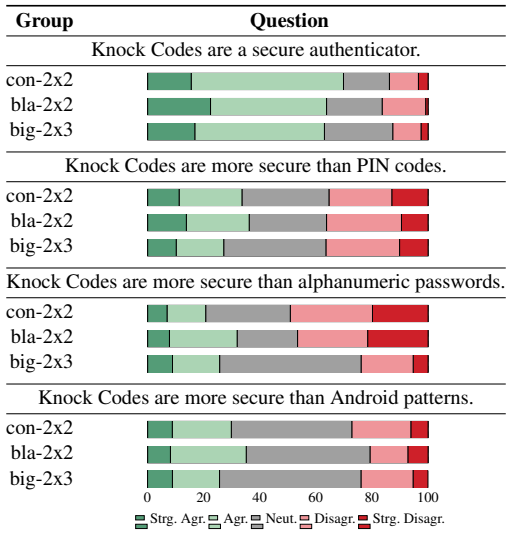

Upon comparing Knock Codes with other forms of security, on average users found
passwords, PINs, and Android patterns to be more secure than Knock Codes (see Figure~\ref{fig:likert2}). Overall, users
found Knock Codes adequately secure, i.e., being difficult to hack, resistant
to smudge attacks and shoulder surfing. However, they were not completely
convinced about Knock Codes' security. Users expressed what they disliked overall,
specifically that they found Knock Codes ``Hard to Remember'' and
``Insecure,'' paving the way for an attacker to easily guess a Knock
Code. They also found the interface provided ``No improvement'' and disliked how it was `` Hard to type-in'' the Knock Codes.

To have a more general opinion of the overall usability of Knock Codes, we employed
the System Usability Scale (SUS). The full Likert responses are found in the Appendices. The average response for the con-2x2 treatment is 69.8, the big-2x3 is 68.1, and the bla-2x2 is 64.3. These scores are generally rated as ``ok'' or ``marginal,'' with only the control treatment potentially offering some above-average usability.


\tabSUSaverages{}

%
%

 
%
%
%
%


\section{Discussion}
\label{sec:discussion}
\vspace{-.1in}

\shep{As reported above, while most participants offered some positive thoughts, their perception of the security of Knock Codes lagged behind other deployed options, and the SUS values for all schemes were ``ok'' or ``marginal.'' There was some positive feedback on Knock Codes which suggests an openness to new designs in mobile authentication, particularly to authentication that can be entered while the phone screen is off. There was also increased perceptions of security from targeted attacks, e.g., via shoulder surfing\cite{deluca2014nowyouseeme,eiband2017understanding,aviv-17-shoulder-surfing-baseline}. It is reasonable to view Knock Codes as offering new design concepts that can ultimately improve mobile authentication.}

\shep{However, we find that Knock Codes, as currently deployed, provide weaker security than other available knowledge-based, mobile unlock methods, such as 4-/6-digit PINs and Android patterns. This is far from the ``perfect security'' promised by LG's advertisement of Knock Codes. As such, we cannot recommend deploying Knock Codes in their current form as compared to alternative authentication options.}

\shep{Our results also indicate that a straightforward improvement like increasing the grid size to 2x3 may offer little or worse security. 
Blocklisting common Knock Codes, on the other hand, does provide more resilience to a throttled attacker, as has been found in password authentication~\cite{habib-17-blocklist} and PINs~\cite{markert-20-pin-blocklist}. 
Yet, blocklisting runs the risk of increasing user frustration during selection, but since selecting a Knock Code is a one-time event, the usability trade-off of adding a blocklist may be {\em extremely worthwhile} if Knock Codes continue to be available to LG users. 
It may also be worthwhile for designers to invest in other methods for improving Knock Code selection, e.g., forcing users to start or end at given quadrants, similar to SysPal~\cite{cho-17-syspal}, or using multi-touch, like chords~\cite{oakley2018personal}.}



\vspace{-.15in}
\section{Conclusion}
\vspace{-.1in}
We performed the first comprehensive user study and security analysis of user-chosen Knock Codes using a three-treatment, between groups study: a control 2x2 treatment, a blocklist 2x2 treatment, and a 2x3 treatment. \shep{We find that Knock Codes provide weaker security than other mobile unlock authentication, such as 4-digit PINs, 6-digit PINs, and Android pattern, and that increasing the grid size offered little (or worse) security outcomes, while the addition of a blocklist of common codes substantially increased the security against a throttled attacker. However, Knock Codes suffered in terms of usability, both in terms of entry/recall time and user perception.}



\vspace{-.15in}
\begin{small}
\paragraph{Acknowledgments}
Thank you Timothy Forman, Maximilian Golla and Genry Krichevsky. This material is based upon work supported by the National Science Foundation under Grants Nos. 1845300 and 1617584; the research training group ``Human Centered Systems Security'' (NERD.nrw) sponsored by the state of North Rhine-Westphalia, Germany; and
the Army Research Laboratory Cooperative Agreement Number W911NF-13-2-0045 (ARL Cyber Security CRA).
\end{small}

\balance
\bibliographystyle{plain}
\bibliography{ref} 

\begin{thebibliography}{10}

\bibitem{amitay-11-iphone-pins}
Daniel Amitay.
\newblock {Most Common iPhone Passcodes}, June 2011.
\newblock
  \url{http://danielamitay.com/blog/2011/6/13/most-common-iphone-passcodes}, as
  of \today.

\bibitem{andriotis2014complexity}
Panagiotis Andriotis, Theo Tryfonas, and George Oikonomou.
\newblock {Complexity Metrics and User Strength Perceptions of the Pattern-Lock
  Graphical Authentication Method}.
\newblock In {\em Conference on Human Aspects of Information Security, Privacy
  and Trust}, HAS~'14, pages 115--126. Springer, Heraklion, Crete, Greece, June
  2014.

\bibitem{andriotis2013pilot}
Panagiotis Andriotis, Theo Tryfonas, George Oikonomou, and Can Yildiz.
\newblock {A Pilot Study on the Security of Pattern Screen-Lock Methods and
  Soft Side Channel Attacks}.
\newblock In {\em ACM Conference on Security and Privacy in Wireless and Mobile
  Networks}, WiSec~'13, pages 1--6, Budapest, Hungary, April 2013. ACM.

\bibitem{aviv2015isbigger}
Adam~J. Aviv, Devon Budzitowski, and Ravi Kuber.
\newblock {Is Bigger Better? Comparing User-Generated Passwords on 3x3 vs. 4x4
  Grid Sizes for Android's Pattern Unlock}.
\newblock In {\em Annual Computer Security Applications Conference}, ACSAC~'15,
  pages 301--310, Los Angeles, California, USA, December 2015. ACM.

\bibitem{aviv-17-shoulder-surfing-baseline}
Adam~J. Aviv, John~T. Davin, Flynn Wolf, and Ravi Kuber.
\newblock {Towards Baselines for Shoulder Surfing on Mobile Authentication}.
\newblock In {\em Annual Conference on Computer Security Applications},
  ACSAC~'17, pages 486--498, Orlando, Florida, USA, December 2017. ACM.

\bibitem{aviv2010smudge}
Adam~J. Aviv, Katherine Gibson, Evan Mossop, Matt Blaze, and Jonathan~M. Smith.
\newblock {Smudge Attacks on Smartphone Touch Screens}.
\newblock In {\em USENIX Workshop on Offensive Technologies}, WOOT~'10, pages
  1--7, Washington, District of Columbia, USA, August 2010. USENIX.

\bibitem{aviv2012accel}
Adam~J. Aviv, Benjamin Sapp, Matt Blaze, and Jonathan~M. Smith.
\newblock {Practicality of Accelerometer Side Channels on Smartphones}.
\newblock In {\em Annual Computer Security Applications Conference}, ACSAC~'12,
  pages 41--50, Orlando, Florida, USA, December 2012. ACM.

\bibitem{bhagavatula2015biometric}
Chandrasekhar Bhagavatula, Blase Ur, Kevin Iacovino, Su~Mon Kywey, Lorrie~Faith
  Cranor, and Marios Savvides.
\newblock {Biometric Authentication on iPhone and Android: Usability,
  Perceptions, and Influences on Adoption}.
\newblock In {\em Workshop on Usable Security}, USEC~'15, San Diego,
  California, USA, February 2015. ISOC.

\bibitem{bonneau-12-entropy}
Joseph Bonneau.
\newblock {The Science of Guessing: Analyzing an Anonymized Corpus of 70
  Million Passwords}.
\newblock In {\em IEEE Symposium on Security and Privacy}, SP~'12, pages
  538--552, San Jose, California, USA, May 2012. IEEE.

\bibitem{bonneau2012birthday}
Joseph Bonneau, S{\"o}ren Preibusch, and Ross Anderson.
\newblock {A Birthday Present Every Eleven Wallets? The Security of
  Customer-Chosen Banking PINs}.
\newblock In {\em Financial Cryptography and Data Security}, FC~'12, pages
  25--40, Kralendijk, Bonaire, February 2012. Springer.

\bibitem{brooke1996sus}
John Brooke.
\newblock {SUS: A Quick and Dirty Usability Scale}.
\newblock {\em Usability Evaluation in Industry}, pages 189--194, 1996.

\bibitem{cai2011touchlogger}
Liang Cai and Hao Chen.
\newblock {TouchLogger: Inferring Keystrokes on Touch Screen from Smartphone
  Motion}.
\newblock In {\em Workshop on Hot Topics in Security}, HotSec~'11, Berkeley,
  California, USA, August 2011. USENIX.

\bibitem{castelluccia-12-adaptive}
Claude Castelluccia, Markus D{\"u}rmuth, and Daniele Perito.
\newblock {Adaptive Password-Strength Meters from Markov Models}.
\newblock In {\em Symposium on Network and Distributed System Security},
  NDSS~'12, San Diego, California, USA, February 2012. ISOC.

\bibitem{cherapau2015impact}
Ivan Cherapau, Ildar Muslukhov, Nalin Asanka, and Konstantin Beznosov.
\newblock {On the Impact of Touch ID on iPhone Passcodes}.
\newblock In {\em Symposium on Usable Privacy and Security}, SOUPS~'15, pages
  257--276, Ottawa, Canada, July 2015. USENIX.

\bibitem{cho-17-syspal}
Geumhwan Cho, Jun~Ho Huh, Junsung Cho, Seongyeol Oh, Youngbae Song, and
  Hyoungshick Kim.
\newblock {SysPal: System-Guided Pattern Locks for Android}.
\newblock In {\em IEEE Symposium on Security and Privacy}, SP~'17, pages
  338--356, San Jose, California, USA, May 2017. IEEE.

\bibitem{clarke2005authentication}
N.L. Clarke and S.M. Furnell.
\newblock {Authentication of Users on Mobile Telephones -- A Survey of
  Attitudes and Practices}.
\newblock {\em Computers \& Security}, 24(7):519--527, October 2005.

\bibitem{lgshares2019}
{Comscore, Inc.}
\newblock {Top OEMs - Share of Smartphone Subscribers 3 Month Avg. Ending Nov.
  2019 vs. 3 Month Avg. Ending Sep. 2019}, September 2019.
\newblock
  \url{https://www.comscore.com/Insights/Rankings#tab_mobile_smartphone_oems},
  as of \today.

\bibitem{cubrilovic-09-rockyou}
Nik Cubrilovic.
\newblock {RockYou Hack: From Bad To Worse}, December 2009.
\newblock
  \url{https://techcrunch.com/2009/12/14/rockyou-hack-security-myspace-facebook-passwords/},
  as of \today.

\bibitem{deluca2014nowyouseeme}
Alexander De~Luca, Marian Harbach, Emanuel {von Zezschwitz}, Max-Emanuel
  Maurer, Bernhard~Ewald Slawik, Heinrich Hussmann, and Matthew Smith.
\newblock {Now You See Me, Now You Don't: Protecting Smartphone Authentication
  from Shoulder Surfers}.
\newblock In {\em ACM Conference on Human Factors in Computing Systems},
  CHI~'14, pages 2937--2946, Toronto, Ontario, Canada, April 2014. ACM.

\bibitem{de2007evaluation}
Alexander De~Luca, Roman Weiss, and Heiko Drewes.
\newblock {Evaluation of Eye-Gaze Interaction Methods for Security Enhanced
  PIN-Entry}.
\newblock In {\em Australasian Computer-Human Interaction Conference},
  OZCHI~'07, pages 199--202, Adelaide, Australia, November 2007. ACM.

\bibitem{deyle2006accessible}
Travis Deyle and Volker Roth.
\newblock {Accessible Authentication via Tactile PIN Entry}.
\newblock {\em Computer Graphics Topics}, 3:24--26, 2006.

\bibitem{eiband2017understanding}
Malin Eiband, Mohamed Khamis, Emanuel von Zezschwitz, Heinrich Hussmann, and
  Florian Alt.
\newblock {Understanding Shoulder Surfing in the Wild:Stories from Users and
  Observers}.
\newblock In {\em ACM Conference on Human Factors in Computing Systems},
  CHI~'17, pages 4254--4265, Denver, Colorado, USA, May 2017. ACM.

\bibitem{forget2010eyegaze}
Alain Forget, Sonia Chiasson, and Robert Biddle.
\newblock {Shoulder-Surfing Resistance with Eye-Gaze Entry inCued-Recall
  Graphical Passwords}.
\newblock In {\em ACM Conference on Human Factors in Computing Systems},
  CHI~'10, pages 1107--1110, Atlanta, Georgia, USA, April 2010. ACM.

\bibitem{golla-18-psm}
Maximilian Golla and Markus D\"{u}rmuth.
\newblock {On the Accuracy of Password Strength Meters}.
\newblock In {\em ACM Conference on Computer and Communications Security},
  CCS~'18, pages 1567--1582, Toronto, Ontario, Canada, October 2018. ACM.

\bibitem{habib-17-blocklist}
Hana Habib, Jessica Colnago, William Melicher, Blase Ur, Sean~M. Segreti, Lujo
  Bauer, Nicolas Christin, and Lorrie~Faith Cranor.
\newblock {Password Creation in the Presence of Blocklists}.
\newblock In {\em Workshop on Usable Security}, USEC~'17, San Diego,
  California, USA, February 2017. ISOC.

\bibitem{harbach-16-the-anatomy}
Marian Harbach, Alexander De~Luca, and Serge Egelman.
\newblock {The Anatomy of Smartphone Unlocking: A Field Study of Android Lock
  Screens}.
\newblock In {\em ACM Conference on Human Factors in Computing Systems},
  CHI~'16, pages 4806--4817, San Jose, California, USA, May 2016. ACM.

\bibitem{harbach-14-hard-lock-life}
Marian Harbach, Emanuel {von Zezschwitz}, Andreas Fichtner, Alexander De~Luca,
  and Matthew Smith.
\newblock {It's a Hard Lock Life: A Field Study of Smartphone (Un)Locking
  Behavior and Risk Perception}.
\newblock In {\em Symposium on Usable Privacy and Security}, SOUPS~'14, pages
  213--230, Menlo Park, California, USA, July 2014. USENIX.

\bibitem{kelley-12-again}
Patrick Kelley, Saranga Kom, Michelle~L. Mazurek, Rich Shay, Tim Vidas, Lujo
  Bauer, Nicolas Christin, Lorrie~Faith Cranor, and Julio L{\'o}pez.
\newblock {Guess Again (and Again and Again): Measuring Password Strength by
  Simulating Password-Cracking Algorithms}.
\newblock In {\em IEEE Symposium on Security and Privacy}, SP~'12, pages
  523--537, San Jose, California, USA, May 2012. IEEE.

\bibitem{klein1990foiling}
Daniel~V. Klein.
\newblock {``Foiling the Cracker'': A Survey Of, and Improvements To, Password
  Security}.
\newblock In {\em UNIX Security Workshop}, UNIX~'90, pages 5--14, Portland,
  Oregon, USA, August 1990. USENIX.

\bibitem{knocklockGplay}
{Knock Lock}.
\newblock {Knock Lock Screen - Applock}, 2020.
\newblock
  \url{https://play.google.com/store/apps/details?id=com.knocklock.applock&hl=en_US},
  as of \today.

\bibitem{krombholz2016use}
Katharina Krombholz, Thomas Hupperich, and Thorsten Holz.
\newblock {Use the Force: Evaluating Force-Sensitive Authentication for Mobile
  Devices}.
\newblock In {\em Symposium on Usable Privacy and Security}, SOUPS~'16, pages
  207--219, Denver, Colorado, USA, July 2016. USENIX.

\bibitem{kuber2010toward}
Ravi Kuber and Shiva Sharma.
\newblock {Toward Tactile Authentication for Blind Users}.
\newblock In {\em ACM SIGACCESS Conference on Computers and Accessibility},
  ASSETS~'10, pages 289--290, Orlando, Florida, USA, October 2010. ACM.

\bibitem{kuber2010feasibility}
Ravi Kuber and Wai Yu.
\newblock {Feasibility Study of Tactile-Based Authentication}.
\newblock {\em International Journal of Human-Computer Studies},
  68(3):158--181, March 2010.

\bibitem{kuber2010tactile}
Ravi Kuber and Wai Yu.
\newblock {Toward Tactile Authentication for Blind Users}.
\newblock In {\em International Conference on Human Haptic Sensing and Touch
  Enabled Computer Applications}, EuroHaptics~'10', pages 314--319, Amsterdam,
  Netherlands, July 2010. Springer.

\bibitem{lindsey1981survey}
William~C. Lindsey and Chak~Ming Chie.
\newblock {A Survey of Digital Phase-Locked Loops}.
\newblock {\em Proceedings of the IEEE}, 69(4):410--431, April 1981.

\bibitem{loge-16-pattern-user-choice}
Marte L{\o}ge, Markus D\"urmuth, and Lillian R{\o}stad.
\newblock {On User Choice for Android Unlock Patterns}.
\newblock In {\em European Workshop on Usable Security}, EuroUSEC~'16,
  Darmstadt, Germany, July 2016. ISOC.

\bibitem{man2003shoulder}
Shushuang Man, Dawei Hong, and Manton Matthews.
\newblock {A Shoulder-Surfing Resistant Graphical Password Scheme -- WIW}.
\newblock In {\em International Conference on Security and Management},
  SAM~'03, pages 105--111, Las Vegas, Nevada, USA, June 2003. CSREA Press.

\bibitem{markert-20-pin-blocklist}
Philipp Markert, Daniel~V. Bailey, Maximilian Golla, Markus D\"{u}rmuth, and
  Adam~J. Aviv.
\newblock {This PIN Can Be Easily Guessed: Analyzing the Security of Smartphone
  Unlock PINs}.
\newblock In {\em IEEE Symposium on Security and Privacy}, SP~'20, pages
  1525--1542, San Francisco, California, USA, May 2020. IEEE.

\bibitem{mazurek-13-guessability-uni}
Michelle~L. Mazurek, Saranga Komanduri, Timothy Vidas, Lujo Bauer, Nicolas
  Christin, Lorrie~Faith Cranor, Patrick~Gage Kelley, Richard Shay, and Blase
  Ur.
\newblock {Measuring Password Guessability for an Entire University}.
\newblock In {\em Conference on Computer and Communications Security}, CCS~'13,
  pages 173--186, Berlin, Germany, October 2013. ACM.

\bibitem{melicher-16-mobile-passwords}
William Melicher, Darya Kurilova, Sean~M. Segreti, Pranshu Kalvani, Richard
  Shay, Blase Ur, Lujo Bauer, Nicolas Christin, Lorrie~Faith Cranor, and
  Michelle~L. Mazurek.
\newblock {Usability and Security of Text Passwords on Mobile Devices}.
\newblock In {\em ACM Conference on Human Factors in Computing Systems},
  CHI~'16, pages 527--539, Santa Clara, California, USA, May 2016. ACM.

\bibitem{oakley2018personal}
Ian Oakley, Jun~Ho Huh, Junsung Cho, Geumhwan Cho, Rasel Islam, and Hyoungshick
  Kim.
\newblock {The Personal Identification Chord: A Four Button Authentication
  System for Smartwatches}.
\newblock In {\em ACM Asia Conference on Computer and Communications Security},
  ASIA~CCS~'18, pages 75--87, Incheon, Republic of Kore, June 2018. ACM.

\bibitem{prabhakar2003biometric}
Salil Prabhakar, Sharath Pankanti, and Anil~K Jain.
\newblock {Biometric Recognition: Security and Privacy Concerns}.
\newblock {\em IEEE Security \& Privacy}, 1(2):63--69, March 2003.

\bibitem{schaub2012password}
Florian Schaub, Ruben Deyhle, and Michael Weber.
\newblock {Password Entry Usability and Shoulder Surfing Susceptibility on
  Different Smartphone Platforms}.
\newblock In {\em International Conference on Mobile and Ubiquitous
  Multimedia}, MUM~'12, pages 13:1--13:10, Ulm, Germany, December 2012. ACM.

\bibitem{song-15-pattern-psm}
Youngbae Song, Geumhwan Cho, Seongyeol Oh, Hyoungshick Kim, and Jun~Ho Huh.
\newblock {On the Effectiveness of Pattern Lock Strength Meters: Measuring the
  Strength of Real World Pattern Locks}.
\newblock In {\em ACM Conference on Human Factors in Computing Systems},
  CHI~'15, pages 2343--2352, Seoul, Republic of Korea, April 2015. ACM.

\bibitem{smartphonemarketshare19}
{Team Counterpoint}.
\newblock {US Smartphone Market Share: By Quarter}, November 2019.
\newblock
  \url{https://www.counterpointresearch.com/us-market-smartphone-share/}, as of
  \today.

\bibitem{tofel-2014-LG-G2-knock-code}
Kevin~C. Tofel.
\newblock {LG G2 and G Flex Phones Getting the Knock Code Wake and Unlock
  Feature}, March 2014.
\newblock
  \url{https://gigaom.com/2014/03/25/lg-g2-and-g-flex-phones-getting-the-knock-code-wake-and-unlock-feature/},
  as of \today.

\bibitem{uellenbeck}
Sebastian Uellenbeck, Markus D\"{u}rmuth, Christopher Wolf, and Thorsten Holz.
\newblock {Quantifying the Security of Graphical Passwords: The Case of Android
  Unlock Patterns}.
\newblock In {\em ACM Conference on Computer and Communications Security},
  CCS~'13, pages 161--172, Berlin, Germany, November 2013. ACM.

\bibitem{agebreakdown2018}
{U.S. Census Bureau, Population Division}.
\newblock {Annual Estimates of the Resident Population by Single Year of Age
  and Sex for the United States: April 1, 2010 to July 1, 2018 , 2018
  Population Estimates}, June 2019.
\newblock
  \url{https://factfinder.census.gov/bkmk/table/1.0/en/PEP/2018/PEPSYASEXN?#},
  as of \today.

\bibitem{censusdata2018}
{U.S. Census Bureau, Population Division}.
\newblock {Annual Estimates of the Resident Population for Selected Age Groups
  by Sex for the United States, States, Counties, and Puerto Rico Commonwealth
  and Municipios: April 1, 2010 to July 1, 2018}, June 2019.
\newblock
  \url{https://factfinder.census.gov/bkmk/table/1.0/en/PEP/2018/PEPAGESEX?#},
  as of \today.

\bibitem{wang2017understanding}
Ding Wang, Qianchen Gu, Xinyi Huang, and Ping Wang.
\newblock {Understanding Human-Chosen PINs: Characteristics, Distribution and
  Security}.
\newblock In {\em ACM Asia Conference on Computer and Communications Security},
  ASIA~CCS~'17, pages 372--385, Abu Dhabi, United Arab Emirates, April 2017.
  ACM.

\bibitem{ye-17-pattern-five-attempts}
Guixin Ye, Zhanyong Tang, Dingyi Fang, Xiaojiang Chen, Kwang~In Kim, Ben
  Taylor, and Zheng Wang.
\newblock {Cracking Android Pattern Lock in Five Attempts}.
\newblock In {\em Symposium on Network and Distributed System Security},
  NDSS~'17, San Diego, California, USA, February 2017. ISOC.

\bibitem{zhou2018patternlistener}
Man Zhou, Qian Wang, Jingxiao Yang, Qi~Li, Feng Xiao, Zhibo Wang, and Xiaofen
  Chen.
\newblock {PatternListener: Cracking Android Pattern Lock Using Acoustic
  Signals}.
\newblock In {\em ACM Conference on Computer and Communications Security},
  CCS~'18, pages 1775--1787, Toronto, Ontario, Canada, October 2018. ACM.

\end{thebibliography}

\clearpage
\nobalance
\appendix
\begin{center}
{\LARGE APPENDICES}
\end{center}
\section{Survey Material}
\label{sec:survey}

\subsection{Main Study}
\label{sec:survey-main}

\setlength\parindent{0pt}
\textbf{1. \hspace{.1em} Device Usage Questions}
\label{survey-device-main}

When referring to ``mobile devices'' throughout this survey, consider these to include smartphones and tablet computers, such as iPhone and Android phones and tablets. Traditional laptop computers, two-in-one computers, like the Microsoft Surface, or e-readers, like the Amazon Kindle, are not considered mobile devices for the purposes of this survey.

\begin{enumerate}
    \item How many mobile devices do you use regularly? (Including phones and tablets, but excluding laptops) \newline
    $\circ$~0 \hspace{.5em}
    $\circ$~1 \hspace{.5em}
    $\circ$~2 \hspace{.5em}
    $\circ$~3 \hspace{.5em}
    $\circ$~4+ \hspace{.5em}
    $\circ$~Prefer not to say

    \item What brand of smartphone do you use? (Select all that apply) \\
    $\square$~Apple \hspace{.5em}
    $\square$~Samsung \hspace{.5em}
    $\square$~LG \hspace{.5em}
    $\square$~Google (Pixel/Nexus) \hspace{.5em}
    $\square$~Motorola \hspace{.5em}
    $\square$~ZTE \hspace{.5em}
    $\square$~I do not own a smartphone \hspace{.5em}
    $\square$~Other: \rule{5em}{.1pt}
  
    \item Select ``No'' as the answer to this questions: \\
    $\circ$~Yes \hspace{.5em}
    $\circ$~No \hspace{.5em}
    $\circ$~Sometimes \hspace{.5em}
    $\circ$~Always
    
    \item Which method(s) do you use to lock your mobile device(s)?(Select all that apply) \\
    $\square$~4-digit PIN \hspace{.5em}
    $\square$~6-digit PIN \hspace{.5em}
    $\square$~PIN of other length  
    $\square$~Android Graphical Pattern \hspace{.5em}
    $\square$~LG Knock Codes \hspace{.5em}
    $\square$~Fingerprint \hspace{.5em}
    $\square$~Face \hspace{.5em}
    $\square$~None \hspace{.5em}
    $\square$~Other: \rule{5em}{.1pt}

\end{enumerate}

\textit{Where indicated, the text and the graphics on the following pages changed depending on the assigned grid size.}\\

\textbf{2. \hspace{.1em} What are Knock Codes?}\\
Knock Codes are an authentication method used to unlock your smartphone, much like a PIN. 
To unlock the phone, the user enters their knock Code by tapping different regions (or quadrants) of a [2x2|2x3] grid on the smartphone display. 
The grid may or may not be displayed at the time of entry, for example, below is a video of someone entering a Knock Code without a grid displayed.

  \begin{center}
  \includegraphics[width=0.8\linewidth]{images/survey/knock-training-video.png}
  \end{center}

As part of this survey, you will be asked to select your own Knock Codes using an on-screen approximation of a smartphone. You will enter your codes by clicking on different regions of a [2x2|2x3] grid with your mouse. Below is an image of the [2x2|2x3] grid and smartphone approximation.

  \begin{center}
    \begin{minipage}{0.4\linewidth}
      \centering
      \includegraphics[width=\linewidth]{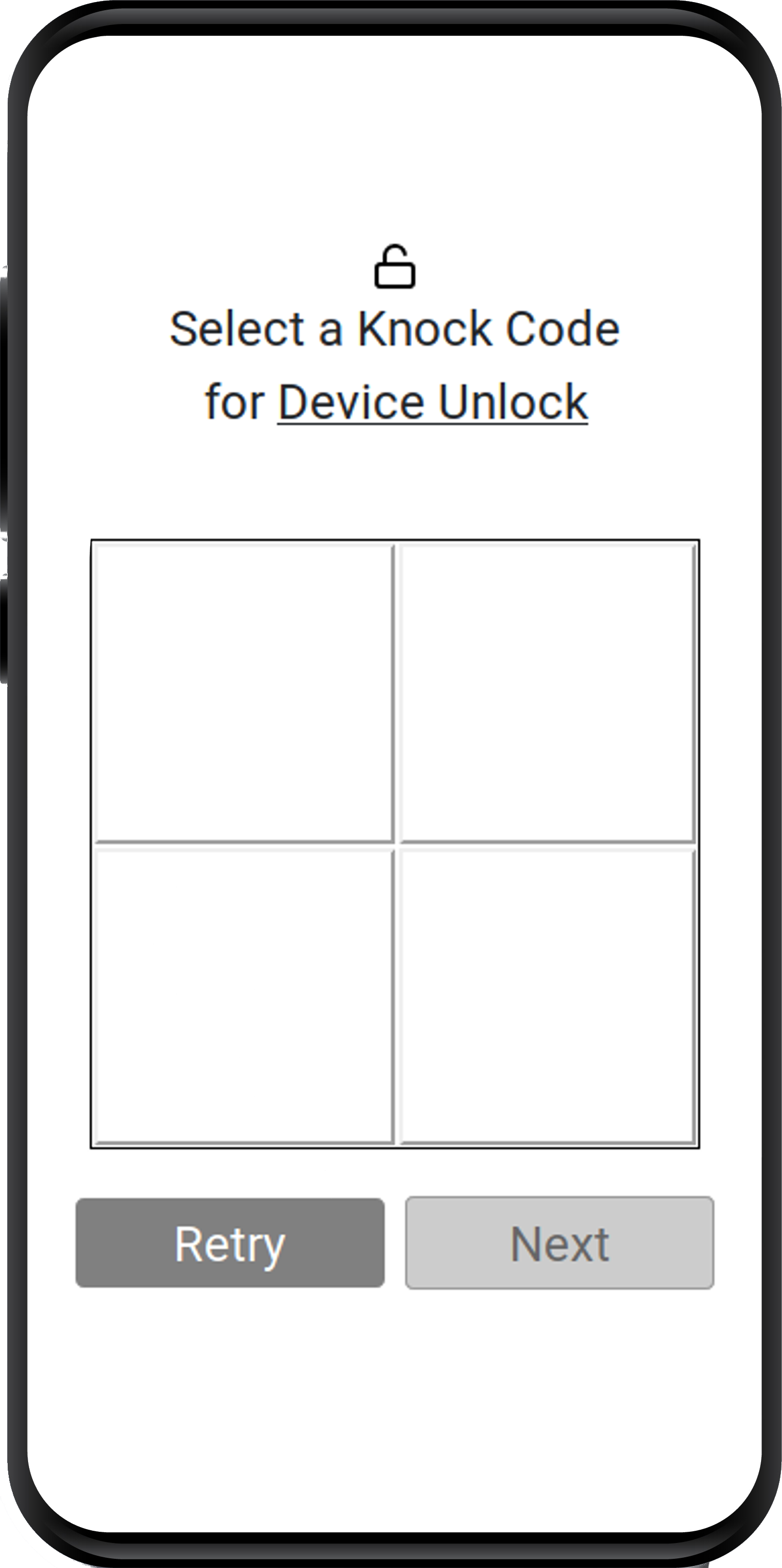}
      {\em con-2x2 \& bla-2x2}
    \end{minipage}
    \quad
    \begin{minipage}{0.4\linewidth}
      \centering
      \includegraphics[width=\linewidth]{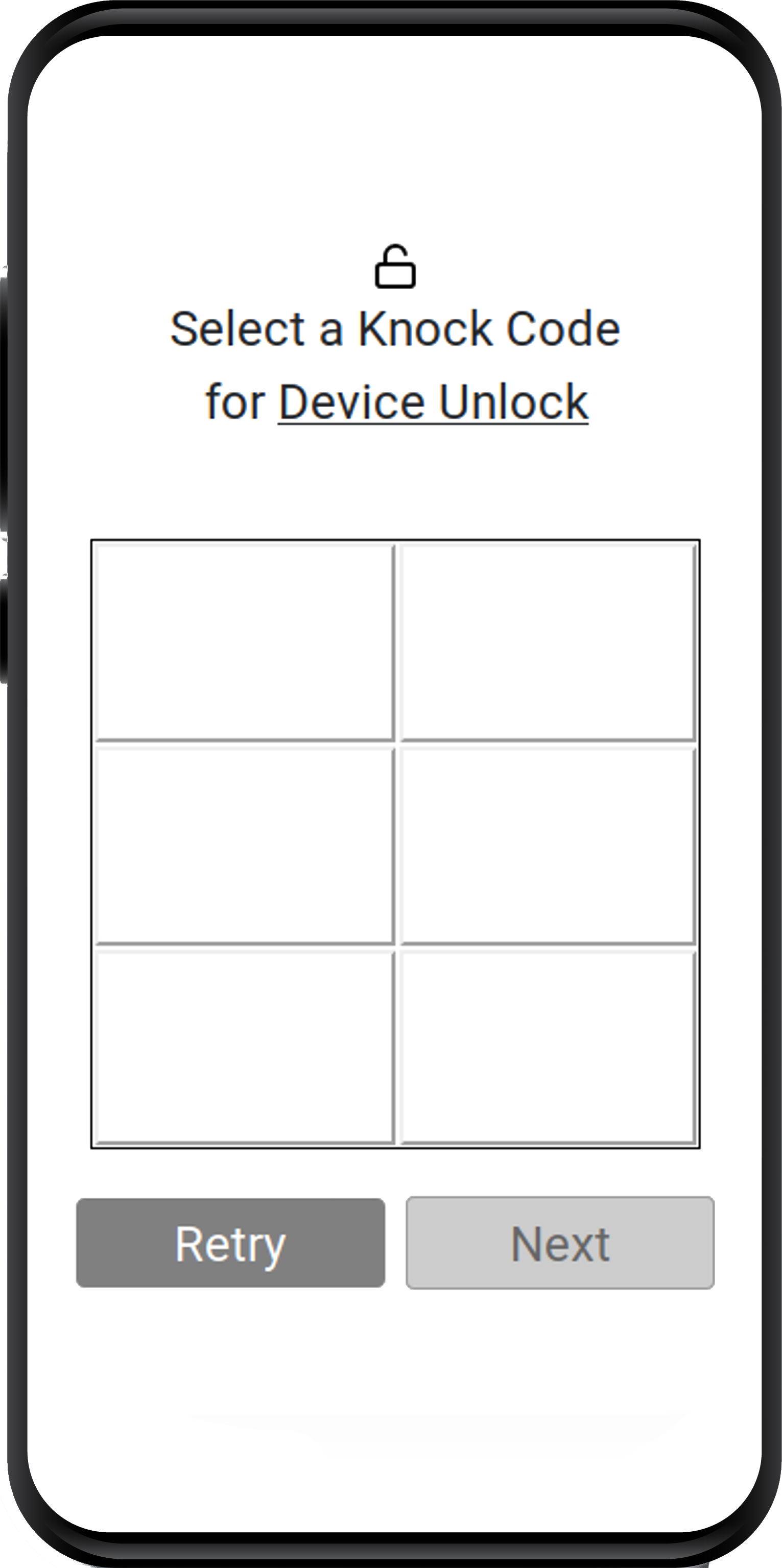}
      {\em big-2x3}
    \end{minipage}
  \end{center}

  There are some rules! When selecting a Knock Code it must:
  \begin{enumerate}
  \item Use \textit{at least} 3 regions of the grid. 
  \item Use \textit{at least} 6 total knocks.  
  \end{enumerate}

On the next page, you will have a chance to practice entering Knock Codes after which you will proceed with the rest of the survey. \\

\textit{Participants performed a practice run of using the interface. After completion, they were given the option to continue.} \\

\textbf{3. \hspace{.1em} Practice} 

  \begin{center}
    \begin{minipage}{0.44\linewidth}
      \centering
      \includegraphics[width=\linewidth]{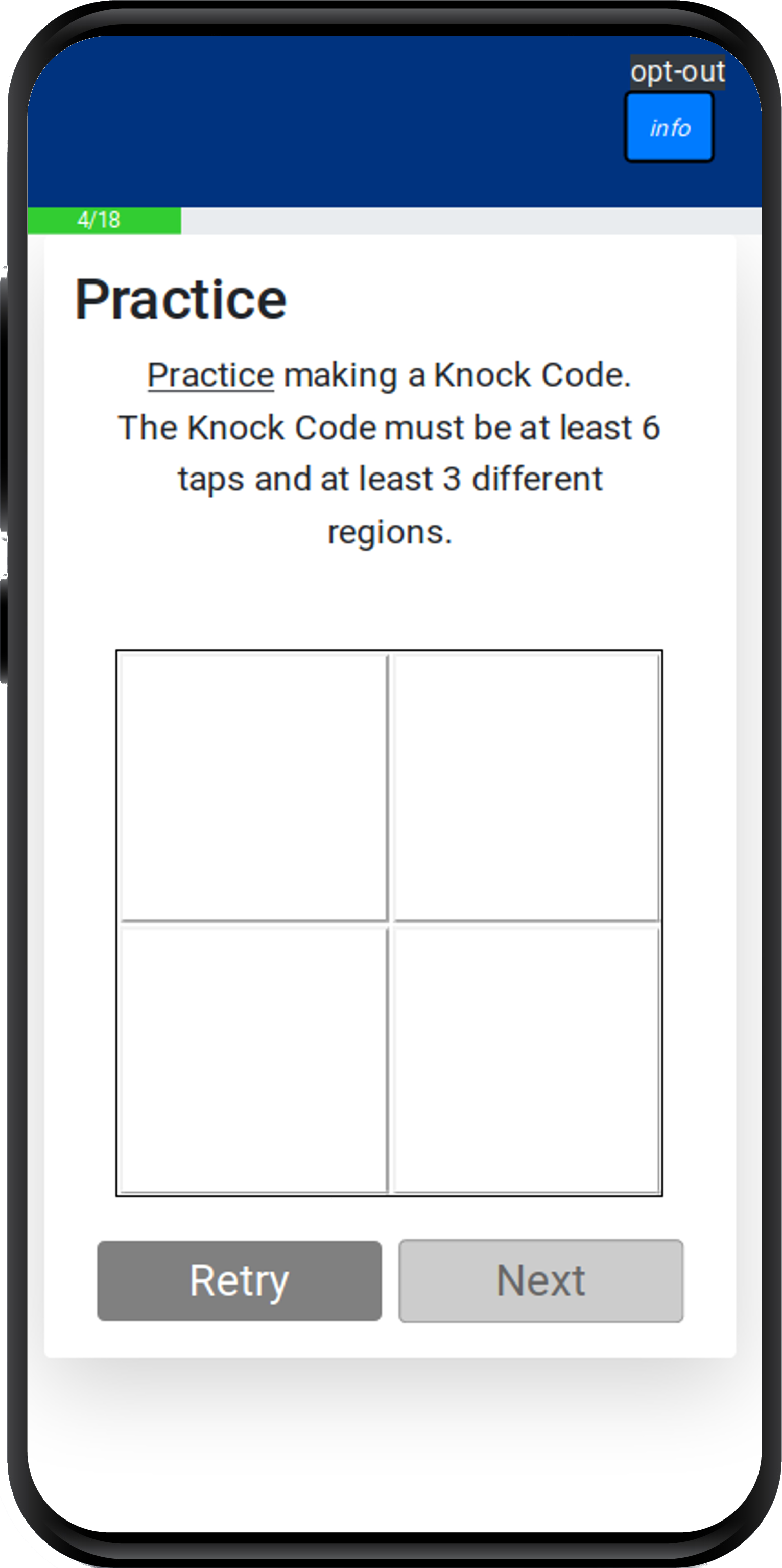}
      {\em con-2x2 \& bla-2x2}
    \end{minipage}
    \quad
    \begin{minipage}{0.44\linewidth}
      \centering
      \includegraphics[width=\linewidth]{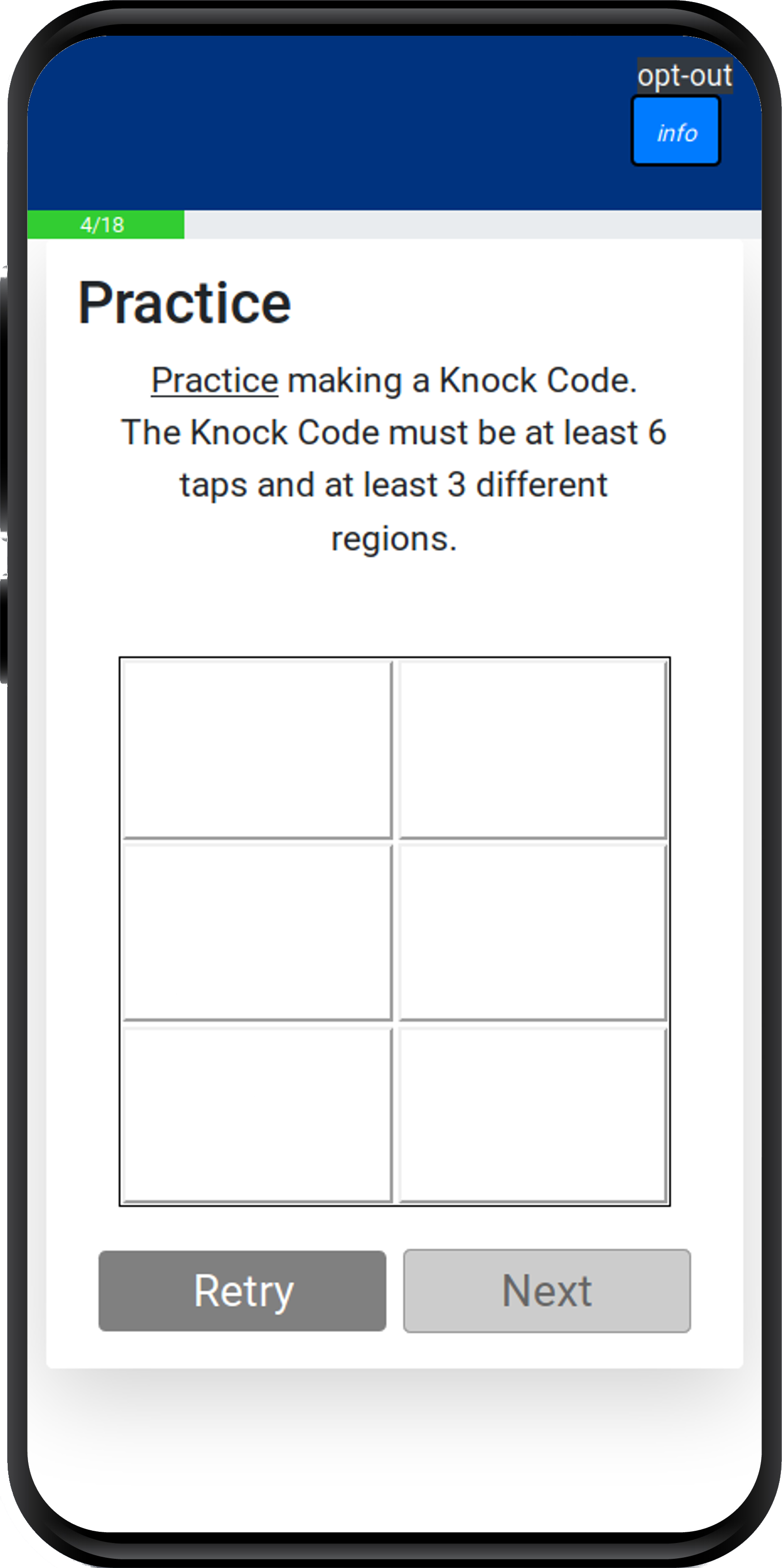}
      {\em big-2x3}
    \end{minipage}
  \end{center}

\textbf{4. \hspace{.1em} Scenarios}\\
For the remainder of this survey, you will be asked to create Knock Codes for different scenarios. 

Importantly, you will need to recall these codes later. So choose something that is secure and memorable. However, we ask that you DO NOT write down your codes or use other aids to help you remember.

\textbf{I understand that I should not write down my codes or use other aids to assist in the survey:} \\
$\circ$~I understand \\

You will be asked to create Knock-Knock Codes for the following scenarios. \\

\textit{All participants were assigned to Device Unlock, and then one of either Banking App or Shopping Cart.} \\

\includegraphics[width=.3cm]{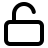} - \textit{Device Unlock} \hspace{.5em} Create a Knock Code you would use to {\bf unlock your smartphone or tablet}. \\
\includegraphics[width=.3cm]{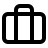} - \textit{Banking App} \hspace{1.1em} Create a Knock Code you would use to secure access to your {\bf mobile banking application}. \\
\includegraphics[width=.3cm]{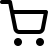} - \textit{Shopping Cart} \hspace{.5em} Create a Knock Code you would use to protect your {\bf Amazon shopping cart}. \\
 
\textbf{I understand that I should not write down my codes or use other aids to assist in the survey:} \\
$\circ$~I understand \\

\textit{Steps 5, 6, and 7 were done twice. First for the Device Unlock, then for the banking or shopping scenario.} \\

\textbf{5. \hspace{.1em} Selection} \\
Select a Knock Code for [SCENARIO] \\

\textbf{6. \hspace{.1em} Confirmation} \\
Confirm the Knock Code for [SCENARIO] \\

\textbf{7. \hspace{.1em} Thinking about the Knock Code you just chose, answer the following questions.} 
 \begin{enumerate}
  \item I feel this Knock Code provides adequate security for this scenario. \\
  $\circ$~Strongly Agree \hspace{.25em}
  $\circ$~Agree \hspace{.25em}
  $\circ$~Neither agree nor disagree \hspace{.25em}
  $\circ$~Disagree \hspace{.25em}
  $\circ$~Strongly disagree \hspace{.25em}
  
  \item It was difficult to choose this Knock Code. \\
  $\circ$~Strongly Agree \hspace{.25em}
  $\circ$~Agree \hspace{.25em}
  $\circ$~Neither agree nor disagree \hspace{.25em}
  $\circ$~Disagree \hspace{.25em}
  $\circ$~Strongly disagree \hspace{.25em}
      
  \item What strategy did you use to make your code {\bf more secure}? \\
  Answer:~\rule{10em}{.1pt}

  \item What strategy did you use to make your code {\bf more memorable}? \\
  Answer:~\rule{10em}{.1pt}

  \end{enumerate}

\textbf{8. \hspace{.1em} Please answer the following questions/prompts.} 

Select your agreement/disagreement with the following statements
 \begin{enumerate}
  \item Knock Codes are a secure authenticator. \\
  $\circ$~Strongly Agree \hspace{.25em}
  $\circ$~Agree \hspace{.25em}
  $\circ$~Neither agree nor disagree \hspace{.25em}
  $\circ$~Disagree \hspace{.25em}
  $\circ$~Strongly disagree \hspace{.25em}

  \item Knock Codes are more secure than PIN codes. \\
  $\circ$~Strongly Agree \hspace{.25em}
  $\circ$~Agree \hspace{.25em}
  $\circ$~Neither agree nor disagree \hspace{.25em}
  $\circ$~Disagree \hspace{.25em}
  $\circ$~Strongly disagree \hspace{.25em} \\
  $\circ$~Do not know what a PIN code is \hspace{.25em}
  
  \item Knock Codes are more secure than alphanumeric passwords. \\
  $\circ$~Strongly Agree \hspace{.25em}
  $\circ$~Agree \hspace{.25em}
  $\circ$~Neither agree nor disagree \hspace{.25em}
  $\circ$~Disagree \hspace{.25em}
  $\circ$~Strongly disagree \hspace{.25em} \\
  $\circ$~Do not know what an alphanumeric passwords is \hspace{.25em}

  \item Knock Codes are more secure than Android patterns. \\
  $\circ$~Strongly Agree \hspace{.25em}
  $\circ$~Agree \hspace{.25em}
  $\circ$~Neither agree nor disagree \hspace{.25em}
  $\circ$~Disagree \hspace{.25em}
  $\circ$~Strongly disagree \hspace{.25em} \\
  $\circ$~Do not know what an Android pattern is \hspace{.25em} 

Provide general feedback on the following questions

  \item What are some aspects you {\bf like} about Knock Codes? (use N/A if you do not wish to answer) \\
  Answer:~\rule{10em}{.1pt}
    
  \item What are some aspects you {\bf do not like} about Knock Codes? (use N/A if you do not wish to answer) \\
  Answer:~\rule{10em}{.1pt}

\end{enumerate}

\textbf{9. \hspace{.1em} Please answer the following questions/prompts.} \\
Select your agreement/disagreement with the following statements
  \begin{enumerate}
  \item I think that I would like to use Knock Codes frequently. \\
  $\circ$~Strongly Agree \hspace{.25em}
  $\circ$~Agree \hspace{.25em}
  $\circ$~Neither agree nor disagree \hspace{.25em}
  $\circ$~Disagree \hspace{.25em}
  $\circ$~Strongly disagree \hspace{.25em}

  \item I found Knock Codes unnecessarily complex. \\
  $\circ$~Strongly Agree \hspace{.25em}
  $\circ$~Agree \hspace{.25em}
  $\circ$~Neither agree nor disagree \hspace{.25em}
  $\circ$~Disagree \hspace{.25em}
  $\circ$~Strongly disagree \hspace{.25em}

  \item I thought Knock Codes were easy to use. \\
  $\circ$~Strongly Agree \hspace{.25em}
  $\circ$~Agree \hspace{.25em}
  $\circ$~Neither agree nor disagree \hspace{.25em}
  $\circ$~Disagree \hspace{.25em}
  $\circ$~Strongly disagree \hspace{.25em} 

  \item I think that I would need the support of a technical person to be able to use Knock Codes. \\
  $\circ$~Strongly Agree \hspace{.25em}
  $\circ$~Agree \hspace{.25em}
  $\circ$~Neither agree nor disagree \hspace{.25em}
  $\circ$~Disagree \hspace{.25em}
  $\circ$~Strongly disagree \hspace{.25em}

  \item I found the various functions in Knock Codes were well integrated. \\
  $\circ$~Strongly Agree \hspace{.25em}
  $\circ$~Agree \hspace{.25em}
  $\circ$~Neither agree nor disagree \hspace{.25em}
  $\circ$~Disagree \hspace{.25em}
  $\circ$~Strongly disagree \hspace{.25em}

  \item I thought there was too much inconsistency in Knock Codes. \\
  $\circ$~Strongly Agree \hspace{.25em}
  $\circ$~Agree \hspace{.25em}
  $\circ$~Neither agree nor disagree \hspace{.25em}
  $\circ$~Disagree \hspace{.25em}
  $\circ$~Strongly disagree \hspace{.25em} 

  \item I would imagine that most people would learn to use Knock Codes very quickly. \\
  $\circ$~Strongly Agree \hspace{.25em}
  $\circ$~Agree \hspace{.25em}
  $\circ$~Neither agree nor disagree \hspace{.25em}
  $\circ$~Disagree \hspace{.25em}
  $\circ$~Strongly disagree \hspace{.25em} 

  \item Select Agree as the answer to this question. \\
  $\circ$~Strongly Agree \hspace{.25em}
  $\circ$~Agree \hspace{.25em}
  $\circ$~Neither agree nor disagree \hspace{.25em}
  $\circ$~Disagree \hspace{.25em}
  $\circ$~Strongly disagree \hspace{.25em} 

  \item I found Knock Codes very cumbersome to use. \\
  $\circ$~Strongly Agree \hspace{.25em}
  $\circ$~Agree \hspace{.25em}
  $\circ$~Neither agree nor disagree \hspace{.25em}
  $\circ$~Disagree \hspace{.25em}
  $\circ$~Strongly disagree \hspace{.25em}

  \item I felt very confident using Knock Codes. \\
  $\circ$~Strongly Agree \hspace{.25em}
  $\circ$~Agree \hspace{.25em}
  $\circ$~Neither agree nor disagree \hspace{.25em}
  $\circ$~Disagree \hspace{.25em}
  $\circ$~Strongly disagree \hspace{.25em} 

  \item I needed to learn a lot of things before I could get going with Knock Codes. \\
  $\circ$~Strongly Agree \hspace{.25em}
  $\circ$~Agree \hspace{.25em}
  $\circ$~Neither agree nor disagree \hspace{.25em}
  $\circ$~Disagree \hspace{.25em}
  $\circ$~Strongly disagree \hspace{.25em} 
      
\end{enumerate}

\textbf{10. \hspace{.1em} Recall} \\
Recall your Knock Code for [SCENARIO] \\

\textbf{11. \hspace{.1em} Demographic Questions}
\label{survey-demo-main}

 \begin{enumerate}
  \item What is your age range: \\
    $\circ$~18-24 \hspace{.25em} 
    $\circ$~25-29 \hspace{.25em} 
    $\circ$~30-34 \hspace{.25em} 
    $\circ$~35-39 \hspace{.25em} 
    $\circ$~40-44 \hspace{.25em} 
    $\circ$~45-49 \hspace{.25em} 
    $\circ$~50-54 \hspace{.25em} 
    $\circ$~55-59 \hspace{.25em} 
    $\circ$~60-64 \hspace{.25em} 
    $\circ$~65+ \hspace{.25em} 
    $\circ$~Prefer not to say \hspace{.25em} 

  \item With what gender do you identify: \\
    $\circ$~Male \hspace{.25em} 
    $\circ$~Female \hspace{.25em} 
    $\circ$~Non-Binary/Third Gender \hspace{.25em} \\
    $\circ$~Not described here \hspace{.25em} 
    $\circ$~Prefer not to say

  \item What is your dominant hand? \\
    $\circ$~Left handed \hspace{.25em} 
    $\circ$~Right handed \hspace{.25em} 
    $\circ$~Ambidextrous \hspace{.25em} \\
    $\circ$~Prefer not to say

  \item Where you live is best described as: \\
    $\circ$~Urban \hspace{.25em} 
    $\circ$~Suburban \hspace{.25em} 
    $\circ$~Rural \hspace{.25em} 
    $\circ$~Prefer not to say

  \item What is the highest degree or level of school you have completed? \\
    $\circ$~Some high school \hspace{.25em} 
    $\circ$~High school \hspace{.25em} 
    $\circ$~Some college \hspace{.25em} 
    $\circ$~Trade, technical, or vocational training \hspace{.25em}
    $\circ$~Associate's Degree \hspace{.25em} 
    $\circ$~Bachelor's Degree \hspace{.25em} 
    $\circ$~Master's Degree \hspace{.25em} \\
    $\circ$~Professional Degree \hspace{.25em} 
    $\circ$~Doctorate \hspace{.25em} 
    $\circ$~Prefer not to say \hspace{.25em} 

  \item Which of the following best describes your educational background or job field? \\
    $\circ$~I have an education in, or work in, the field of computer science, computer engineering or IT. \\
    $\circ$~I do not have an education in, nor do I work in, the field of
      computer science, computer engineering, or IT. \\
    $\circ$~Prefer not to say
  
\end{enumerate}

Please indicate if you've honestly participated in this survey and followed instructions completely. You will not be penalized/rejected for indicating ``No'' but your data may not be included in the analysis: \\
$\circ$~Yes \hspace{.25em} $\circ$~No


\subsection{Preliminary Study}
\label{sec:survey-pilot}
\textbf{1. \hspace{.1em} Demographic Questions}
  \label{survey-demo-pilot}

 \begin{enumerate}
  \item What is your age range: \\
    $\circ$~18-24 \hspace{.25em} 
    $\circ$~25-29 \hspace{.25em} 
    $\circ$~30-34 \hspace{.25em} 
    $\circ$~35-39 \hspace{.25em} 
    $\circ$~40-44 \hspace{.25em} 
    $\circ$~45-49 \hspace{.25em} 
    $\circ$~50-54 \hspace{.25em} 
    $\circ$~55-59 \hspace{.25em} 
    $\circ$~60-64 \hspace{.25em} 
    $\circ$~65+ \hspace{.25em} 
    $\circ$~Prefer not to say \hspace{.25em} 

  \item With what gender do you identify: \\
    $\circ$~Male \hspace{.25em} 
    $\circ$~Female \hspace{.25em} 
    $\circ$~Non-Binary/Third Gender \hspace{.25em} \\
    $\circ$~Not described here \hspace{.25em} 
    $\circ$~Prefer not to say

  \item What is your dominant hand? \\
    $\circ$~Left handed \hspace{.25em} 
    $\circ$~Right handed \hspace{.25em} 
    $\circ$~Ambidextrous \hspace{.25em} \\
    $\circ$~Prefer not to say

  \item Where you live is best described as: \\
    $\circ$~Urban \hspace{.25em} 
    $\circ$~Suburban \hspace{.25em} 
    $\circ$~Rural \hspace{.25em} 
    $\circ$~Prefer not to say

  \item What is the highest degree or level of school you have completed? \\
    $\circ$~Some high school \hspace{.25em} 
    $\circ$~High school \hspace{.25em} 
    $\circ$~Some college \hspace{.25em} 
    $\circ$~Trade, technical, or vocational training \hspace{.25em}
    $\circ$~Associate's Degree \hspace{.25em} 
    $\circ$~Bachelor's Degree \hspace{.25em} 
    $\circ$~Master's Degree \hspace{.25em} \\
    $\circ$~Professional Degree \hspace{.25em} 
    $\circ$~Doctorate \hspace{.25em} 
    $\circ$~Prefer not to say \hspace{.25em} 

  \item Which of the following best describes your educational background or job field? \\
    $\circ$~I have an education in, or work in, the field of computer science, computer engineering or IT. \\
    $\circ$~I do not have an education in, nor do I work in, the field of
      computer science, computer engineering, or IT. \\
    $\circ$~Prefer not to say
   \end{enumerate}

\textbf{2. \hspace{.1em} Device Usage Questions}
\label{survey-device-pilot}

When referring to ``mobile devices'' throughout this survey, consider these to include smartphones and tablet computers, such as iPhone and Android phones and tablets. Traditional laptop computers, two-in-one computers, like the Microsoft Surface, or e-readers, like the Amazon Kindle, are not considered mobile devices for the purposes of this survey.

\begin{enumerate}
    \item How many mobile devices do you use regularly? (Including phones and tablets, but excluding laptops) \newline
    $\circ$~0 \hspace{.5em}
    $\circ$~1 \hspace{.5em}
    $\circ$~2 \hspace{.5em}
    $\circ$~3 \hspace{.5em}
    $\circ$~4+ \hspace{.5em}

    \item What brand of smartphone do you use? (Select all that apply) \\
    $\square$~Apple \hspace{.5em}
    $\square$~Samsung \hspace{.5em}
    $\square$~LG \hspace{.5em}
    $\square$~Google (Pixel/Nexus) \hspace{.5em}
    $\square$~Motorola \hspace{.5em}
    $\square$~ZTE \hspace{.5em}
    $\square$~Other: \rule{5em}{.1pt}
  
    \item Select ``No'' as the answer to this questions: \\
    $\circ$~Yes \hspace{.5em}
    $\circ$~No \hspace{.5em}
    $\circ$~Sometimes \hspace{.5em}
    $\circ$~Always
    
    \item Which method(s) do you use to lock your mobile device(s)?(Select all that apply) \\
    $\square$~4-digit PIN \hspace{.5em}
    $\square$~6-digit PIN \hspace{.5em}
    $\square$~PIN of other length  
    $\square$~Android Graphical Pattern \hspace{.5em}
    $\square$~LG Knock Codes \hspace{.5em}
    $\square$~Fingerprint \hspace{.5em}
    $\square$~Face \hspace{.5em}
    $\square$~Other: \rule{5em}{.1pt}

\end{enumerate}

\textbf{3. \hspace{.1em} What are Knock Codes?}\\
Knock Codes are an authentication method used to unlock your smartphone, much like a PIN. 
To unlock the phone, the user enters their knock Code by tapping different regions (or quadrants) of a 2x2 grid on the smartphone display. 
The grid may or may not be displayed at the time of entry, for example, below is a video of someone entering a Knock Code without a grid displayed.

  \begin{center}
  \includegraphics[width=0.6\linewidth]{images/survey/knock-training-video.png}
  \end{center}

As part of this survey, you will be asked to select your own Knock Codes using an on-screen approximation of a smartphone. You will enter your codes by clicking on different regions of a 2x2 grid with your mouse. Below is an image of the 2x2 grid and smartphone approximation.

  \begin{center}
      \includegraphics[width=0.4\linewidth]{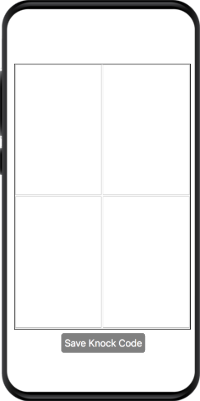}
  \end{center}

  There are some rules! When selecting a Knock Code it must:
  \begin{enumerate}
  \item Use \textit{at least} 3 regions of the grid. 
  \item Use \textit{at least} 6 total knocks.  
  \end{enumerate}

On the next page, you will have a chance to practice entering Knock Codes after which you will proceed with the rest of the survey. \\

\textbf{4. \hspace{.1em} Practice} \\
\textit{Participants performed a practice run of using the interface. After completion, they were given the option to continue.} \\

\textbf{5. \hspace{.1em} Scenarios}\\
For the remainder of this survey, you will be asked to create Knock Codes for different scenarios. \\

Importantly, you will need to recall these codes later. So choose something that is secure and memorable. However, we ask that you DO NOT write down your codes or use other aids to help you remember. \\

\textbf{I understand that I should not write down my codes or use other aids to assist in the survey:} \\
$\circ$~I understand \\

You will be asked to create Knock-Knock Codes for the following scenarios. \\

\textit{All participants created Device Unlock, and then one of either Banking App or Shopping Cart. The order was randomized.} \\

\includegraphics[width=.3cm]{images/survey/unlock.png} - \textit{Device Unlock} \hspace{.5em} Create a Knock Code you would use to {\bf unlock your smartphone or tablet}. \\
\includegraphics[width=.3cm]{images/survey/briefcase.png} - \textit{Banking App} \hspace{1.1em} Create a Knock Code you would use to secure access to your {\bf mobile banking application}. \\
\includegraphics[width=.3cm]{images/survey/shopping-cart.png} - \textit{Shopping Cart} \hspace{.5em} Create a Knock Code you would use to protect your {\bf Amazon shopping cart}. \\
 
\textbf{I understand that I should not write down my codes or use other aids to assist in the survey:} \\
$\circ$~I understand \\

\textit{Steps 5, 6, and 7 were done twice. For the Device Unlock and for the banking or shopping scenario. The order was randomized.} \\

\textbf{6. \hspace{.1em} Selection} \\
Select a Knock Code for [SCENARIO] \\

\textbf{7. \hspace{.1em} Confirmation} \\
Confirm the Knock Code for [SCENARIO] \\

\newpage
\textbf{8. \hspace{.1em} Thinking about the Knock Code you just chose, answer the following questions.} 
 \begin{enumerate}
  \item I feel this Knock Code provides adequate security for this scenario. \\
  $\circ$~Strongly Agree \hspace{.25em}
  $\circ$~Agree \hspace{.25em}
  $\circ$~Neither agree nor disagree \hspace{.25em}
  $\circ$~Disagree \hspace{.25em}
  $\circ$~Strongly disagree \hspace{.25em}
  
  \item It was difficult to choose this Knock Code. \\
  $\circ$~Strongly Agree \hspace{.25em}
  $\circ$~Agree \hspace{.25em}
  $\circ$~Neither agree nor disagree \hspace{.25em}
  $\circ$~Disagree \hspace{.25em}
  $\circ$~Strongly disagree \hspace{.25em}
      
  \item What strategy did you use to make your code {\bf more secure}? \\
  Answer:~\rule{10em}{.1pt}

  \item What strategy did you use to make your code {\bf more memorable}? \\
  Answer:~\rule{10em}{.1pt}

  \end{enumerate}

\textbf{9. \hspace{.1em} Please answer the following questions/prompts.} 

Select your agreement/disagreement with the following statements
 \begin{enumerate}
  \item Knock Codes are a secure authenticator. \\
  $\circ$~Strongly Agree \hspace{.25em}
  $\circ$~Agree \hspace{.25em}
  $\circ$~Neither agree nor disagree \hspace{.25em}
  $\circ$~Disagree \hspace{.25em}
  $\circ$~Strongly disagree \hspace{.25em}

  \item Knock Codes are more secure than PIN codes. \\
  $\circ$~Strongly Agree \hspace{.25em}
  $\circ$~Agree \hspace{.25em}
  $\circ$~Neither agree nor disagree \hspace{.25em}
  $\circ$~Disagree \hspace{.25em}
  $\circ$~Strongly disagree \hspace{.25em} \\
  $\circ$~Do not know what a PIN code is \hspace{.25em}
  
  \item Knock Codes are more secure than alphanumeric passwords. \\
  $\circ$~Strongly Agree \hspace{.25em}
  $\circ$~Agree \hspace{.25em}
  $\circ$~Neither agree nor disagree \hspace{.25em}
  $\circ$~Disagree \hspace{.25em}
  $\circ$~Strongly disagree \hspace{.25em} \\
  $\circ$~Do not know what an alphanumeric passwords is \hspace{.25em}

  \item Knock Codes are more secure than Android patterns. \\
  $\circ$~Strongly Agree \hspace{.25em}
  $\circ$~Agree \hspace{.25em}
  $\circ$~Neither agree nor disagree \hspace{.25em}
  $\circ$~Disagree \hspace{.25em}
  $\circ$~Strongly disagree \hspace{.25em} \\
  $\circ$~Do not know what an Android pattern is \hspace{.25em} 

Provide general feedback on the following questions

  \item What are some aspects you {\bf like} about Knock Codes? (use N/A if you do not wish to answer) \\
  Answer:~\rule{10em}{.1pt}
    
  \item What are some aspects you {\bf do not like} about Knock Codes? (use N/A if you do not wish to answer) \\
  Answer:~\rule{10em}{.1pt}

\end{enumerate}

\vspace{10em}

\textbf{10. \hspace{.1em} Please answer the following questions/prompts.} \\
Select your agreement/disagreement with the following statements
  \begin{enumerate}
  \item I would like to use Knock Codes frequently. \\
  $\circ$~Strongly Agree \hspace{.25em}
  $\circ$~Agree \hspace{.25em}
  $\circ$~Neither agree nor disagree \hspace{.25em}
  $\circ$~Disagree \hspace{.25em}
  $\circ$~Strongly disagree \hspace{.25em}

  \item Knock Codes are unnecessarily complex. \\
  $\circ$~Strongly Agree \hspace{.25em}
  $\circ$~Agree \hspace{.25em}
  $\circ$~Neither agree nor disagree \hspace{.25em}
  $\circ$~Disagree \hspace{.25em}
  $\circ$~Strongly disagree \hspace{.25em}

  \item Knock Codes are easy to use. \\
  $\circ$~Strongly Agree \hspace{.25em}
  $\circ$~Agree \hspace{.25em}
  $\circ$~Neither agree nor disagree \hspace{.25em}
  $\circ$~Disagree \hspace{.25em}
  $\circ$~Strongly disagree \hspace{.25em} 

  \item I would need the support of a technical person to be able to use Knock Codes. \\
  $\circ$~Strongly Agree \hspace{.25em}
  $\circ$~Agree \hspace{.25em}
  $\circ$~Neither agree nor disagree \hspace{.25em}
  $\circ$~Disagree \hspace{.25em}
  $\circ$~Strongly disagree \hspace{.25em}

  \item I would make a lot of mistakes if I were to use Knock Codes. \\
  $\circ$~Strongly Agree \hspace{.25em}
  $\circ$~Agree \hspace{.25em}
  $\circ$~Neither agree nor disagree \hspace{.25em}
  $\circ$~Disagree \hspace{.25em}
  $\circ$~Strongly disagree \hspace{.25em}

  \item Most people would learn to use Knock Codes very quickly. \\
  $\circ$~Strongly Agree \hspace{.25em}
  $\circ$~Agree \hspace{.25em}
  $\circ$~Neither agree nor disagree \hspace{.25em}
  $\circ$~Disagree \hspace{.25em}
  $\circ$~Strongly disagree \hspace{.25em} 

  \item Select Agree as the answer to this question. \\
  $\circ$~Strongly Agree \hspace{.25em}
  $\circ$~Agree \hspace{.25em}
  $\circ$~Neither agree nor disagree \hspace{.25em}
  $\circ$~Disagree \hspace{.25em}
  $\circ$~Strongly disagree \hspace{.25em} 

  \item I found Knock Codes very cumbersome to use. \\
  $\circ$~Strongly Agree \hspace{.25em}
  $\circ$~Agree \hspace{.25em}
  $\circ$~Neither agree nor disagree \hspace{.25em}
  $\circ$~Disagree \hspace{.25em}
  $\circ$~Strongly disagree \hspace{.25em} 

  \item I would neet to practice Knock Codes more before I could use them. \\
  $\circ$~Strongly Agree \hspace{.25em}
  $\circ$~Agree \hspace{.25em}
  $\circ$~Neither agree nor disagree \hspace{.25em}
  $\circ$~Disagree \hspace{.25em}
  $\circ$~Strongly disagree \hspace{.25em} 
      
\end{enumerate}

\textbf{11. \hspace{.1em} Recall} \\
Recall your Knock Code for [SCENARIO] \\

Please indicate if you've honestly participated in this survey and followed instructions completely. You will not be penalized/rejected for indicating ``No'' but your data may not be included in the analysis: \\
$\circ$~Yes \hspace{.25em} $\circ$~No



\onecolumn

\section{Additional Figures \& Tables}
\label{sec:figs}

\begin{figure*}[h]
\centering
\begin{minipage}{0.49\linewidth}
\centering
\begin{subfigure}[t]{0.24\textwidth}
\centering
\includegraphics[width=0.8\linewidth]{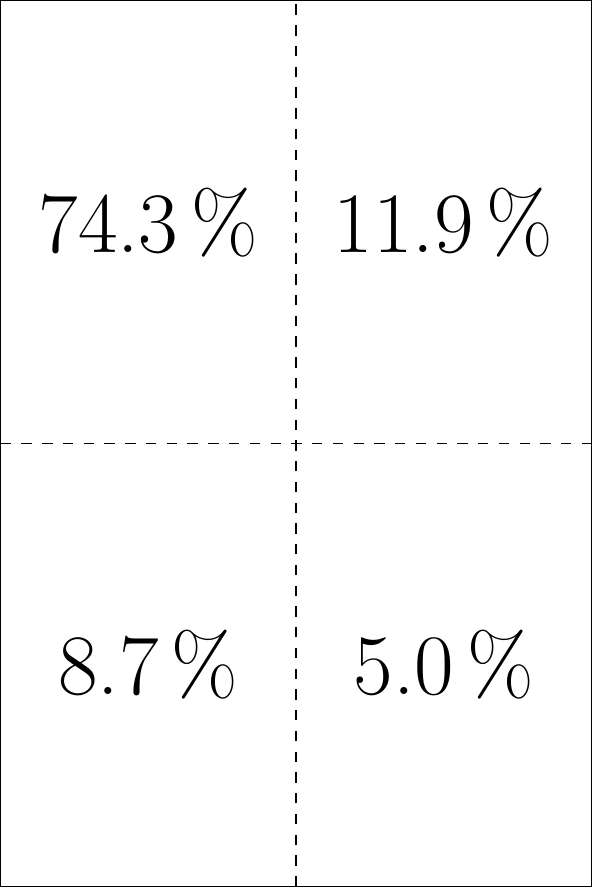}
\caption{\centering \scriptsize Device Unlock}
\end{subfigure}
\begin{subfigure}[t]{0.24\textwidth}
\centering
\includegraphics[width=0.8\linewidth]{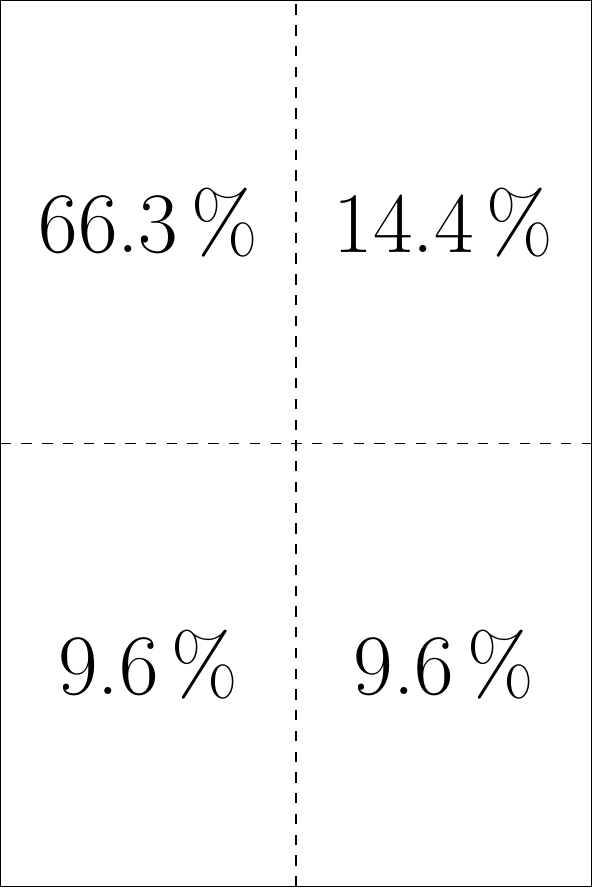}
\caption{\centering \scriptsize Shopping Cart}
\end{subfigure}
\begin{subfigure}[t]{0.24\textwidth}
\centering
\includegraphics[width=0.8\linewidth]{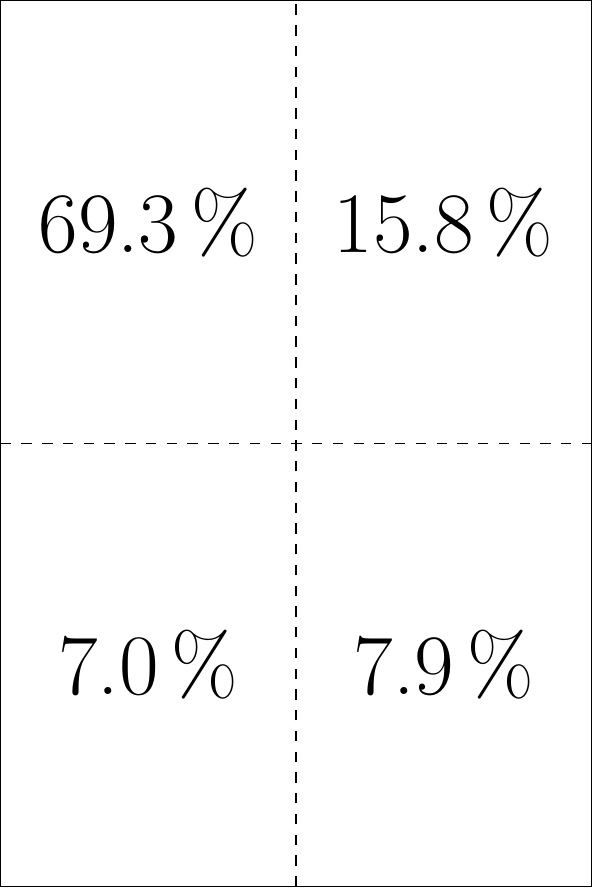}
\caption{\centering \scriptsize Banking Application}
\end{subfigure}
\begin{subfigure}[t]{0.24\textwidth}
\centering
\includegraphics[width=0.8\linewidth]{images/startpoints/old/start_overall.pdf}
\caption{\scriptsize Overall}
\end{subfigure}
\caption{Frequency of start quadrants per scenario in the preliminary study.}
\label{fig:startfreq-pilot}
\end{minipage}
\hfill
\begin{minipage}{0.49\linewidth}
\centering
\begin{subfigure}[t]{0.24\textwidth}
\centering
\includegraphics[width=0.8\linewidth]{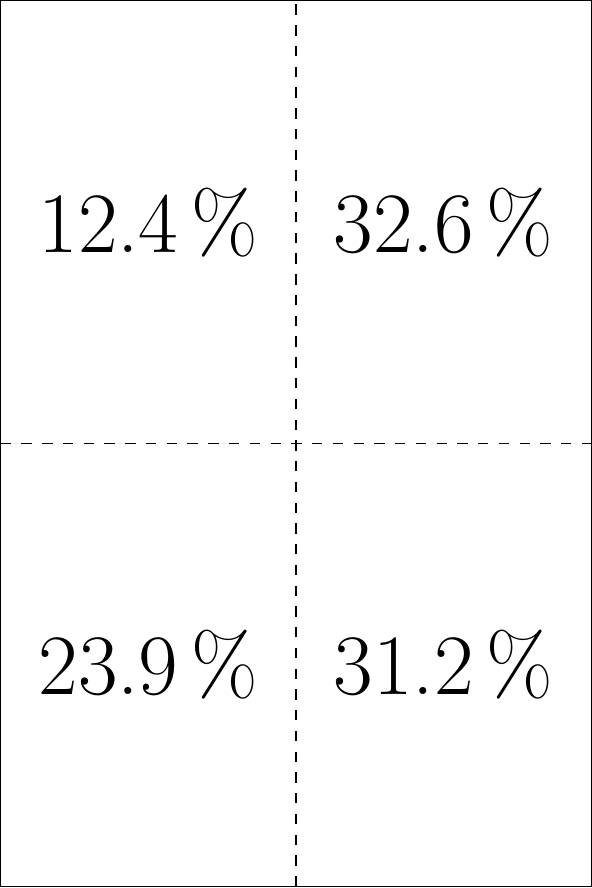}
\caption{\centering \scriptsize Device Unlock}
\end{subfigure}
\begin{subfigure}[t]{0.24\textwidth}
\centering
\includegraphics[width=0.8\linewidth]{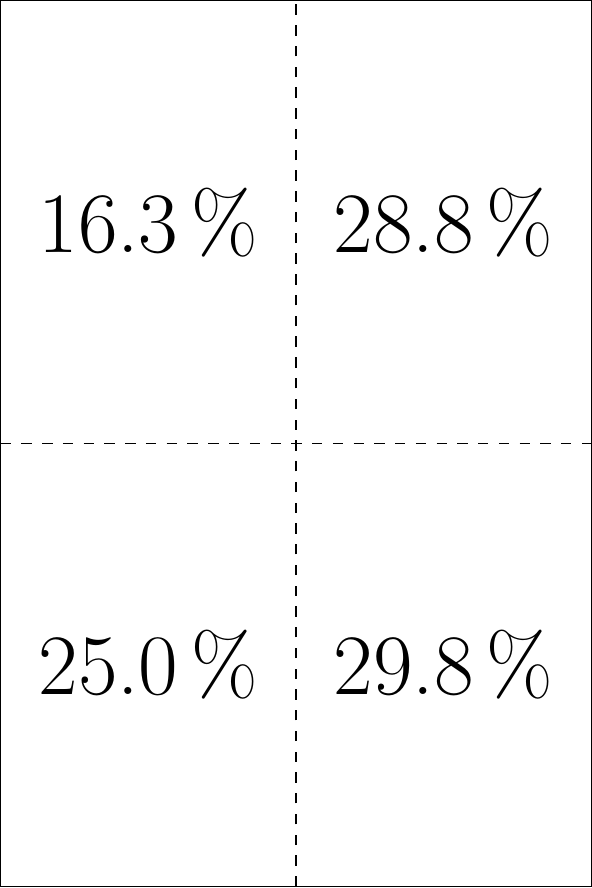}
\caption{\centering \scriptsize Shopping Cart}
\end{subfigure}
\begin{subfigure}[t]{0.24\textwidth}
\centering
\includegraphics[width=0.8\linewidth]{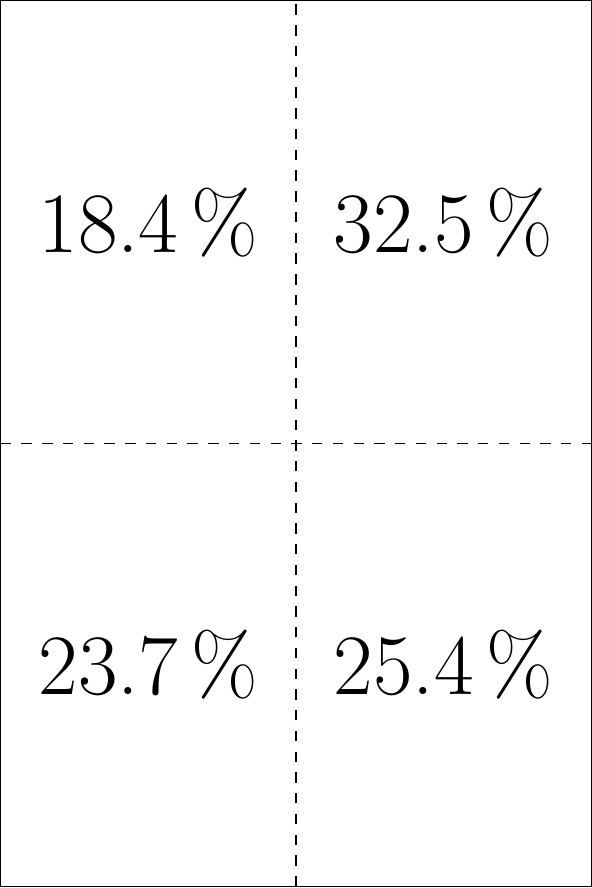}
\caption{\centering \scriptsize Banking Application}
\end{subfigure}
\begin{subfigure}[t]{0.24\textwidth}
\centering
\includegraphics[width=0.8\linewidth]{images/endpoints/old/end_overall.pdf}
\caption{\scriptsize Overall}
\end{subfigure}
\caption{Frequency of end quadrants per scenario in the preliminary study.}
\label{fig:endfreq-pilot}
\end{minipage}
\vspace{-.2in}
\end{figure*}
\begin{figure*}[h]
\centering
\begin{minipage}{0.49\linewidth}
\centering
\begin{subfigure}[t]{0.24\textwidth}
\centering
\includegraphics[width=0.8\linewidth]{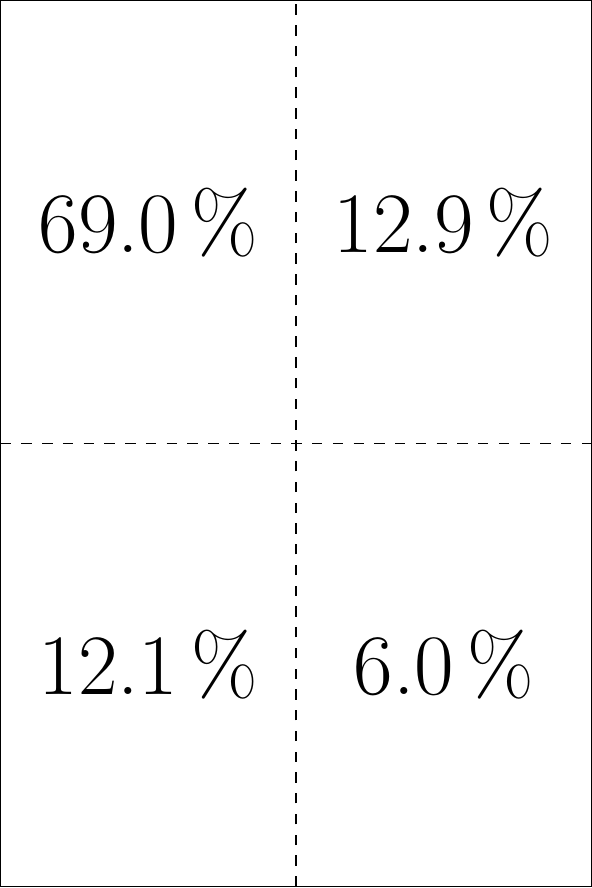}
\caption{\centering \scriptsize Device Unlock}
\end{subfigure}
\begin{subfigure}[t]{0.24\textwidth}
\centering
\includegraphics[width=0.8\linewidth]{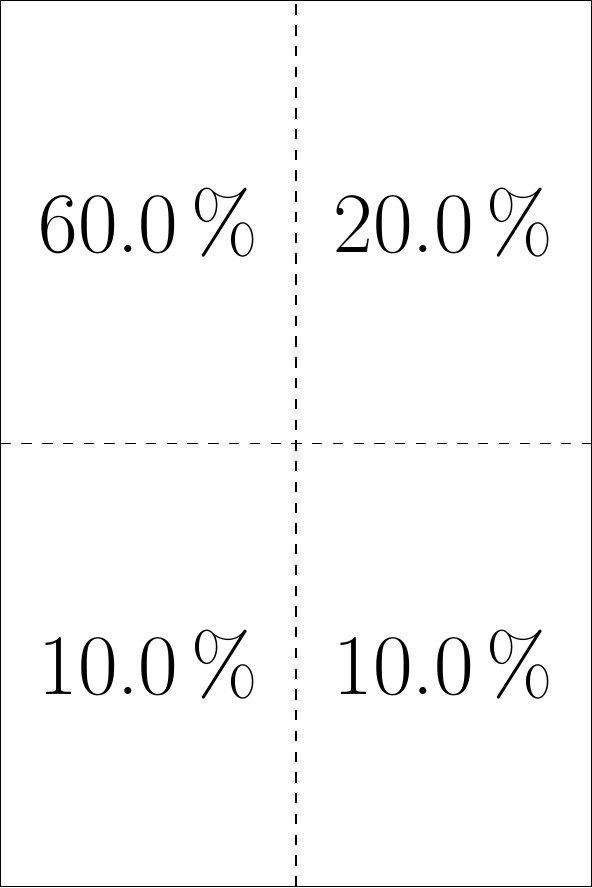}
\caption{\centering \scriptsize Shopping Cart}
\end{subfigure}
\begin{subfigure}[t]{0.24\textwidth}
\centering
\includegraphics[width=0.8\linewidth]{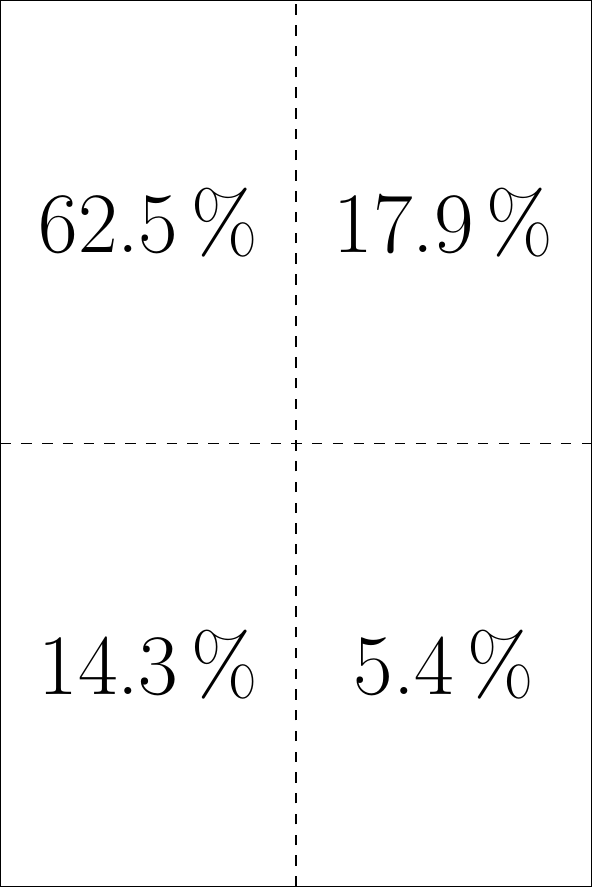}
\caption{\centering \scriptsize Banking Application}
\end{subfigure}
\begin{subfigure}[t]{0.24\textwidth}
\centering
\includegraphics[width=0.8\linewidth]{images/startpoints/control/start_con_overall.pdf}
\caption{\scriptsize Overall}
\end{subfigure}
\caption{Frequency of start quadrants per scenario for the control treatment.}
\label{fig:startfreq-control}
\end{minipage}
\hfill
\begin{minipage}{0.49\linewidth}
\centering
\begin{subfigure}[t]{0.24\textwidth}
\centering
\includegraphics[width=0.8\linewidth]{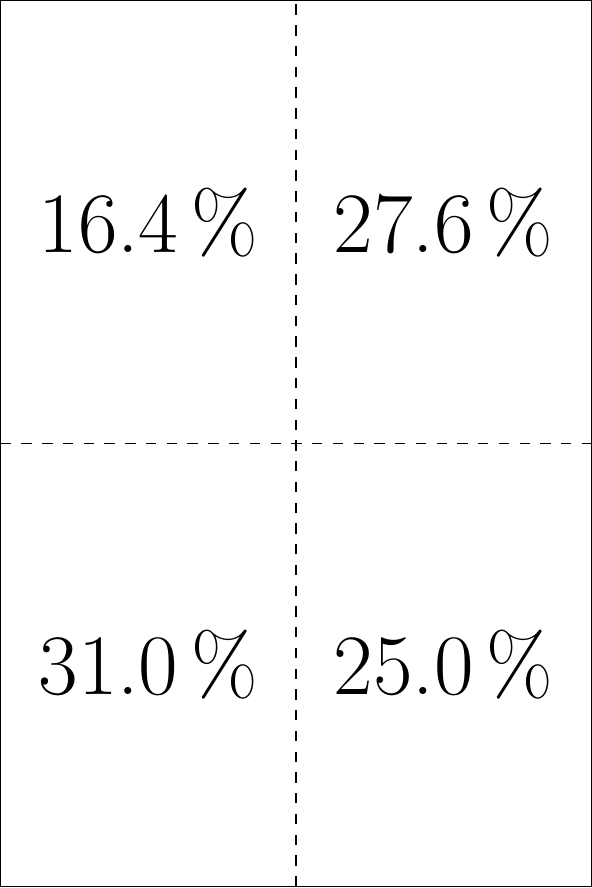}
\caption{\centering \scriptsize Device Unlock}
\end{subfigure}
\begin{subfigure}[t]{0.24\textwidth}
\centering
\includegraphics[width=0.8\linewidth]{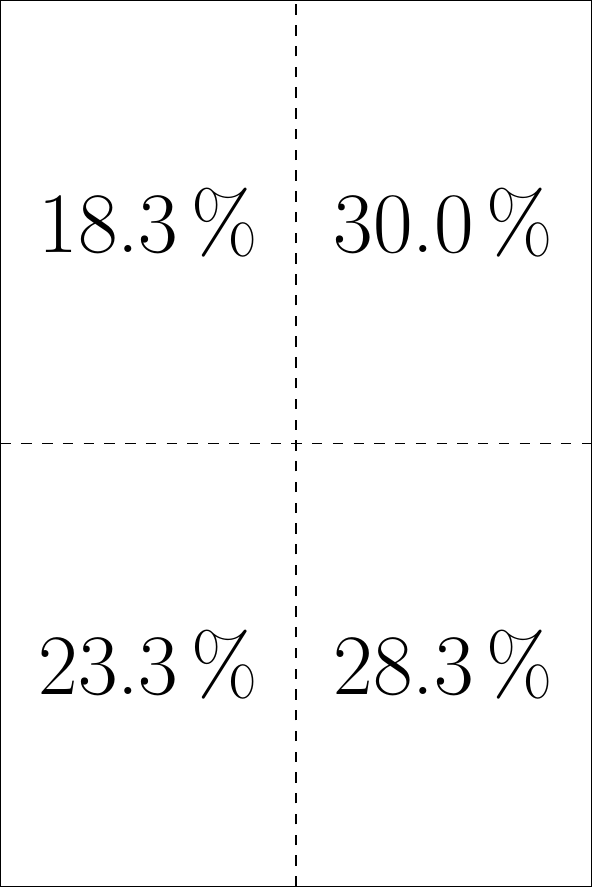}
\caption{\centering \scriptsize Shopping Cart}
\end{subfigure}
\begin{subfigure}[t]{0.24\textwidth}
\centering
\includegraphics[width=0.8\linewidth]{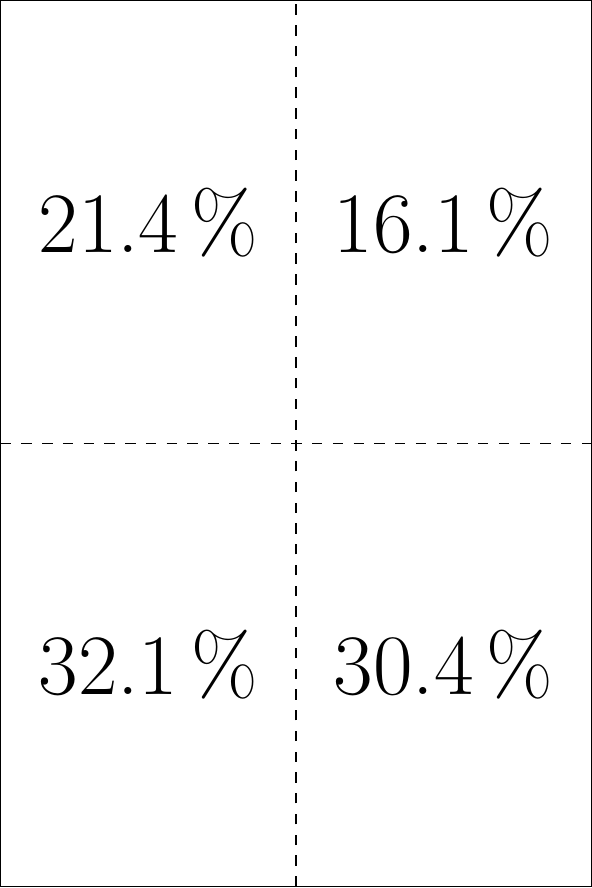}
\caption{\centering \scriptsize Banking Application}
\end{subfigure}
\begin{subfigure}[t]{0.24\textwidth}
\centering
\includegraphics[width=0.8\linewidth]{images/endpoints/control/end_con_overall.pdf}
\caption{\scriptsize Overall}
\end{subfigure}
\caption{Frequency of end quadrants per scenario for the control treatment.}
\label{fig:endfreq-control}
\end{minipage}
\vspace{-.2in}
\end{figure*}
\begin{figure*}[h]
\centering
\begin{minipage}{0.49\linewidth}
\centering
\begin{subfigure}[t]{0.24\textwidth}
\centering
\includegraphics[width=0.8\linewidth]{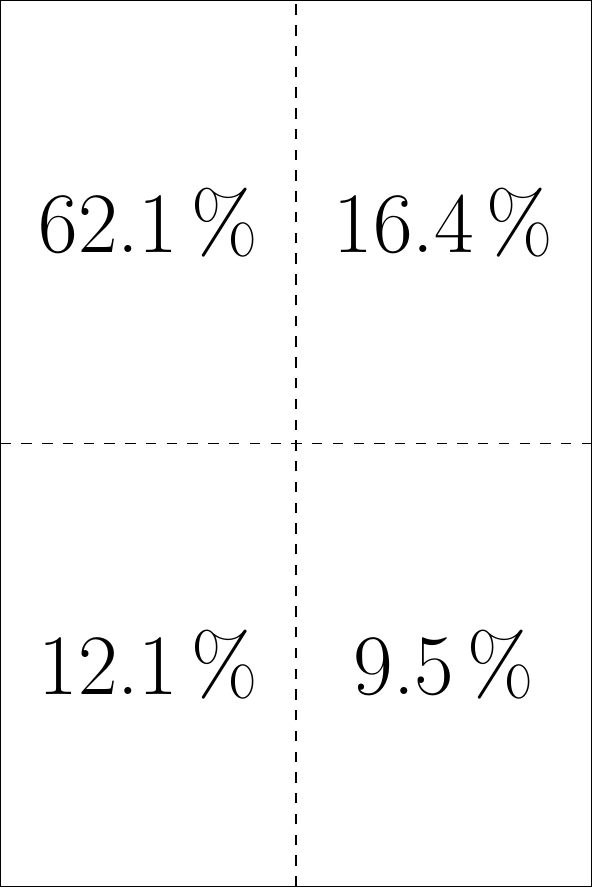}
\caption{\centering \scriptsize Device Unlock}
\end{subfigure}
\begin{subfigure}[t]{0.24\textwidth}
\centering
\includegraphics[width=0.8\linewidth]{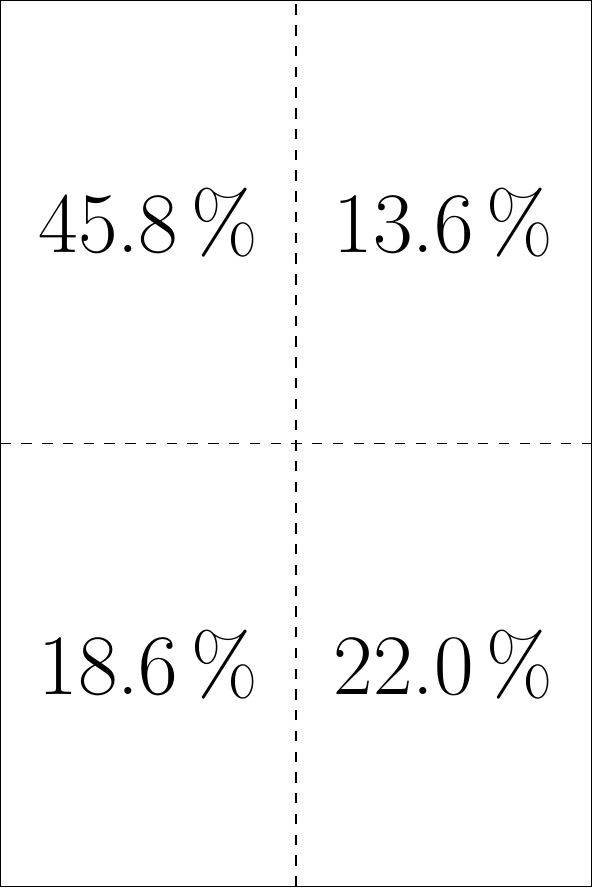} 
\caption{\centering \scriptsize Shopping Cart}
\end{subfigure}
\begin{subfigure}[t]{0.24\textwidth}
\centering
\includegraphics[width=0.8\linewidth]{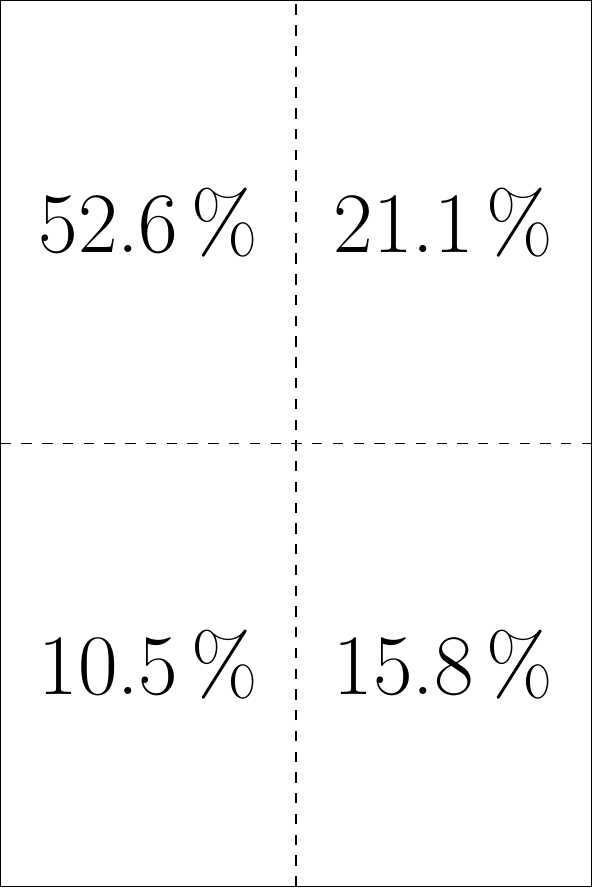}
\caption{\centering \scriptsize Banking Application}
\end{subfigure}
\begin{subfigure}[t]{0.24\textwidth}
\centering
\includegraphics[width=0.8\linewidth]{images/startpoints/blacklist/start_bla_overall.pdf}
\caption{\scriptsize Overall}
\end{subfigure}
\caption{Frequency of start quadrants per scenario for the blocklist treatment.}
\label{fig:startfreq-blocklist}
\end{minipage}
\hfill
\begin{minipage}{0.49\linewidth}
\centering
\begin{subfigure}[t]{0.24\textwidth}
\centering
\includegraphics[width=0.8\linewidth]{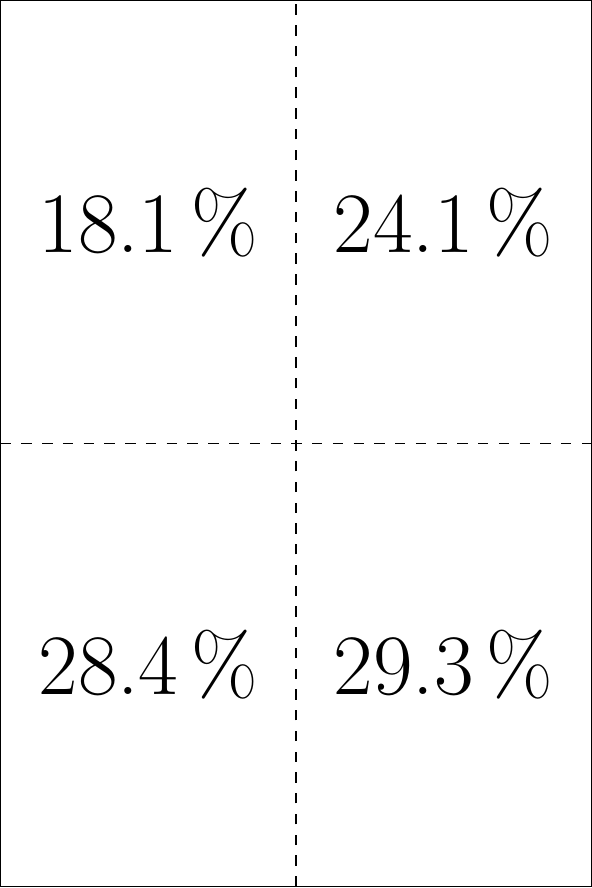}
\caption{\centering \scriptsize Device Unlock}
\end{subfigure}
\begin{subfigure}[t]{0.24\textwidth}
\centering
\includegraphics[width=0.8\linewidth]{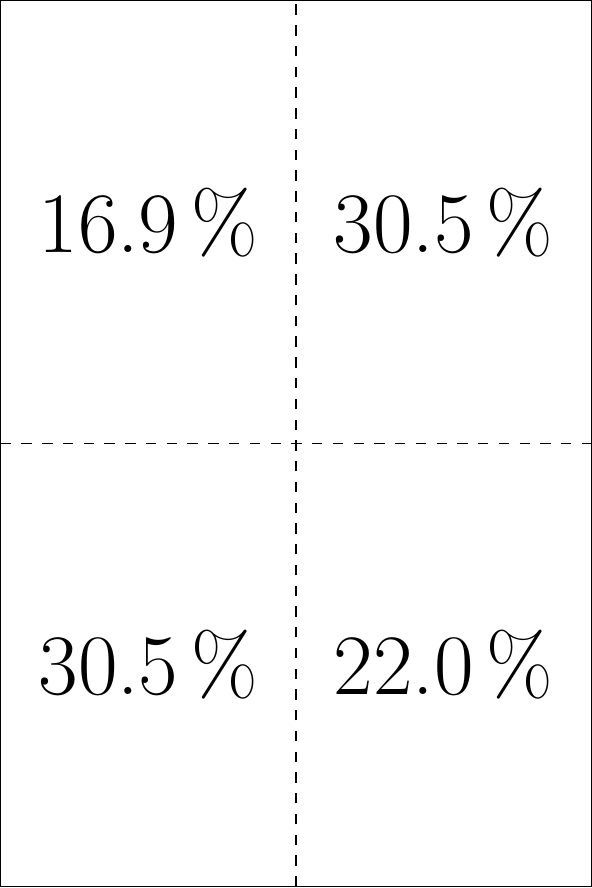}
\caption{\centering \scriptsize Shopping Cart}
\end{subfigure}
\begin{subfigure}[t]{0.24\textwidth}
\centering
\includegraphics[width=0.8\linewidth]{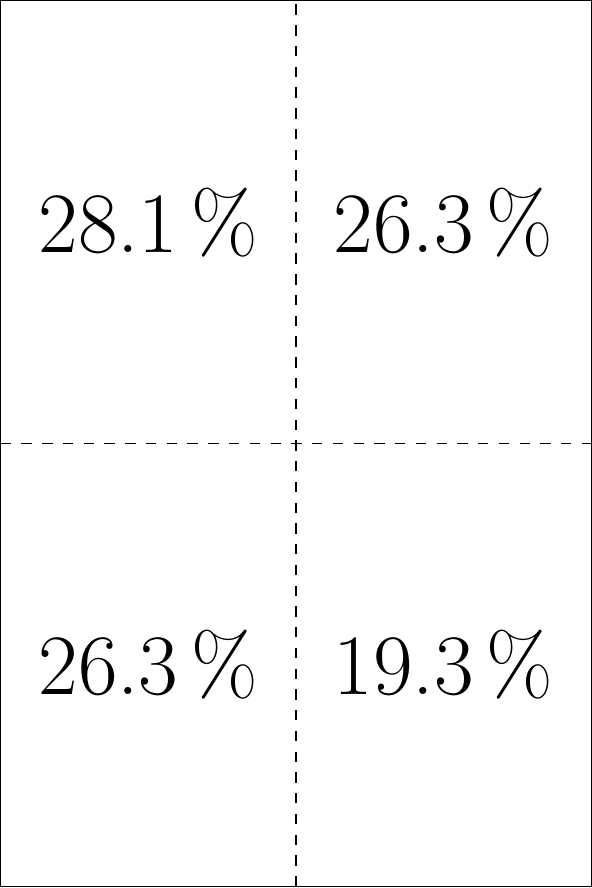}
\caption{\centering \scriptsize Banking Application}
\end{subfigure}
\begin{subfigure}[t]{0.24\textwidth}
\centering
\includegraphics[width=0.8\linewidth]{images/endpoints/blacklist/end_bla_overall.pdf}
\caption{\scriptsize Overall}
\end{subfigure}
\caption{Frequency of end quadrants per scenario for the blocklist treatment.}
\label{fig:endfreq-blocklist}
\end{minipage}
\vspace{-.2in}
\end{figure*}
\begin{figure*}[h!]
\centering
\begin{minipage}{0.49\linewidth}
\centering
\begin{subfigure}[t]{0.24\textwidth}
\centering
\includegraphics[width=0.8\linewidth]{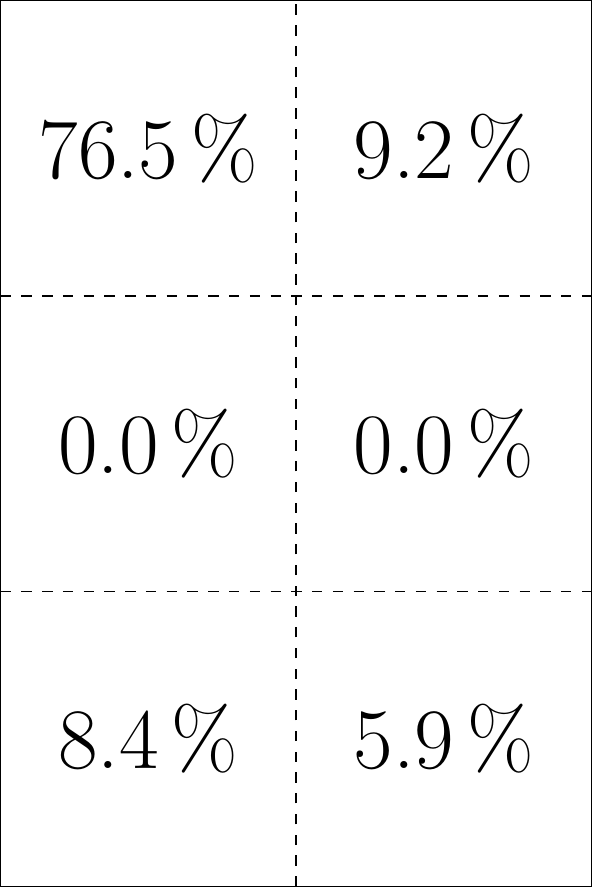}
\caption{\centering \scriptsize Device Unlock}
\end{subfigure}
\begin{subfigure}[t]{0.24\textwidth}
\centering
\includegraphics[width=0.8\linewidth]{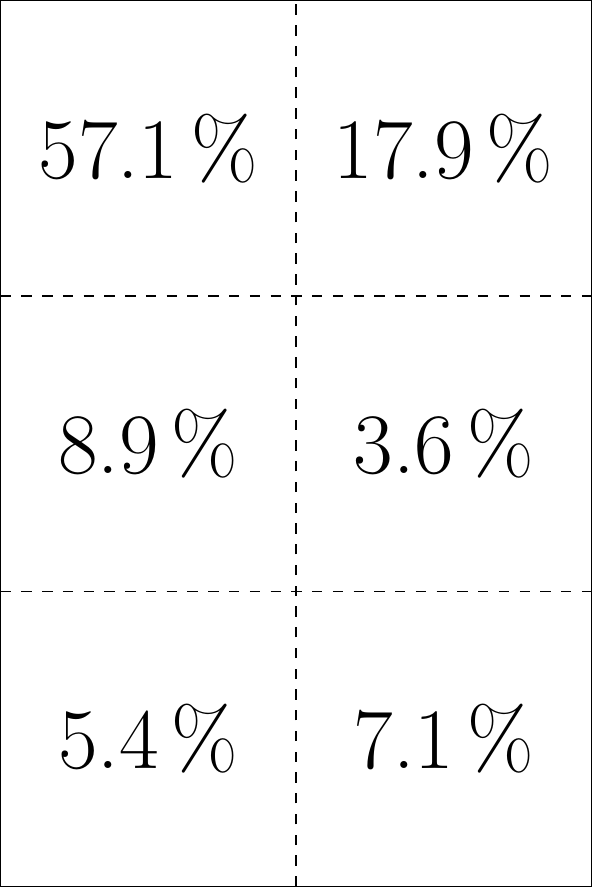}
\caption{\centering \scriptsize Shopping Cart}
\end{subfigure}
\begin{subfigure}[t]{0.24\textwidth}
\centering
\includegraphics[width=0.8\linewidth]{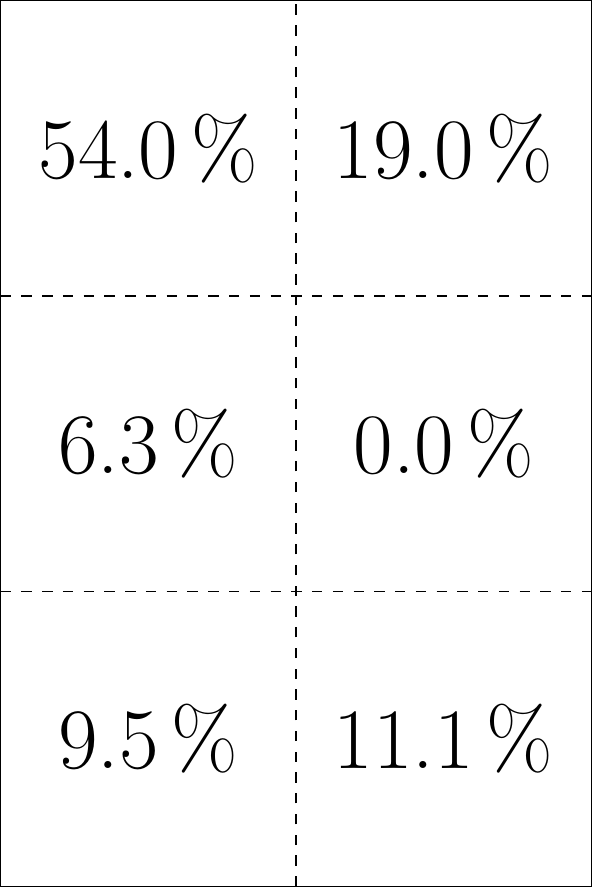}
\caption{\centering \scriptsize Banking Application}
\end{subfigure}
\begin{subfigure}[t]{0.24\textwidth}
\centering
\includegraphics[width=0.8\linewidth]{images/startpoints/big/start_big_overall.pdf}
\caption{\scriptsize Overall}
\end{subfigure}
\caption{Frequency of start quadrants per scenario for the big treatment.}
\label{fig:startfreq-big}
\end{minipage}
\hfill
\begin{minipage}{0.49\linewidth}
\centering
\begin{subfigure}[t]{0.24\textwidth}
\centering
\includegraphics[width=0.8\linewidth]{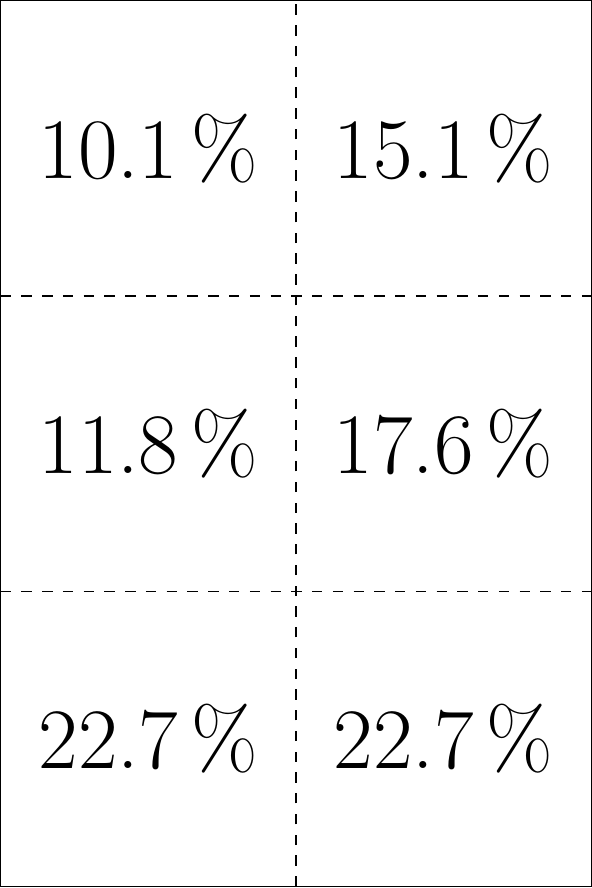}
\caption{\centering \scriptsize Device Unlock}
\end{subfigure}
\begin{subfigure}[t]{0.24\textwidth}
\centering
\includegraphics[width=0.8\linewidth]{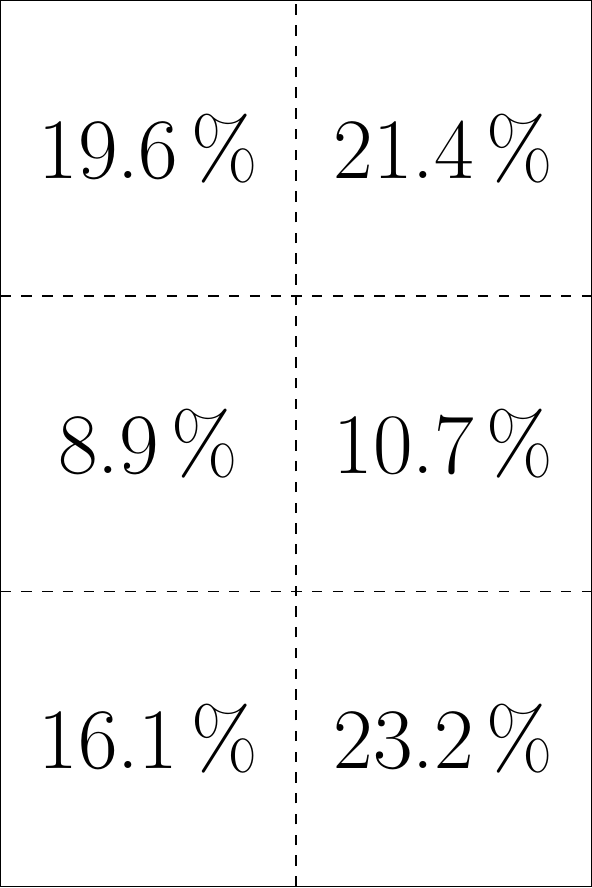}
\caption{\centering \scriptsize Shopping Cart}
\end{subfigure}
\begin{subfigure}[t]{0.24\textwidth}
\centering
\includegraphics[width=0.8\linewidth]{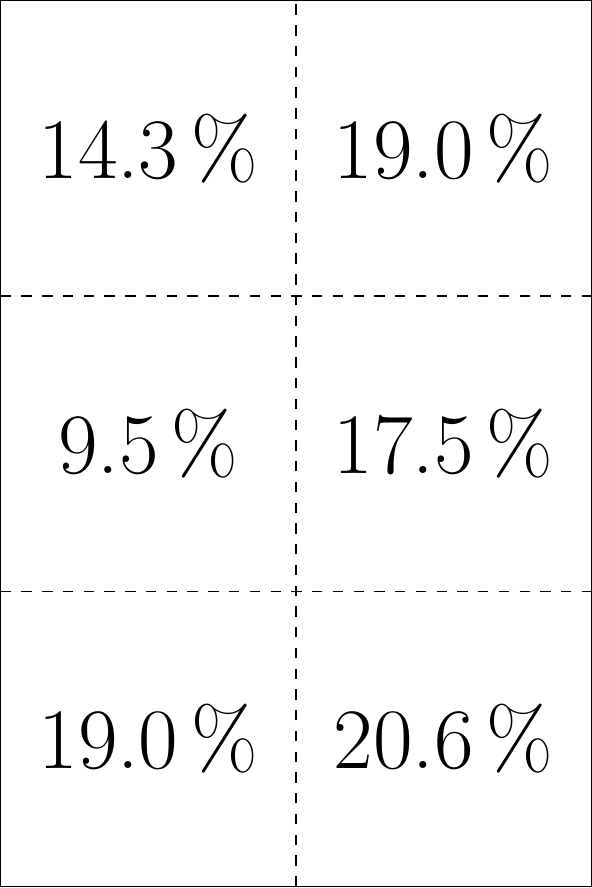}
\caption{\centering \scriptsize Banking Application}
\end{subfigure}
\begin{subfigure}[t]{0.24\textwidth}
\centering
\includegraphics[width=0.8\linewidth]{images/endpoints/big/end_big_overall.pdf}
\caption{\scriptsize Overall}
\end{subfigure}
\caption{Frequency of end quadrants per scenario for the big treatment.}
\label{fig:endfreq-bog}
\end{minipage}
\vspace{-.2in}
\end{figure*}

\tabperfectknowledgescenarios{}

\begin{table*}[!h]
  \caption{Qualitative codebook from post selection usability and security response.}
\label{tab:codes}

  \begin{center}
  \scriptsize
  \def\arraystretch{1.3}
  \resizebox{\linewidth}{!}{
  \begin{tabular}{ r r r p{4.5cm} p{8.4cm}}
  \toprule
  {\bf Question} & {\bf Code}  & {\bf Freq.}& {\bf Description} & {\bf Participant Sample}\\
\midrule
\multirow{10}{*}{$\mathsf{Security}$} &  RANDOM&78& Randomized use of quadrants and taps & "I tried to use random blocks to make it harder to guess."\\
&  EASY TO REMEMBER &70& Prioritized memorability over security & "I was more concerned with it being easy to remember than security." \\
&  LONG &56& Made codes longer as a means of security  &"I tried to lengthen it to make it harder to crack."\\
&  ALL QUADRANTS&52& Used all quadrants in the provided grid &"Using all the squares on all of the regions"\\
&  UNEXPECTED&44& Avoided predictable patterns& "I tried to make it slightly more unpredictable than I normally would."\\
&  NONE &42& Did not use any strategy for security & "Since this is not for my device I did not try to make it that secure. If it were my device I would write it down and it would be extensive." \\
&  MULTIPLE QUADRANTS & 40 & Used a variety of quadrants, not necessarily all & "I tried to use multiple squares more than once to make it more secure."\\
&  HARD TO GUESS &38& Chose a code that is difficult to guess&"Something I didn't think anyone could guess."\\
&  DIFFERENT&37& Using a different code than the first one &"I needed it to be drastically different then [sic] the first code."\\
\midrule
\multirow{10}{*}{$\mathsf{Memorability}$} &  PATTERN & 104 & Visualized a sequence or pattern  & "Not overly random but three blocks of two patterns"\\
&  SIMPLE & 100 &Used simple methods  & "I used something that wasn't to [sic] complicated" \\
&  EASY TO REMEMBER & 77 &Focused on overall memorability & "Something easy for me to remember but hard for someone else" \\
& DIRECTIONAL& 76 & Went in a specific sequence or order & "I used a specific direction as my way to remember like opening a box or lifting a lid."\\
&  REPEATED &55& Tapped  same quads multiple times  & "I started at the top left quadrant and went clockwise."\\
&  PERSONAL &52 &Associated code with something personal  & "I assigned numbers to the quadrants and input a date I'd remember."\\
&  NONE& 51&Had no strategy &"Didn't use one."\\
&  VARIATION& 40& Altered previous codes  &"I used a combination that was similar to my other code but with a Twist."\\
& SHAPE& 38 &Followed a specific shape & "I patterned it off of a shape I would remember. In this case it was an underlined x."\\
&  GAME & 18 & Used or made a game out of the sequence &"I tried to imagine a song pattern like Simon says."\\
\midrule
\multirow{6}{*}{$\mathsf{Like}$} &  EASY&75 &Found usability to be simple/straightforward& "Simple to input doesn't need much screen confirmation."\\
&  HARD TO GUESS&42& Considered it a complex authentication & "I like how you can switch the codes up to many different patterns. It really makes it harder for people to guess what it is."\\
&  DISCREET & 40&   Liked that it was/can be hidden and discrete& "You can be surreptitious and lock or unlock things without seeming like you are."\\
&  QUICK &39 & Found it to be efficient and quick &"It seems very convenient it can be quick and it gets old typing in my pin so much."\\
&  FUN &32& Found it fun to use& "I like that they are unique and I like entering them it is enjoyable."\\
\midrule
\multirow{7}{*}{$\mathsf{Dislike}$} &  HARD TO REMEMBER &124& Found it difficult to recall codes& "It's seems hard to remember the different patterns"\\
&  INSECURE &90& Found it to be a less complex authentication& "Same thing as a pin without the numbers and with less combination possibilities."\\
&  HARD TO TYPE &19& Found it difficult to input& "I could easily forget or tap the wrong location especially if there is no grid. Also it doesnt seem as fast as using a pattern to unlock like I currently do.."\\
&  NONE&16& Had no issues & "Can't think of anything I overly dislike."\\
&  NOT AN IMPROVEMENT&7& Considered it not better than other existing authentication methods & "There is absolutely no reason to use them for me or most people. They are hard to remember and not any different from a pin code."\\
\bottomrule
  \end{tabular}}

  \smallskip
  
$\mathsf{Security}$: What strategy did you use to make your code {\bf more memorable}? \\
$\mathsf{Memorability}$: What strategy did you use to make your code {\bf more secure}? \\
$\mathsf{Like}$: What are some aspects you {\bf like} about Knock Codes? \\
$\mathsf{Dislike}$: What are some aspects you {\bf do not like} about Knock Codes? \\
\end{center}
\vspace{-.2in}
\end{table*}

\input{appendix/likertfigure}



\end{document}